\documentclass{aa}  

\usepackage{graphicx}
\usepackage[dvipsnames]{xcolor}
\usepackage{txfonts}
\usepackage{color}
\usepackage{soul}

\begin{document}

   \title{The Fornax Deep Survey with VST.}

   \subtitle{VIII. Connecting the accretion history with the cluster density}

   \author{M. Spavone
          \inst{1}
          \and
          E. Iodice\inst{1,2}
        \and 
        G. van de Ven\inst{3}
        \and 
        J.~Falc\'on-Barroso\inst{4,5}
        \and 
        M.~A.~Raj\inst{1}
        \and 
        M.~Hilker\inst{2}
        \and 
        R.~P.~Peletier\inst{6}
        \and 
        M.~Capaccioli\inst{7}
        \and 
        S.~Mieske\inst{8}
        \and 
        A.~Venhola\inst{9}
        \and 
        N.~R.~Napolitano\inst{10}
        \and 
        M.~Cantiello\inst{11}
        \and 
        M.~Paolillo\inst{7,1,12}
        \and 
        P.~Schipani\inst{1}
          }

   \institute{INAF-Astronomical Observatory of Capodimonte,
 Salita Moiariello 16, 80131, Naples, Italy\\
              \email{marilena.spavone@inaf.it}
         \and
             European Southern Observatory, Karl-Schwarzschild-Strasse 2, D-85748 Garching bei Muenchen, Germany
             \and
             Department of Astrophysics, University of Vienna, Tuerkenschanzstrasse 17, A-1180 Vienna, Austria
             \and 
             Instituto de Astrof\'{\i}sica de Canarias, C/ V\'{\i}a L\'actea, s/n, E-38205, La Laguna, Tenerife, Spain
             \and
            Departamento de Astrof\'{\i}sica, Universidad de La Laguna, E-38200 La Laguna, Tenerife, Spain
             \and
             Kapteyn Astronomical Institute, University of Groningen, PO Box 72, 9700 AV Groningen, The Netherlands
             \and
             University of Naples ``Federico II'', C.U. Monte Sant'Angelo, Via Cinthia, 80126, Naples, Italy
             \and
             European Southern Observatory, Alonso de Cordova 3107, Vitacura, Santiago, Chile
             \and
             Division of Astronomy, Department of Physics, University of Oulu, Oulu, Finland
             \and 
             School of Physics and Astronomy, Sun Yat-sen University Zhuhai Campus, Daxue Road 2, 519082 -Tangjia, Zhuhai, Guangdong, P.R. China
             \and
             INAF-Astronomical Abruzzo Observatory, Via Maggini, 64100, Teramo, Italy
             \and
             INFN, Sezione di Napoli, Napoli 80126, Italy
             }

   \date{??; ??}

 \abstract{}{}{}{}{} 
 
  \abstract
   {This work is based on deep multi-band ({\it g, r, i}) data from the Fornax Deep Survey (FDS) with the VLT Survey Telescope (VST). We analyse the surface brightness profiles of the 19 bright Early Type Galaxies ($m_{B}\leq 15$ mag) inside the virial radius of the Fornax cluster ($R_{vir} \sim 0.7$ Mpc), in the mass range $8\times 10^{8} \leq M_{\ast} \leq 1.2\times 10^{11} M_{\odot}$.}
   {The main aim of this work is to identify signatures of accretion onto galaxies by studying the presence of outer stellar halos, and understand their nature and occurrence. Our analysis also provides a new and accurate estimate of the intra-cluster light inside the virial radius of Fornax.}
   {We performed multi-component fits to the azimuthally averaged surface brightness profiles available for all sample galaxies. This allows to quantify the relative weight of all components in the galaxy structure that contribute to the total light. In addition, we derived the average $g-i$ colours in each component identified by the fit, as well as the azimuthally averaged $g-i$ colour profiles, to correlate them with the stellar mass of each galaxy and the location inside the cluster.}
   {We find that in the most massive ($10^{10} \leq M \leq 10^{11}$~M$_\odot$) and reddest ETGs the fraction of light in, probably accreted, halos (50\%-90\%) is much larger than in the other galaxies. All of them are located in the high-density region of the cluster ($\leq 0.4 R_{vir}\sim 0.3$~Mpc), belonging to the north-south clump. Less-massive galaxies ($10^{9} \leq M \leq 10^{10}$~M$_\odot$) have an accreted mass fraction lower than 30\%, bluer colours and reside in the low-density regions of the cluster. The colour profiles of the ETGs with the largest accreted mass fraction tend to flatten in the galaxy's outskirts, i.e. beyond the transition radius from the central in-situ to the ex-situ accreted component. Inside the virial radius of the cluster ($\sim$ 0.7 Mpc), the total luminosity of the intra-cluster light, compared with the total luminosity of all cluster members (bright galaxies as well as dwarfs), is about 34\%.}
   {Inside the Fornax cluster there is a clear correlation 
between the amount of accreted material in the stellar halos of galaxies and the 
density of the environment in which those galaxies reside. 
By comparing this quantity with theoretical predictions and previous 
observational estimates, there is a clear indication that the driving factor 
for the accretion process is the total stellar mass of the galaxy, in
agreement with the hierarchical accretion scenario. Massive galaxies in the north-south clump, with the largest accreted mass fractions,
went through pre-processing in a group environment before this group merged with the main cluster early on.
At the present epoch of the Fornax assembly history, they 
are the major contribution to the stellar density in the core of the cluster. }


   \maketitle
%

\section{Introduction}

Exploring the low surface brightness (LSB) universe is one of the most challenging tasks of the era of the deep imaging and spectroscopic surveys. 
It is a crucial ingredient to map the mass assembly of galaxies at all scales (from galaxies to clusters) and all environments (in the groups of galaxies as well as in rich clusters), and to constrain their evolution within the Lambda-Cold Dark Matter paradigm. 
Clusters of galaxies are expected to grow over time by accreting smaller groups. During the infall process, the material stripped from the galaxy outskirts builds up the intra-cluster light \citep[ICL; see][]{Napolitano2003,Contini2014,Cui2014,Pillepich2018,DeMaio2018,Contini2019,Henden2019}. 
The ICL is the fossil record of all past interactions and mergers. It is a diffuse and typically very faint component ($\mu_g \geq 28$~mag/arcsec$^2$) that grows over time with the infalling of galaxies 
in the potential well of the brightest cluster galaxy (BCG, \citealt{Mihos2015}). 
The imprint of mass assembly in the BGGs resides in their extended stellar halos. 
This is an extended \citep[$\geq 100$ ~kpc;][]{Pillepich2018} and faint ($\mu_g \geq 26 - 27$ 
mag/arcsec$^2$) component made of stars stripped from satellite galaxies, in the form of streams and 
tidal tails, with multiple stellar components and complex kinematics \citep[see][for reviews]{Duc2017,Mihos2017}.

Semi-analytic models combined with cosmological simulations give detailed predictions 
about the structure and stellar populations of stellar halos, the ICL formation and 
the amount of substructures in various kinds of environment \citep{Oser2010,Cooper2013,Cooper2015,Cook2016,Pillepich2018,Monachesi2019}. 
New sets of simulations have recently been able to reproduce the 
faint features in galaxy outskirts at levels comparable to the deep observations
\citep[i.e. 29-33 mag/arcsec$^2$,][]{Pop2018,Mancillas2019}.

In the last two decades, deep imaging surveys enabled a huge progress in the study of 
the mass assembly in different types of environments, by providing an extensive analysis 
of the light and colour distribution of galaxies out to the regions of stellar halos and 
intra-cluster space 
\citep{Ferrarese2012,vanDokkum2014,Duc2015,Munoz2015,Capaccioli2015,Trujillo2016,Merritt2016,Mihos2017, Iodice2016}.
Investigations of mass assembly in the galaxies' outskirts have been conducted also by means of stellar 
kinematics and population properties \citep[e.g.][]{Coccato2010, Coccato2011, Ma2014, Barbosa2018, 
Veale2018, Hilker2018, Greene2019} and kinematics of discrete tracers like globular clusters (GCs) and planetary 
nebulae (PNe) \citep[e.g.][]{Coccato2013, Longobardi2013, Spiniello2018, Hartke2018, Prole2019, Napolitano2002, Pota2018}.
The main goal of the works cited above is to provide a set of observables that can be directly compared with theoretical predictions. 
For that deep and large-scale multi-band imaging is required to {\it i)} detect and characterise the 
substructures (like tidal tails or streams) in the stellar halos, through morphology and colours; 
{\it ii)} estimate the total accreted stellar mass by fitting the light distribution; 
{\it iii)} estimate the total luminosity of the ICL in the galaxy environment, i.e. group or cluster. 
Furthermore, deep spectroscopy in the galaxy outskirts is required to accurately constrain the age and metallicity 
of the stellar halos. 


Clusters and groups of galaxies are excellent sites to study the mass assembly 
and the build up of the ICL. 
The extensive multi-wavelength observations available for the Fornax galaxy cluster made 
this target one of the best environments where galaxy structure, stellar population and LSB 
features, including the ICL, can be studied in great detail and used to trace the assembly history of 
the cluster \citep[see][]{Iodice2019,Iodice2019a}.
Fornax is the second most massive galaxy concentration within 20 Mpc, after the Virgo 
cluster, and has a virial mass of $M_{\rm vir}= 7 \times 10^{13}$~M$_\odot$ \citep{Drinkwater2001}. 

In the optical wavelength range, HST data \citep{Jordan2007}, DECam data from the Next Generation Fornax Cluster Survey \citep{Munoz2015}, and 
the Fornax Deep Survey (FDS) with the VLT Survey Telescope \citep[VST,][]{Venhola2017, Iodice2016} are 
the deepest and widest datasets mapping the Fornax cluster out to the virial radius 
\citep[$R_{vir} \sim 0.7$ Mpc,][]{Drinkwater2001}. 
With FDS we {\it i)} mapped the surface brightness around the BCGs NGC~1399 and NGC~1316 out to 
an unprecedented distance of about $\sim 200$~kpc ($R\sim6R_e$) and down to $\mu_g \simeq 29-31$ 
mag/arcsec$^2$  \citep{Iodice2016,Iodice2017}; 
{\it ii)} traced the spatial distribution of candidate GCs inside a large fraction of the cluster core \citep{Dabrusco2016,Cantiello2018};
{\it iii)} studied the galaxy outskirts, detected ICL and faint ($\mu_g \simeq28-30$ mag/arcsec$^2$) features in the intra-cluster region in the core of the cluster \citep{Iodice2016, Iodice2017b, Iodice2019, Raj2019} and in the outskirts of NGC~1316 \citep{Iodice2017}; 
{\it iv)} provided the largest size and magnitude limited catalogue of dwarf galaxies in the cluster \citep{Venhola2017,Venhola2018}.

The deep integral-field spectroscopic observations, provided by the Fornax3D (F3D) project \citep{Sarzi2018,Iodice2019a}, acquired with MUSE at the VLT, give a complementary, unique and complete 
data-set for the brightest galaxies ($m_B \leq15$~mag) inside the virial radius of the Fornax cluster.
In addition, new data from ALMA \citep{Zabel2019}, neutral
hydrogen data from the MeerKAT survey \citep{Serra2016} and integral field observations of dwarfs from the SAMI Fornax cluster dwarf galaxy survey (Scott et al. submitted) will provide a complete census of the cool 
interstellar medium in Fornax. 

\begin{figure*}
    \centering
    \includegraphics[width=\hsize]{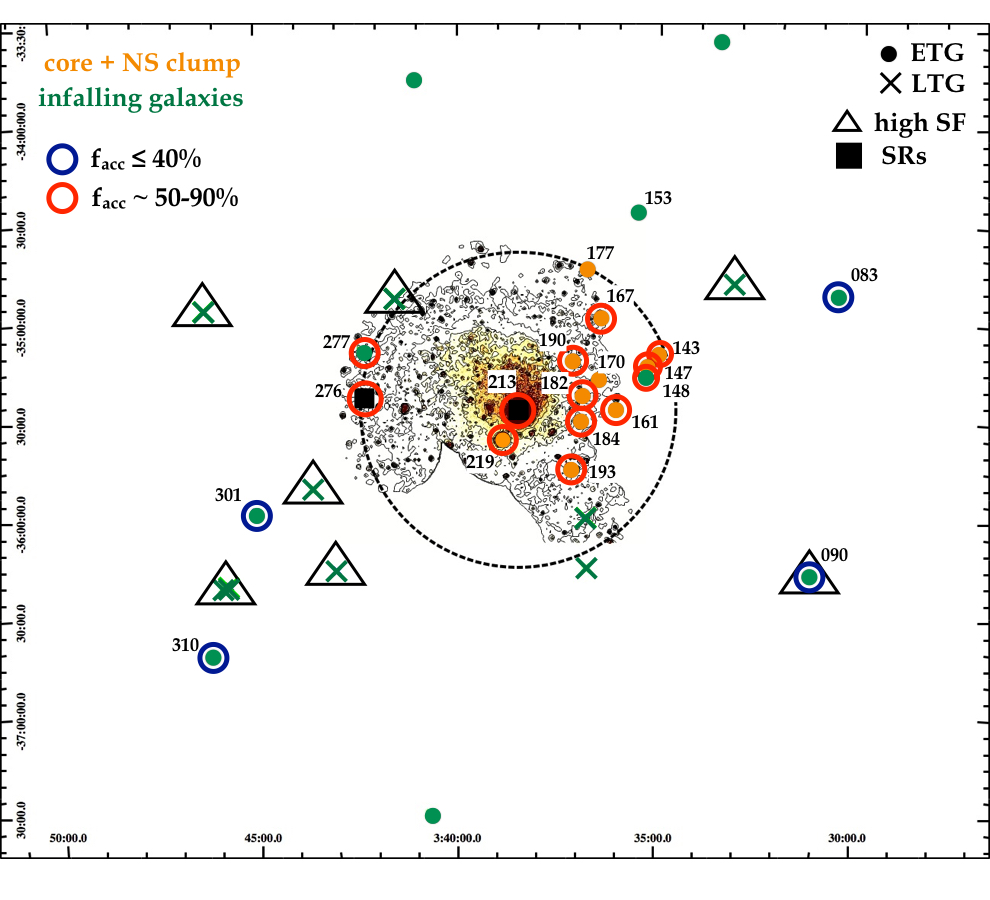}
    \caption{Distribution of the brightest ($m_B \leq 16$~mag) ETGs (circles) and LTGs (crosses) of the Fornax cluster inside the virial radius, projected onto the sky plane. All ETGs studied in this paper are labelled with their FCC number (see Tab.~\ref{tabfit3comp}). The right ascension and declination (J2000.0) are given in degrees on the horizontal and vertical axes of the field of view, respectively. The background image (and contours) is the X-ray emission in the energy range 0.4--1.3 KeV as measured by XMM-Newton \citep{Frank2013}. The dashed circle indicates the transition from the high-to-low density region of the cluster at 0.3 Mpc~$\simeq 0.4 R_{\rm vir}$. The orange and green symbols represent the galaxies residing in the North-South clump and the infalling galaxies, respectively (see text for details). Galaxies with high star formation (black open triangles) are also shown \citep{Iodice2019a}. The two slow rotators (SRs), FCC 213 and FCC 276, are marked with black squares. Open blue and red circles indicates the fraction of the accreted mass estimated in this work. For the three edge-on lenticular galaxies FCC153, FCC170, FCC177 the accreted fraction could not be determined}
    \label{fig:Fornax_groups}
\end{figure*}

The wealth of data available for the Fornax cluster has confirmed that it has a complex structure, 
indicative of continuing mass assembly, previously suggested by \citet{Drinkwater2001} and \citet{Scharf2005}.
By combining the structural properties of the galaxies
from FDS imaging with the spectroscopic data from F3D (i.e., morphology, colours,  
kinematics, and stellar population) in the two-dimensional projected phase-space, 
the cluster shows three well-defined groups of galaxies: the {\it core}, the {\it north-south clump} 
and the {\it infalling galaxies} (see Fig.~\ref{fig:Fornax_groups}),
apart from the southwest merging group centred on NGC~1316 \citep{Drinkwater2001}. 
Galaxies in each group have different light and colour distributions, 
kinematics, and stellar populations properties \citep{Iodice2019a}.\\
The core is dominated by the brightest and massive cluster member NGC~1399, 
which is one of only two slow-rotators inside the virial radius. 
This coincides with the peak of the X-ray emission \citep{Paolillo2002}.\\
The galaxies in the NS-clump are the reddest and most metal-rich galaxies of the sample, with
stellar masses in the range $0.3-9.8 \times 10^{10}$~M$_{\odot}$, and resides in the high-density 
region of the cluster (at a cluster-centre distance of $R_{\rm proj} \leq 0.4 R_{vir} \sim 0.3$~Mpc), 
where the X-ray emission is still detected (see Fig.~\ref{fig:Fornax_groups}).
The bulk of the gravitational interactions between galaxies 
takes place in this region, where the intra-cluster baryons (i.e. diffuse light, GCs, and PNe) are found \citep{Cantiello2018, Spiniello2018, Iodice2019}.
All galaxies in the NS-clump are fast-rotators ETGs,
with many of them showing distinct nuclear components and
kinematically decoupled cores, and two out of a total of three showing ionised-gas emission in
their centres. On average, galaxies populating this group show the largest differences 
between kinematic and photometric position angles. The stellar populations in the outskirts of
galaxies in this clump have lower metallicity than the bright central regions, 
which is an indication that the mass assembly of metal-poor satellites has continued longer in the outskirts.
\\
The infalling galaxies  are  distributed nearly symmetrically around the core, 
in the low-density region of the cluster ($R_{\rm proj} \geq 0.4 R_{vir} \sim 0.3$~Mpc).
The majority are late-type galaxies (LTGs) with ongoing star formation and signs of interaction with the environment and/or 
minor merging events, in the form of tidal tails and disturbed molecular gas \citep{Zabel2019, Raj2019}.
In this region, galaxies have on average lower [$M$/H] and [Mg/Fe] with respect to galaxies in the NS-clump \citep{Iodice2019a}.

By exploiting even further the FDS images, in this work we add a new piece to the puzzle built above: 
we study the light and colour distribution
of all brightest ETGs inside the viral radius of the Fornax cluster in order to constrain the 
accreted mass fraction as function of cluster density.

Theoretical predictions suggest that the brightest ETGs at the centre of groups and clusters are 
made by an inner stellar component formed {\it in-situ} and an accreted {\it ex-situ} 
component, which grows during the mass assembly.
The {\it ex-situ} component is made up by a relaxed component, which is completely merged with the 
{\it in-situ} component, and by an un-relaxed component, which is the outer stellar envelope. 
Simulations show that in the surface-brightness radial profile of simulated galaxies 
there is evidence of {\it inflection} in the region of the stellar halos, corresponding 
to variations in the ratio between the accreted relaxed and the accreted 
un-relaxed components \citep{Cooper2010,Deason2013,Amorisco2017}. 
This distance from the galaxy centre where the inflection occurs 
is the {\it transition radius} ($R_{\rm tr}$) used to 
characterise stellar halos. Massive galaxies with a high accreted mass fraction have small 
$R_{\rm tr}$ \citep{Cooper2010,Cooper2013}. 
The un-relaxed component of the stellar envelope appears as a change in the 
slope at larger radii of the surface brightness profiles.
In this context, the study of the surface brightness profiles of galaxies at the faintest 
levels is potentially one of the main ``tools'' to quantify the contribution of the accreted mass, 
which becomes particularly efficient when the outer stellar envelope starts 
to be dominant beyond the transition radius
\citep{Iodice2016,Iodice2017b, Spavone2017, Spavone2018, Iodice2019b}.
To this aim, in this paper we used the {\it r}-band azimuthally averaged surface
brightness profiles of ETGs inside the virial radius of the Fornax cluster \citep{Iodice2019}, 
to perform a multi-component fit to estimate the relative contribution of the accreted component with respect to that formed in-situ.

\section{Data}\label{sec:data}

The data used in this work are part of the Fornax Deep Survey (FDS),
which is a joint project based on Guaranteed Time Observation of the
{\it FOCUS} (P.I. R. Peletier) and {\it VEGAS} (P.I. E. Iodice,
\citep{Capaccioli2015}) surveys.
 FDS data consist of exposures in {\it u}, {\it g}, 
{\it r} and {\it i} bands obtained with OmegaCAM at the VLT Survey Telescope \citep[VST][]{Schipani2012,Kuijken2011}), 
located at the European Southern Observatory in Cerro Paranal (Chile).
Observations were acquired in visitor mode, in dark time for the {\it u}, 
{\it g} and {\it r} bands and in grey time for the {\it i} band. 
FDS covers a 26 square degrees of the Fornax cluster, out to the virial radius
\citep[$R_{vir}\sim 0.7$~Mpc,][]{Drinkwater2001}.
The FDS observations, data quality and reduction were extensively described in 
the published FDS papers from 
\citet{Iodice2016,Iodice2017,Iodice2019, Venhola2017, Venhola2018}.
In this work we analysed the bright ($m_B\leq 15$~mag) ETGs presented by \citet{Iodice2019},
inside the virial radius of cluster, covered by the inner 9
square degrees around the core. \citet{Iodice2019} performed the surface photometry of all ETGs in this sample, providing total magnitudes, effective radii, integrated ($g-i$ and $g-r$) colors and mass-to-light ratios for all galaxies.

\subsection{Correction for the PSF}\label{sec:psf}

The faint outskirts of galaxies can be contaminated by light scattered from the bright core by
the telescope and the atmosphere, which creates artificial
halos. In order to account for this effect, the point spread function (PSF) must be
mapped out to a radial distance that is at least comparable to the extent of the
galaxian halo \citep{Capaccioli1983}. For VST we derived an extended PSF as described in
Appendix B of \citet{Capaccioli2015}. 

As shown in \citet{Capaccioli2015}, the effect of the scattered light is very different
at the same surface brightness level between galaxies with large and small
angular extent, and which are more or less flattened. In particular, it as been found 
that the surface brightness profiles of smallest galaxies were
marginally affected ($\sim$ 0.2 mag) by the extended PSF at
$\mu_{g} \sim\ 29$ mag/arcsec$^{2}$, while galaxies with a large angular
extent and no bright central nuclei were not significantly affected ($\sim0.05$ mag).

In this work, due to presence of both small and very elongated galaxies in
the studied sample, the amount of scattered light and its effect on
the surface brightness profiles needs to be estimated. To this aim, we
deconvolved each galaxy in our sample for the VST PSF shown in
\citet{Capaccioli2015} , by using the Lucy-Richardson (hereafter RL) algorithm \citep{Lucy1974,
Richardson1972}. 

\begin{figure*}
\centering
\includegraphics[width=18cm, angle=0]{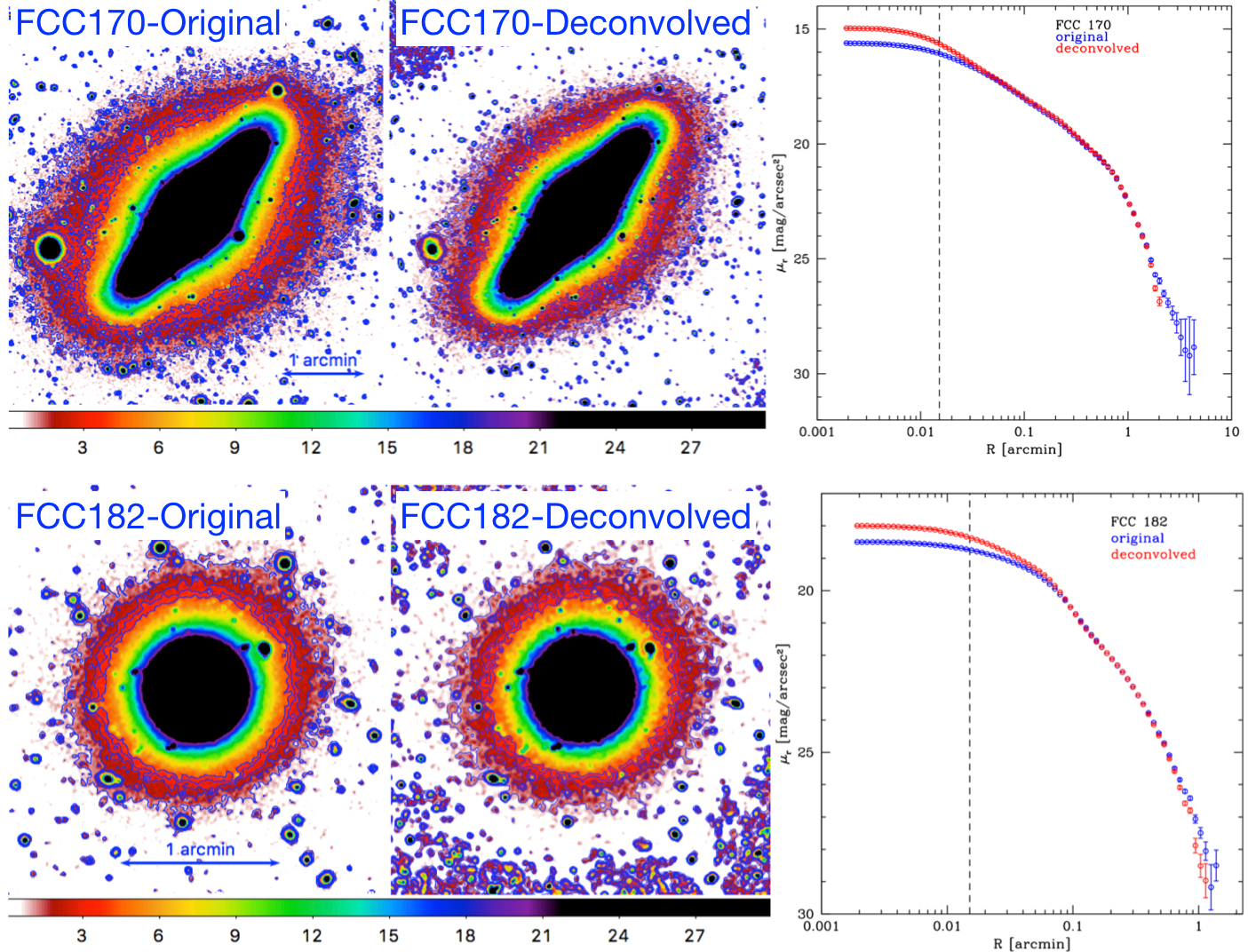}
\caption{{\it Top panel}: Original (left) and deconvolved (middle) images of
  the flattest galaxy in our sample, FCC170. In the right panel we plot the original (blue) and deconvolved (red) surface brightness profiles of FCC170. {\it Bottom panel}: Original (left) and deconvolved (middle) images of
  the smallest galaxy in our sample, FCC182. In the right panel we plot the original (blue) and deconvolved (red) surface brightness profiles of FCC182.}\label{fig:psf}

\end{figure*}

As expected, the flatter is the galaxy, the greater is the effect of
the PSF deconvolution on the surface brightness profile. Moreover,
smaller round galaxies are also more affected than larger ones. The effect of the PSF on the distribution of light from the
galaxy is particularly relevant below a surface brightness of 27 mag/arcsec$^{2}$ (r-band),
but it is however not greater than 0.2 mag for all the analysed galaxies. In Fig. \ref{fig:psf} we show an
example of the result of the deconvolution on the flattest (top panels)
and the smallest (bottom panels) galaxy in our sample. 

The effect of the PSF on the surface brightness profiles of the galaxies
is illustrated in the rightmost panels of Fig. \ref{fig:psf}. Beyond 27 mag/arcsec$^{2}$
the deconvolved profiles start to deviate from the original ones. At radial distances greater
than 1 arcmin, the difference between the original and the PSF-corrected profiles is around 0.1 mag/arcsec$^{2}$.

\begin{figure}
\includegraphics[width=9cm, angle=0]{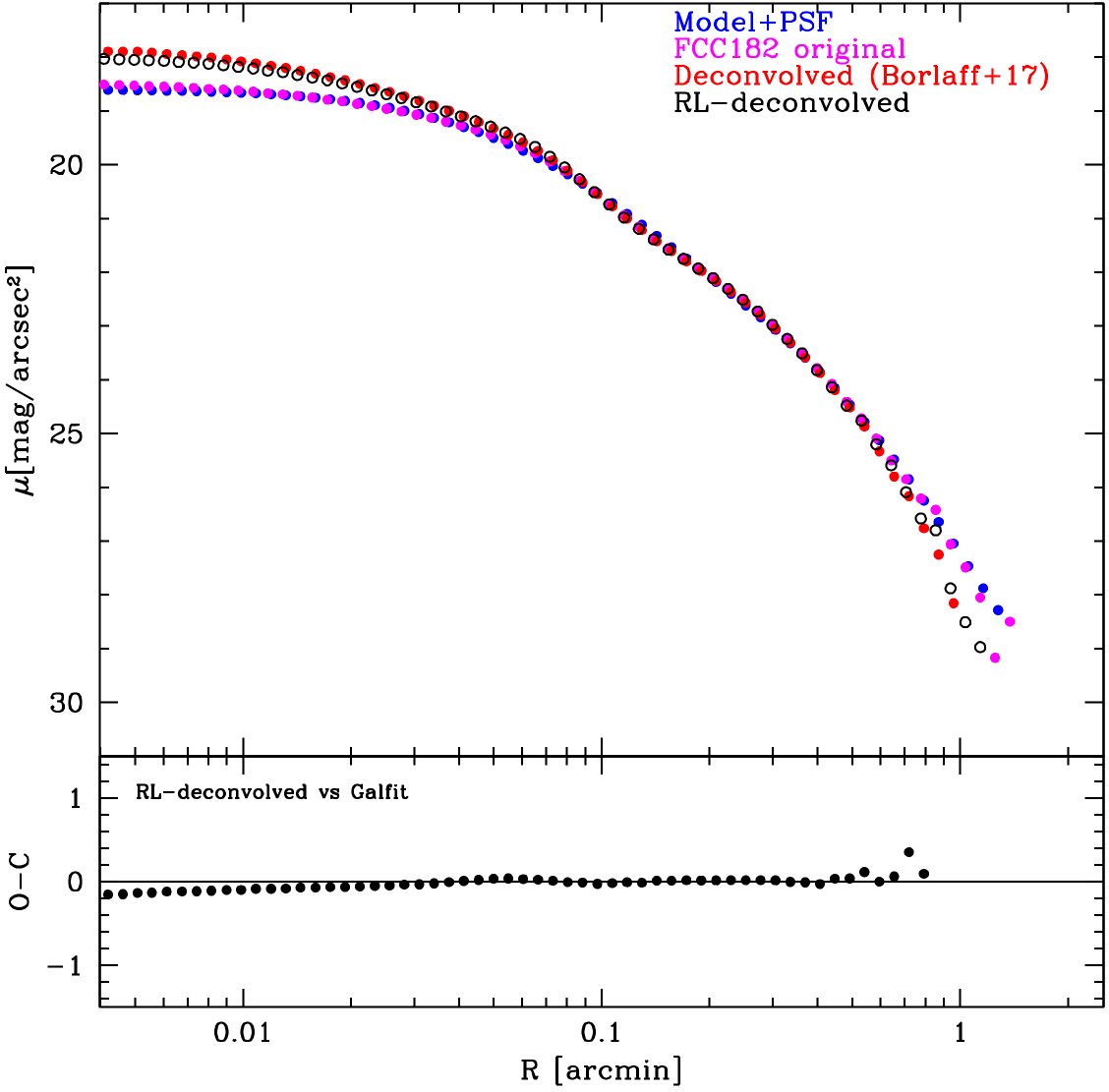}
\caption{{\it Top panel}: Deconvolved SB profile using RL algorithm (black) and the method by \citet{Borlaff2017} (red) over-plotted on the original (magenta) and the PSF-convolved model (blue) profiles for FCC 182. {\it Bottom panel}: residuals between the profile deconvolved with the RL algorithm and the one deconvolved with the technique by \citet{Borlaff2017}.}\label{borlaff}

\end{figure}

In order to test the robustness and the reliability of our deconvolution technique, we compare the performance of the RL algorithm with the method from \citet{Borlaff2017}. These authors estimate the deconvolved image by using the following operations:
\begin{equation}
    Residuals = Image\ raw - PSF \ast Model_{GALFIT}
\end{equation}
\begin{equation}
    Deconvolved\ Image = Model_{GALFIT} + Residuals
\end{equation}

where $PSF \ast Model_{GALFIT}$ is the 2D model (with best-fitted parameters) convolved with the adopted PSF, obtained from GALFIT3.0 \citep{Peng2010}, and $Model_{GALFIT}$ is the 2D model obtained using the best fit parameters of the galaxy from $PSF \ast Model_{GALFIT}$, that is, the model of the galaxy without PSF convolution.
We applied this technique to galaxies of different sizes. Here we show the results of this test for one of the smaller galaxies in our sample, FCC 182,  since the effect of the deconvolution is much more evident for galaxies with small angular dimensions. The result of this comparison is shown in Fig. \ref{borlaff}, from which it is clear that our technique of deconvolution (RL algorithm) is fully consistent with the above method. The same test has been also performed by \citet{Raj2019} for LTGs in FDS, achieving similar result.

\section{Multi-component fit of the light distribution}\label{sec:fit}

In the last decade, a big effort was made to model the surface
brightness profiles of the brightest cluster/group members in order to estimate 
the contribution of the outer envelope to the total light. In this context,
there are several observational and theoretical papers in literature
showing that this component appears as an extended additional exponential-like profile to the
azimuthally averaged surface brightness distribution of many ETGs 
 \citep{Seigar2007, Donzelli2011, Arnaboldi2012, Iodice2016, Spavone2017,
  Spavone2018}. 

\citet{Spavone2017}, following the predictions of numerical simulations
\citep{Cooper2013,Cooper2015}, described the surface brightness profiles
of a sample of ETGs as the superposition of different components: an inner S{\'e}rsic profile
representing the (sub-dominant) in situ component in the central regions,
another S{\'e}rsic profile \citep{Sersic, Caon1993} representing the (dominant) superposition of the
relaxed, phase-mixed accreted components, and an outer diffuse component (the envelope)
representing unrelaxed accreted material (streams and other coherent
concentrations of debris). The latter component does not contribute any significant surface
density to the brighter regions of the galaxy. In order to mitigate
the degeneracy between the parameters, \citet{Spavone2017} fixed $n \sim
2$ for the in-situ component of their three-component fits, following
the predictions of \citet{Cooper2013} for massive galaxies. 

As explained in \citet{Spavone2017}, since we are interested in the study of the stellar distribution in the outer envelopes of our sample galaxies, we do not use the $\chi^{2}$ statistics in our fitting procedure. We adopt the same approach described by \citet{Seigar2007} and perform least-square fits using a Levenberg–Marquardt algorithm, in which the function to be minimised is the rms scatter, defined as $\Delta=\sqrt{ \frac{\sum_{i=1}^{m}
    \delta_{i}^{2}}{m}}$, where $m$ is the number of data points
and $\delta_{i}$ is the {\it i}th residual. In all the fit presented
below, the innermost seeing-dominated regions, indicated with
dashed lines, were excluded. For further details on the adopted fitting procedure see \citet{Spavone2017}.

Since the ETGs in our sample are in the mass range $8\times 10^{8} - 1.2\times 10^{11}$~M$_{\odot}$, which is partly covered by the available numerical simulations ($10^{10} < M_{\ast} \leq 10^{13}$~M$_{\odot}$), the surface brightness profiles have  been fitted with multi component models adopting the fitting technique presented in \citet{Spavone2017}, with the only difference that here we do not fix the S{\'e}rsic index for the first component of the fit, since we have no theoretical predictions about this index for less massive galaxies. 

Eleven out of nineteen galaxies in the studied sample have masses in the mass range covered by simulations by \citet{Cooper2013}. Six of them already have a S{\'e}rsic index for the innermost component in the range 1.5-2.5, as predicted by simulations. For the remaining five galaxies we checked how much their accreted mass fractions are affected by fixing $n_{1}=2 \pm 0.5$. We found that the accreted mass fractions change of 1\%-2\%, while the rms scatter increase of 5\%-20\%. So, leaving $n_{1}$ free reduces the scatter without biasing our results.
Results of the fit are shown in Fig. \ref{fig:fit_all}. 
The best-fitting parameters are reported in Tab. \ref{tabfit3comp}.

\begin{figure*}
\hspace{-0.cm}
 \includegraphics[width=8.8cm, angle=-0]{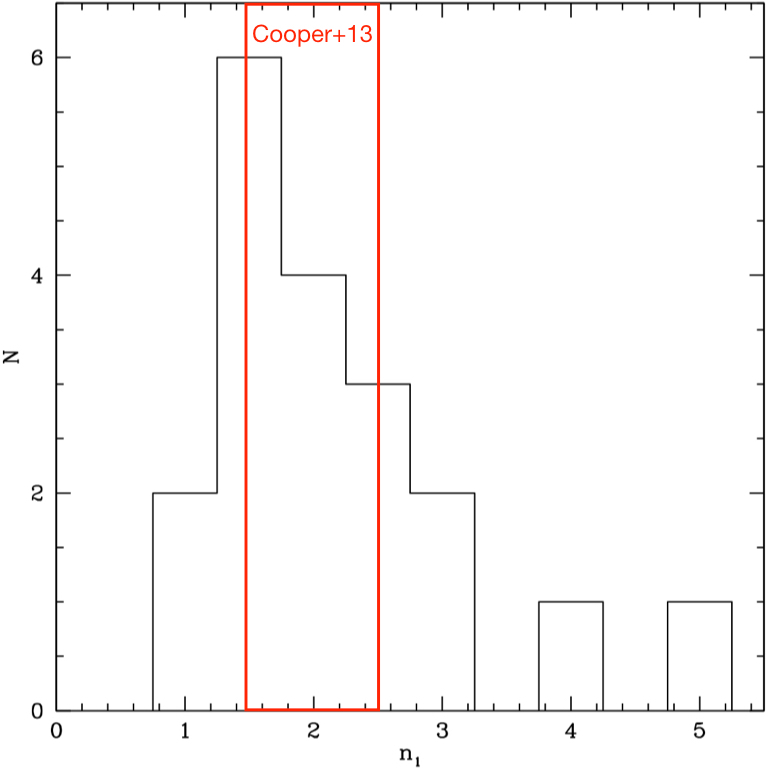}
  \includegraphics[width=9cm, angle=-0]{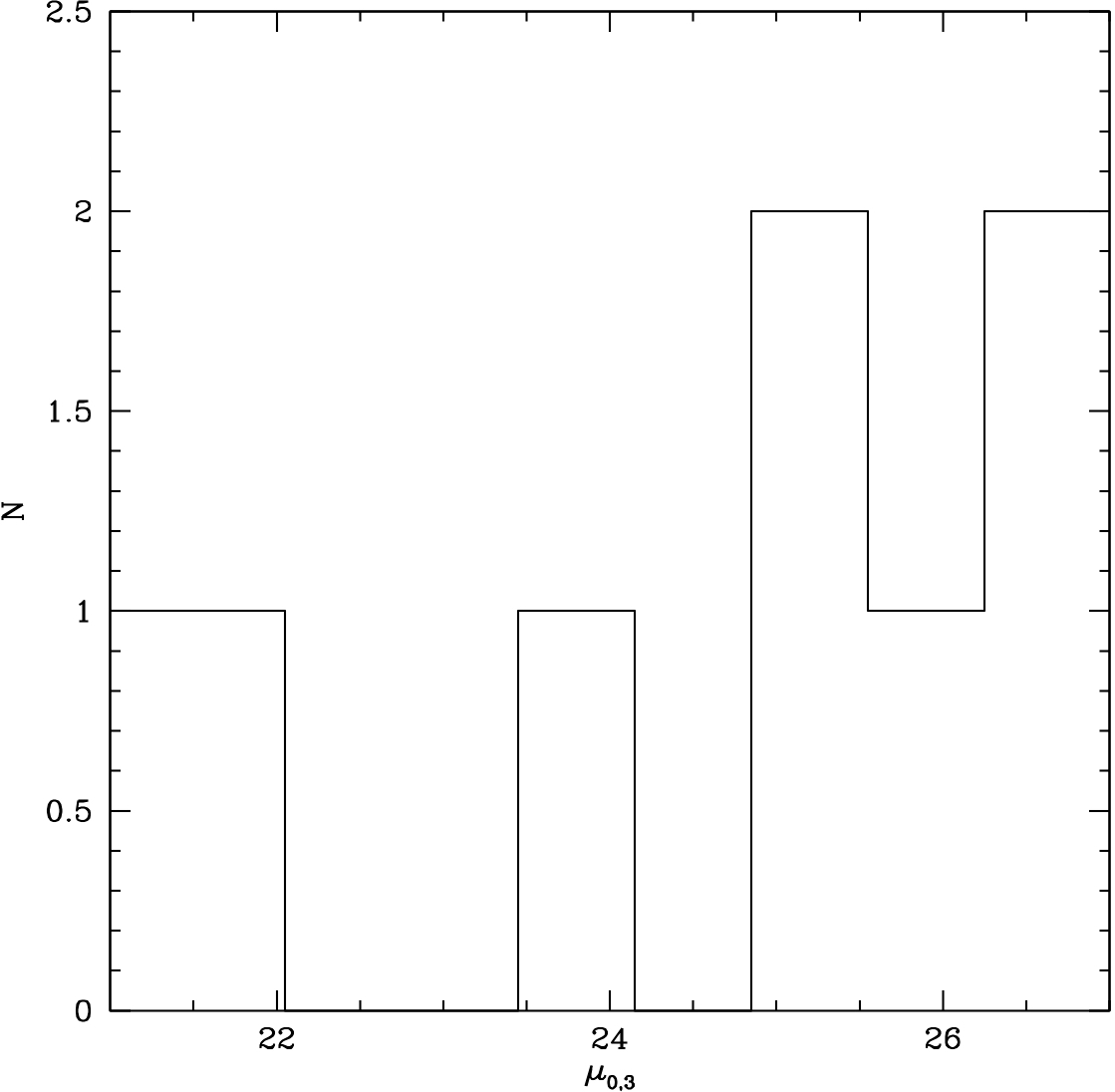}
\caption{{\it Left panel} - Distribution of the S{\'e}rsic index of the innermost
  component of our fit for the 19 analysed galaxies. The red region brackets the range of $n_{1}$ in the simulations by \citet{Cooper2013}. {\it Right panel} - Distribution of the central surface brightness of the 8 galaxies in the sample having a third exponential component ($\mu_{0,3}$).}\label{M_n1}
\end{figure*}

Although most of the galaxies analysed in this work are 
less massive than those studied in simulations, 
it is worth noticing that the
S{\'e}rsic index for the in-situ component for most of them, falls in
the range predicted by \citet{Cooper2013} ($n = 2\pm0.5$), as shown in the left panel of
Fig. \ref{M_n1}. In the right panel of the same figure we plot the distribution of the central surface brightness ($\mu_{0,3}$) of the third component of the fit for galaxies which have been fitted with a third exponential component. For most of them, $\mu_{0,3}$ is fainter than 25 mag/arcec$^{2}$ and the average value of the scale length $r_{h,3}$ is about 165 arcsec. By comparing these values with the typical ones obtained for a large sample of disk galaxies from the SDSS data, we found that in the $\mu_{0}-r_{h}$ plane they are all located outside the region where typical disks are found, with the only exceptions of FCC190 and FCC167 (see Figure 1 in \citet{vdKruit2011} and references therein). This suggest these are different components and they are more similar to extended halos that are found to be exponential.

\begin{table*}
\setlength{\tabcolsep}{0.8pt}
\tiny 
\caption{Best-fitting structural parameters for the 19 bright ETGs inside the virial radius of the Fornax cluster considered in this work.} \label{tabfit3comp}
\vspace{12pt}
\begin{tabular}{lccccccccccccccccc}
\hline\hline
Object & $\mu_{e1}$ &$r_{e1}$&$n_{1}$& $\mu_{e2}$ &$r_{e2}$&$n_{2}$&$\mu_{0,3}$&$r_{h,3}$&$f_{h,T}$&log $M_{\ast}$&$R_{tr1}$&$R_{tr2}$&$R_{e,r}$&$M_{r}$&{\it (g-i)}\\ 
    & [mag/arcsec$^{2}$] &[arcsec]& & [mag/arcsec$^{2}$] & [arcsec]&&[mag/arcsec$^{2}$] & [arcsec]& &[$M_{\odot}$]&[kpc]&[kpc]&[kpc]&[mag]&[mag]\\
  (1)  & (2) &(3)& (4) & (5) & (6)& (7) &(8) & (9)&(10) &(11)&(12)&(13)&(14)&(15)&(16)\\
\hline \vspace{-7pt}\\
FCC083  & 20.85$\pm$0.20 & 20.29$\pm$0.24  & 3.91$\pm$0.04&23.73$\pm$0.24&71.59$\pm$1.97&1.05$\pm$0.02&-& -&47\%$\pm$ 5\%&10.30$\pm$0.05&5.58&-&3.30&-20.56&1.3\\
FCC090  & 20.28$\pm$0.80 & 5.00$\pm$0.15  & 1.64$\pm$0.07&23.94$\pm$0.17&25.00$\pm$0.40&1.00$\pm$0.21& &-&0&8.91$\pm$0.02&1.36&-&1.14&-18.83&0.78\\
FCC143 & 18.95$\pm$0.05 & 3.00$\pm$0.90 &2.03$\pm$0.02&21.73$\pm$0.02&12.90$\pm$0.02&1.33$\pm$0.03& 26.53$\pm$0.07 &52.46$\pm$0.30&60\%$\pm$9\%&9.45$\pm$0.01&0.56&6.18&1.03&-18.77&0.97\\ 
FCC147 & 18.88$\pm$0.27 & 5.01$\pm$0.75  &3.09$\pm$0.44&21.44$\pm$0.01&33.77$\pm$0.70&2.69$\pm$0.24& - &-&81\%$\pm$7\%&10.38$\pm$0.03&0.38&-&2.36&-20.96&0.99\\
FCC148  & 19.63$\pm$0.55 & 5.03$\pm$0.49  & 3.11$\pm$1.05&20.98$\pm$0.61&25.27$\pm$2.59&0.96$\pm$0.21&24.88$\pm$1.11 &46.67$\pm$14.51&88\%$\pm$9\%&9.76$\pm$0.04&0.47&11&2.73&-19.79&0.86\\
FCC153  & 19.11$\pm$0.61 & 15.28$\pm$4.11& 1.04$\pm$0.14&22.23$\pm$0.11&40.68$\pm$3.68&0.50$\pm$0.40&- &-&-&9.88$\pm$0.01&4.29&-&2.00&-19.89&0.77\\ 
FCC161  & 19.66$\pm$0.54 & 5.00$\pm$0.29  & 5.00$\pm$0.31&20.84$\pm$0.58&26.65$\pm$3.94&1.58$\pm$0.55& 25.29$\pm$0.71&109.37$\pm$18.91&91\%$\pm$8\%&10.42$\pm$0.03&0.27&13.89&2.76&-21.02&1.04\\
FCC167  & 19.83$\pm$0.41 & 23.94$\pm$5.04  & 2.70$\pm$0.68&21.43$\pm$0.74&56.86$\pm$8.32&0.40$\pm$0.96& 21.43$\pm$0.81&44.55$\pm$14.72&63\%$\pm$7\%&10.99$\pm$0.05&4.32&12.33&5.80&-22.36&1.03\\
FCC170  & 18.58$\pm$0.53 & 6.22$\pm$1.94  & 2.73$\pm$1.19&20.36$\pm$0.82&27.63$\pm$3.23&0.93$\pm$0.35& - &-&-&10.35$\pm$0.02&0.86&-&1.69&-20.71&1.07\\
FCC177  & 19.06$\pm$0.19 & 2.00$\pm$0.26  & 1.35$\pm$0.45&20.63$\pm$0.02&27.04$\pm$0.12&0.86$\pm$0.08&- &-&-&9.93$\pm$0.02&0.21&-&3.48&-19.71&1.08\\
FCC182  & 19.77$\pm$0.16 & 3.00$\pm$0.62  & 0.92$\pm$0.10&22.21$\pm$0.22&12.54$\pm$0.66&1.25$\pm$0.14&- &-&65\%$\pm$5\%&9.18$\pm$0.02&0.47&-&0.94&-17.88&1.03\\
FCC184  & 18.05$\pm$0.57 & 5.19$\pm$0.58  & 1.91$\pm$0.39&21.27$\pm$0.34&37.02$\pm$7.18&1.53$\pm$0.90&25.98$\pm$0.26 &229.09$\pm$3.23&75\%$\pm$11\%&10.67$\pm$0.01&0.93&17.40&3.22&-21.43&1.14\\
FCC190  & 20.07$\pm$0.02 & 4.02$\pm$1.50  & 1.44$\pm$0.04&21.12$\pm$0.03&13.10$\pm$0.23&0.54$\pm$0.01& 21.32$\pm$0.02 &18.32$\pm$0.10&85\%$\pm$7\%&9.73$\pm$0.03&0.47&2.36&1.80&-19.28&1.04\\ 
FCC193  & 18.96$\pm$0.32 & 9.12$\pm$1.87  & 2.27$\pm$0.27&22.80$\pm$0.75&54.36$\pm$9.61&1.28$\pm$0.62&26.59$\pm$0.83 &387.26$\pm$33.98&60\%$\pm$6\%&10.52$\pm$0.04&2.82&20.35&2.90&-20.93&1.17\\
FCC219  & 18.46$\pm$0.03 &12.48$\pm$1.66  & 2.13$\pm$0.02&24.54$\pm$0.02&291.58$\pm$1.94&3.23$\pm$0.04&-&-&67\%$\pm$14\%&11.10$\pm$0.01&3.79&-&15.77&-22.95&0.96\\
FCC276  & 17.55$\pm$0.25& 3.00$\pm$0.29  & 1.85$\pm$0.24&20.87$\pm$0.23&32.81$\pm$1.10&1.75$\pm$0.26&23.68$\pm$0.31&99.45$\pm$2.02&87\%$\pm$4\%&10.26$\pm$0.03&0.46&12.54&4.24&-21.31&0.64\\
FCC277  & 20.14$\pm$0.15 & 7.64$\pm$0.32  & 1.66$\pm$0.13&21.88$\pm$0.16&20.88$\pm$0.23&1.38$\pm$0.17&- &-&60\%$\pm$8\%&9.53$\pm$0.01&1.04&-&1.28&-19.24&0.77\\
FCC301  & 19.15$\pm$0.10 & 5.85$\pm$0.11  & 1.30$\pm$0.07&23.56$\pm$0.11&26.53$\pm$2.15&0.60$\pm$0.07&- &-&26\%$\pm$1\%&9.30$\pm$0.05&1.80&-&1.12&-18.82&0.70\\
FCC310  & 21.12$\pm$0.23 & 18.62$\pm$2.26  & 1.62$\pm$0.22&25.71$\pm$0.93&61.41$\pm$5.15&0.20$\pm$0.3& -&-&14\%$\pm$6\%&9.73$\pm$0.02&7.99&-&3.43&-19.70&0.77\\

\hline
\end{tabular}

\tablefoot{Columns 2, 3, and 4 report effective surface brightness and effective
radius for the inner component of each fit. Columns
5, 6, 7 list the same parameters for the second component, whereas columns 8 and 9 list the
central surface brightness and scale length for the outer exponential component. Column 10 gives the accreted mass fraction derived from our fit. Column 11 reports the total stellar mass, while columns 12 and 13 report the transition radii derived by the intersection between the first and the second component of the fit and the second and the third component, respectively. In columns 14, 15 and 16 there are effective radii, absolute magnitudes in the {\it r} band and average {\it (g-i)} colours from \citet{Iodice2019}}
\end{table*}

\section{Results: the accreted mass fraction and ICL}\label{sec:results}

Following \citet{Spavone2017}, we used the multi-component fits to derive the relative contribution of the accreted component with respect
to the total galaxy light, $f_{h,T}$. 
The accreted mass fraction $f_{h,T}$ is the contribution to the total light of the phase-mixed accreted components plus the envelope (see Sec.~\ref{sec:fit}). 
 The $f_{h,T}$ values are given in Tab.~\ref{tabfit3comp} for all galaxies in the sample, except for FCC153, FCC170 and FCC177. These are the only three edge-on S0 galaxies of our sample. Kinematical studies show that these galaxies possess a bulge and a disk component (see e.g. \citealt{Pinna2019b, Pinna2019a, Bedregal2011, Iodice2019a}). For these objects, in the azimuthally averaged surface brightness profile the exponential halo would be completely superposed to the exponential disk, and so they would be impossible to disentangle with the photometry alone. For FCC090 the second component of the fit is a S{\'e}rsic with n=1, that is an exponential with a central surface brightness of 22.12 mag/arcsec$^{2}$, which is typical of disks. This component is reasonably mapping the exponential disk of the galaxy, classified as an S0, and not the accreted material. 

\begin{figure}
\hspace{-0.cm}
 \includegraphics[width=9cm, angle=-0]{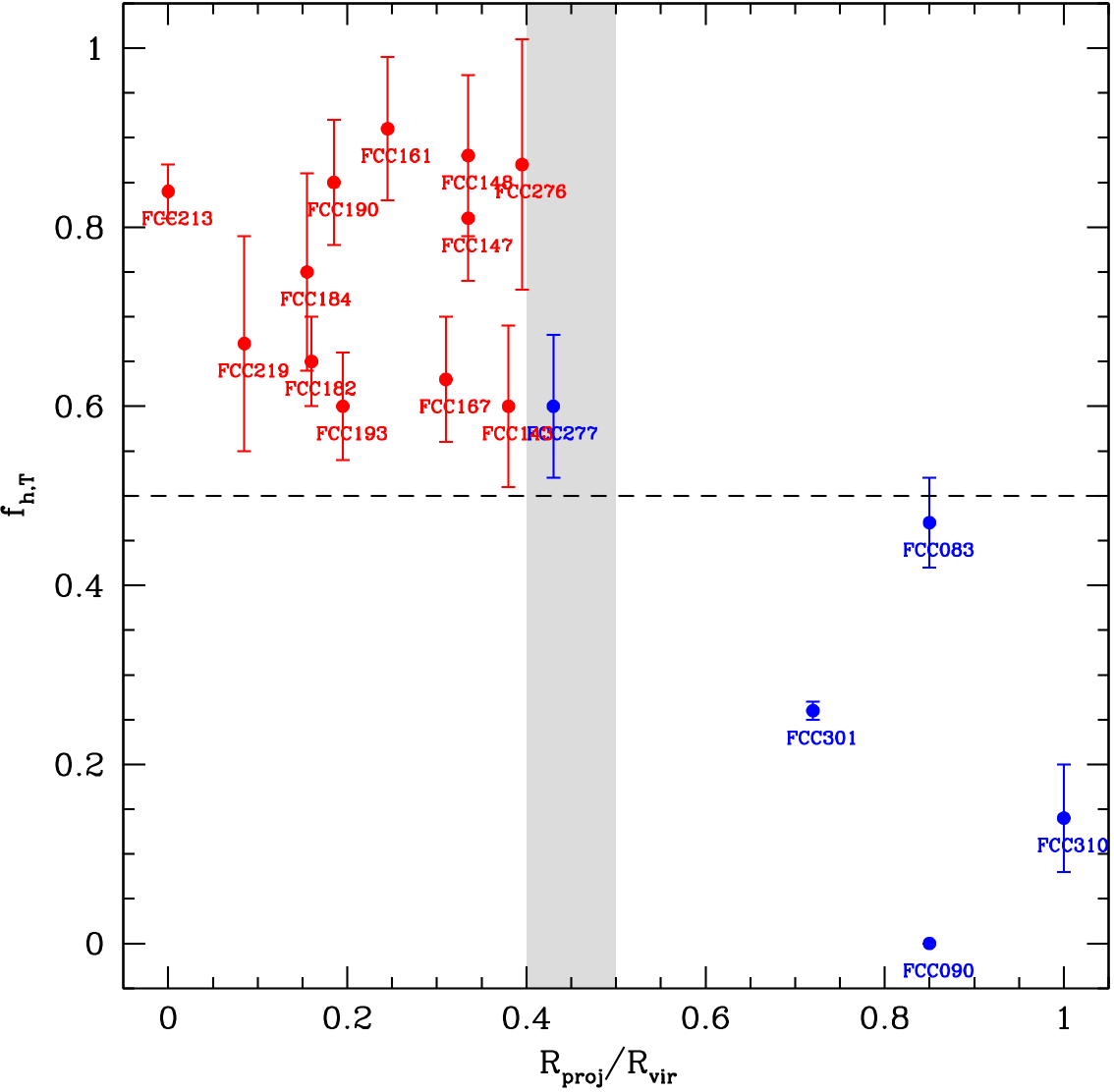}
 \includegraphics[width=9.2cm, angle=-0]{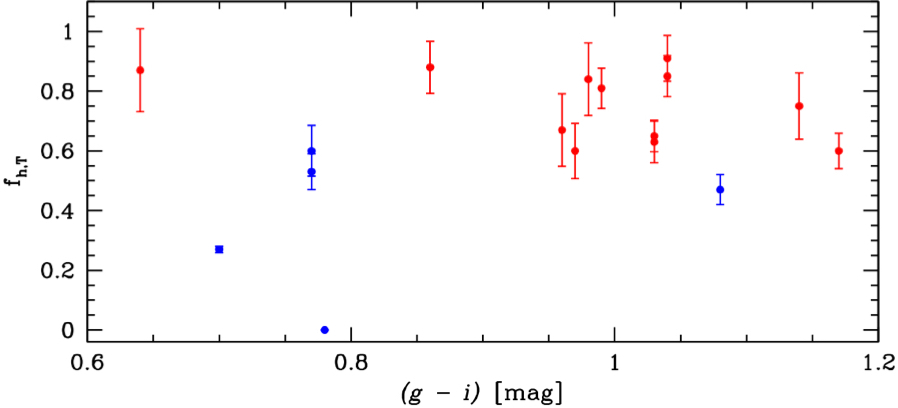}
\caption{{\it (Top panel)} - Accreted mass fraction as a function the projected distance from the cluster centre, normalised to the virial radius of the cluster ($R_{vir} = 0.7$ Mpc, \citealt{Drinkwater2001}). Red dots are for galaxies in the high density region of the cluster, while blue dots are for galaxies in the low density region.
The vertical grey region marks the range of cluster-centric
distance where a clear separation occurs: galaxies located inside
$\sim$ 0.8-1 degree from the core of the cluster show a higher outer envelope mass fraction. {\it (Bottom panel)} - Accreted mass fraction as a function of the {\it (g - i)} average colour for galaxies in the low (blues dots) and high (red dots) density regions of the cluster. On average, redder galaxies
  have higher accreted mass fractions. FCC~277 is in the transition region from high-to-low density region, with a higher accreted mass fraction 
  but comparable colours with galaxies in the low-density region. }\label{MvsDist}
\end{figure}

For all the other galaxies, we found that the total accreted mass fraction ranges between 27\% and 98\%. FCC~219 and FCC~213 are the two brightest members in the core centre, and they are so close in projection that their extended stellar envelopes are photometrically indistinguishable. As described in \citet{Iodice2019}, the surface photometry of FCC~219 was derived on the residual image where the 2D model of FCC~213 is subtracted off. Therefore, the stellar envelope that we expect to be present in FCC~219 was partly counted in the extended stellar envelope of FCC~213. As a consequence, according to the stellar mass of FCC~219, the accreted mass fraction we estimate for this galaxy should be considered as a lower limit. Anyway, the less extended stellar envelope in FCC~219 could be consistent with the scenario previously proposed  \citep{Forbes1998,Bekki2003,Sheardown2018}, where the outer halo of FCC~219 could have been stripped during a past interaction with FCC~213. This is also supported by the low specific frequency of its globular clusters (GCs) system, indicating that it has lost many GCs, and thus probably also outer stellar light, into the intracluster population around FCC~213.
In Fig. \ref{MvsDist} we plot the accreted mass fraction $f_{h,T}$ as a function of the projected
distance from the cluster centre ($R_{proj}/R_{vir}$, see Tab.1 in \citealt{Iodice2019a}). A clear separation occurs at a
cluster-centric distance of about 0.8-1 degree ($R_{proj}/R_{vir} \sim$ 0.4-0.5): galaxies located inside 1
degree have a higher mass fraction in the
outer envelope ($\geq$ 55\%) than those beyond this radius. 
This distance corresponds to the transition region from the high-to-low density region of the cluster (where $\Sigma \leq 40$~galaxies Mpc$^{-2}$), found by \citet{Iodice2019},  
at R $\sim 0.4 R_{vir} (\sim 0.3$~Mpc). As shown by \citet{Iodice2019}, at this radius a decrease in the X-ray emission also occurs. 

The majority of the massive ETGs, which are the reddest objects, are located in the high density region of the cluster (see Fig. 12 in \citealt{Iodice2019}), which have higher accreted mass fractions than the galaxies in the low-density regions.
Bottom panel of Fig. \ref{MvsDist} shows that, on average, the accreted mass fraction also correlates with   
{\it (g - i)} colour. 
We found that for redder galaxies the contribution of accreted stars to the total mass 
is higher. Two exceptions are FCC~276 and FCC~083. 
FCC~276 is the brightest elliptical galaxy on the eastern side of the cluster,
at $R_{\rm proj}=0.4 R_{vir}$ from FCC~213 (NGC~1399), in the transition region from high 
to low density. The location of this galaxy in the cluster reflects in its properties. 
FCC~276 has bluer colours ($g-r\sim0.3$~mag), comparable colours to the ETGs in the low-density region of the cluster, but a high accreted mass fraction ($f_{h,T}=0.87$), 
consistent with that typical for all the others ETGs in the high 
density region of the cluster, suggesting that this galaxy might be in a transition phase 
of mass assembly. In particular the 
bluer colours are related to a distinct kinematical component in the centre
($R\la10$~arcsec), which has $v_{\rm max}\sim50$~km~s$^{-1}$ and $\sigma\sim180$~km~s$^{-1}$
\citep{Carollo1997,Iodice2019,Iodice2019b}.\\
FCC~083 is located on the western side of the cluster at a distance $R_{\rm proj}=0.85 R_{vir}$
from the cluster centre. It is among the brightest and more massive ETGs in the low-density 
region of the cluster, with quite red colours ($g-r\sim0.67$~mag) and accreted mass fraction comparable with ETGs in the high density region of the cluster with similar stellar mass (see also Fig.~\ref{col_lowhigh}).

Fitting the light distribution of all brightest ETGs inside the virial radius 
of the Fornax cluster, we can provide a new and more accurate estimate of the ICL in this region of the cluster. The total amount of ICL includes the contributions from 
the stellar envelope in all galaxies ($R > R_{tr2}$), plus the luminosity of the patch of diffuse 
light previously detected in the core from \citet{Iodice2017} and of the stellar
bridge connecting NGC~1399 and FCC~184 \citep{Iodice2016}. In the {\it r} band, the total luminosity of the ICL is $(2.6\pm0.1) \times 10^{11} L_{\odot}$. 
Taking into account the total magnitude of all the bright galaxies inside the virial radius, from \citet{Iodice2019} and \citet{Raj2019}, as well as that of the dwarf 
galaxies from \citet{Venhola2017}, the total luminosity of the Fornax cluster is 
$L_{tot}=(7.6\pm 0.09) \times 10^{11} L_{\odot}$. Therefore, the total fraction of ICL inside 
the virial radius of the Fornax cluster is $M_{ICL}/M^{\ast}_{tot} \sim 0.34\pm 0.20$.
Since it is not easy to separate the ICL from the stars that are bound to the BCG, in literature the ratio $(M_{ICL}+M_{BCG})/M^{\ast}_{tot}$ is also used to measure the ICL fraction \citep[see e.g.][]{Gonzalez2005}. By adopting this approach, for the Fornax cluster we obtain a fraction of $(M_{ICL}+M_{BCG})/M^{\ast}_{tot} \sim 0.65 \pm 0.30$. In Sec. \ref{disc} we will compare these values with both similar studies for other clusters and with theoretical predictions.


\subsection{Colour gradients in the stellar halos}\label{color}

\begin{figure}
\centering
\hspace{-0.cm}
 \includegraphics[width=9cm, angle=-0]{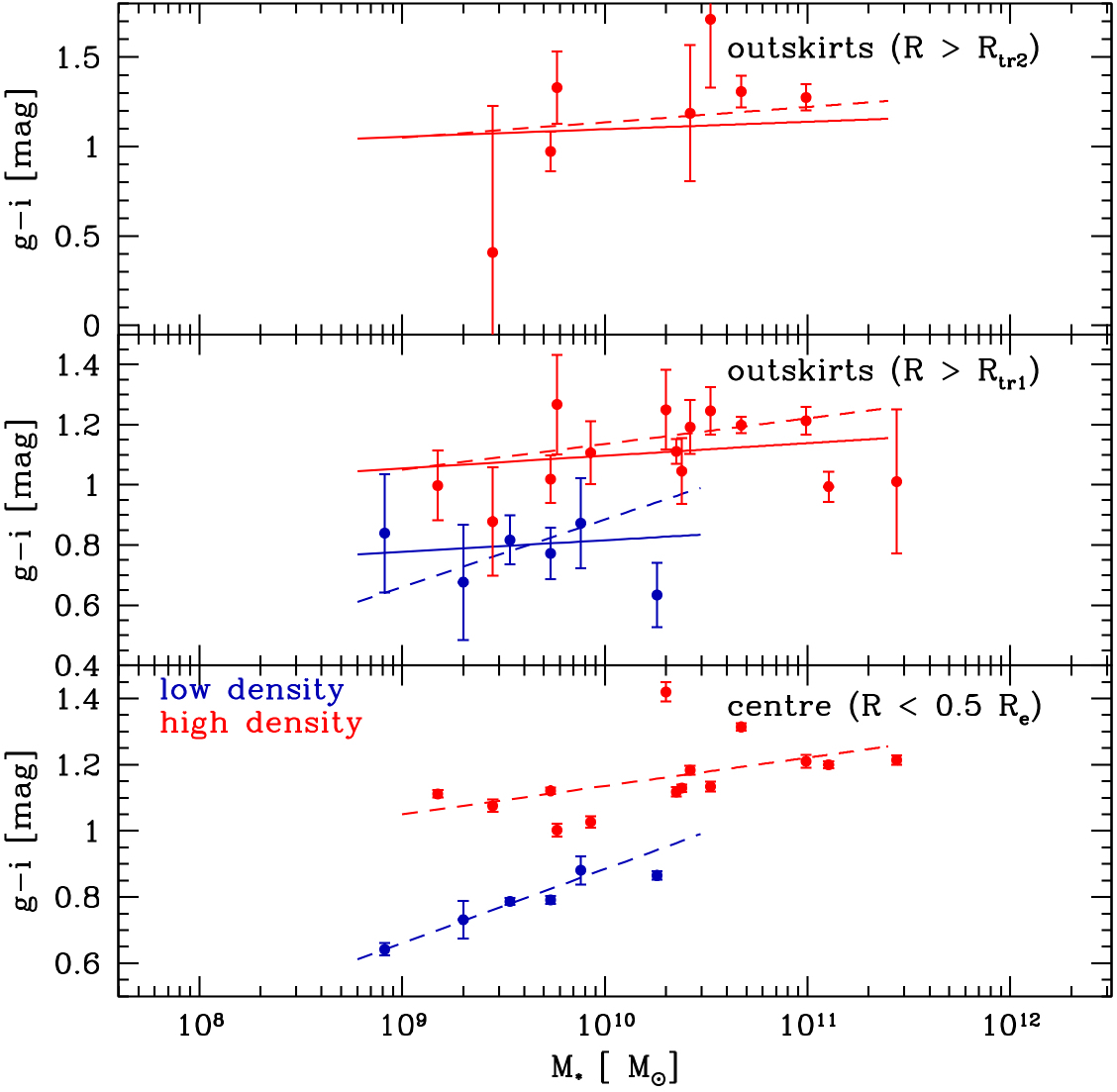}
\caption{{\it (g - i)} average colours for the central parts ($R<0.5R_{e}$, bottom panel), for $R>R_{tr1}$ (middle panel), and for $R>R_{tr2}$ (top panel)
of the FDS ETGs located in the high (red circles) and low-density region (blue circles) of the cluster as a function of the total stellar mass. The red and blue dashed lines in all the panels are the least-square fits of the values for the central parts ($R<0.5R_{e}$) of the ETGs in the high and low-density regions of the cluster, respectively. The continuous red and blue lines in the middle panel are the least-square fits of the values for $R>R_{tr1}$, while the dashed and the continuous red lines in the top panel are the least-square fit of the values for $R<0.5R_{e}$ and $R>R_{tr1}$, respectively.}\label{col_inout}
\end{figure}

In Fig. \ref{col_inout} we show the average {\it (g-i)} colours of ETGs, 
derived in three galaxy regions: the bright central part inside 0.5$R_{e}$, outside the first transition radius $R > R_{tr1}$ and outside the second transition radius $R>R_{tr2}$ (see Tab. \ref{tabfit3comp}). The bottom panel of this figure shows that, on average, the bright central parts of galaxies in the high density region of the cluster are redder than those of galaxies in the low density region, as already shown by \citet{Iodice2019}. 
Moreover, \citet{Iodice2019a}, using the Fornax 3D (F3D) spectroscopic data, 
found that the central parts of ETGs in the high density region of Fornax are more metal rich than those in the low density region, at fixed galaxy mass, suggesting that the difference in 
colours found from the photometry is due to a difference in metallicity.

The middle panel of Fig.~\ref{col_inout} shows the average {\it (g-i)} colours derived for $R>R_{tr1}$. The outskirts of the most massive ETGs ($M\geq 10^{10} M_{\odot}$) 
in the high density region are bluer than the central parts, as can be clearly seen by comparing the least-square fit of the values inside 0.5$R_{e}$ (red dashed line) 
with that for $R>R_{tr1}$ (red continuous line). Taking errors into account, there is no appreciable difference in colours between inner parts and
outskirts, for the ETGs in the low density regions.

\begin{figure}
\centering
\hspace{-0.cm}
 \includegraphics[width=9cm, angle=-0]{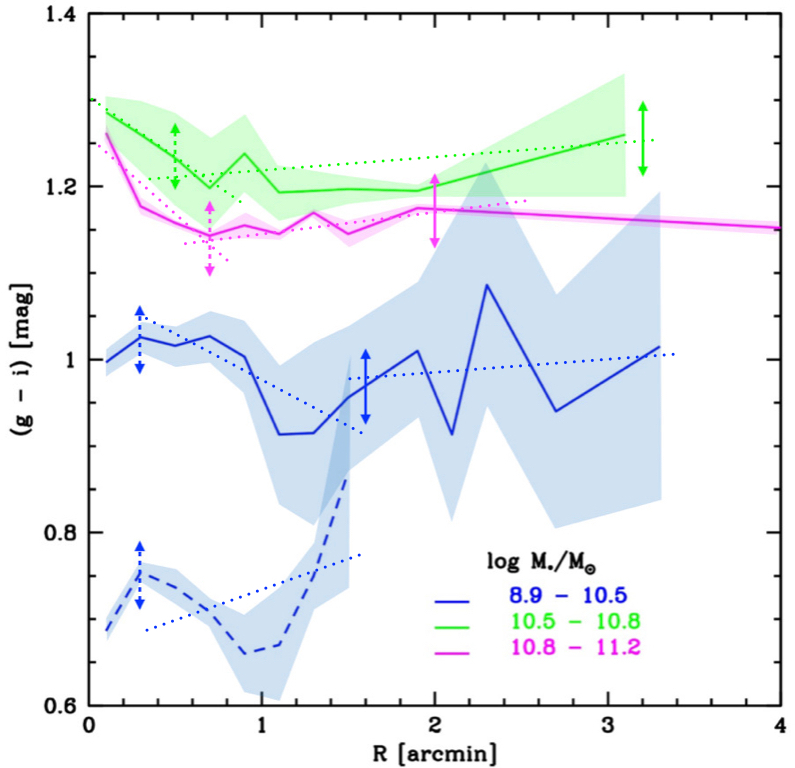}
\caption{Running mean of the average {\it (g - i)} colour profiles of the ETGs in Fornax, in three different mass bins: $8.9 \leq log M_{\ast}/M_{\odot}\leq 10.5$ (blue), $10.5 \leq log M_{\ast}/M_{\odot}\leq 10.8$ (green) and $10.8 \leq log M_{\ast}/M_{\odot}\leq 11.2$ (magenta). Blue dashed line is for galaxies in the low density region, while the blue continuous line is for that in the high density region of the cluster. The dashed and the continuous vertical arrows mark the position of the average transition radii, $R_{tr1}$ and $R_{tr2}$ respectively, in each mass bin. The dotted lines represent the fits to the colour gradients between transition radii.}\label{col_lowhigh}
\end{figure}

As presented in Sec.~\ref{sec:fit}, for the most massive ETGs ($M\geq 10^{10} M_{\odot}$) in our sample, by fitting the surface brightness profile, we identified a third outer component that dominates the light beyond a second transition radius (see Tab.~\ref{tabfit3comp}). 
All these galaxies are located in the high-density region of the cluster and, by comparing the average colours in their centres  ($R < 0.5 R_{e}$) and for $R>R_{tr1}$, it seems that the colour distribution remains quite constant (see top panel of Fig.~\ref{col_inout}).
The same conclusion can be drawn by looking at the the average {\it (g - i)} colour
profiles shown in Fig.~\ref{col_lowhigh}. They are derived in three different bins of stellar mass: low-mass galaxies with $8.9\leq log M_{\ast}/M_{\odot} \leq 10.5$, intermediate-mass galaxies with $10.5\leq log M_{\ast}/M_{\odot} \leq 10.8$ and high-mass galaxies with $10.8\leq log M_{\ast}/M_{\odot} \leq 11.2$.

In Fig. \ref{col_lowhigh},  the average values of $R_{tr1}$ and $R_{tr2}$ are plotted as dashed and continuous arrows, respectively, on each colour profile. 
At each transition radius corresponds a change in the colour trend. 
For the intermediate and massive galaxies, colours in the outer regions (i.e. for $R \geq R_{tr1}$) are bluer than those in the central ones, and 
for $R > R_{tr1}$ and $R > R_{tr2}$ the colour profiles tend to flatten, confirming the lack of a significant colour gradient. All galaxies in these two mass bins are located in the high-density region of the cluster. It is also worth to note that 
the galaxies in these two mass bins show a clear increase, toward redder colours ($g-i \sim 1.1 - 1.3$~mag), in the centre, for  $R\leq R_{tr1}$. 

The low-mass ETGs, both in the high and in the low 
density region of the cluster, exhibit bluer colours in the central regions 
($R \leq R_{tr1}$), as already pointed out by \citet{Iodice2019}.

\section {Observations versus Simulations}
The main aim of this section is to provide a direct comparison between the observables we derived 
from the surface photometry (i.e. light and colour distribution) of the ETGs inside the virial
radius of the Fornax cluster and the theoretical predictions for the same quantities.

\subsection{Shape of the light profiles} 

According to \citet{Cooper2015}, the surface brightness profile of an ETG should be well 
described by the superposition of {\it i)} an inner S{\'e}rsic profile, where
$n\sim2$, representing the (sub-dominant) in-situ  component  in  the  central  regions,  
{\it ii)} a second  S{\'e}rsic  profile, describing the 
(dominant) ``relaxed'' accreted component, 
{\it iii)} and an outer diffuse envelope, being the  ``unrelaxed'' accreted material, which 
has an exponential decline and a central surface 
brightness $\mu_{0} \sim 26$ mag/arcsec$^{2}$ in the {\it r} band, dominates
the light distribution at smaller transition radius for the most massive galaxies ($\geq 10^{12} M_{\odot}$). 
The stellar population of the in-situ component is expected to be quite similar to the dominant 
``relaxed'' accreted component. As they are well mixed together, the sum of these two components 
is expected to have a smooth distribution with only faint features to suggest they are distinct.
The outer envelope is made by `streams' and other  coherent  concentrations  of  debris, tracing the latest phase of the mass assembly. This component contributes only for a small fraction ($\leq$ 20\%) to the total accreted mass.

The fit of the surface brightness profiles presented in Sec.~\ref{sec:fit} are largely consistent
with the theoretical predictions summarised above. In particular, we found that
the dominant component to the light distribution is the ``relaxed'' accreted, well reproduced 
by the second S{\'e}rsic law. In the majority of ETGs, the in-situ component, 
which is also fitted with a S{\'e}rsic law, has $n\simeq 1-3$, with a distribution peak 
at $n\sim2$ (see Fig.~\ref{M_n1}). 
The outer envelope in all galaxies that show this component is well fitted by an 
exponential law, with an average central surface brightness of $\mu_{0,3}\sim 25$ 
mag/arcsec$^{2}$ and scale radius of $r_{h,3} \sim 165$ arcsec ($\sim$ 16 kpc, see 
Tab.~\ref{tabfit3comp}). Therefore, this component has comparable central surface brightness, but smaller scale radius, to those found in similar observational studies,   
mainly based on BCGs, where the outer envelope has 
$24 \leq \mu_{0} \leq 26$ mag/arcsec$^{2}$ and $30 \leq r_{h} \leq 100$ kpc, in the {\it r} band \citep{Seigar2007, Donzelli2011, Spavone2017}. 
The less extended envelope (i.e. smaller $r_{h,3}$) in the ETGs in Fornax is expected 
since they are less massive galaxies than the BGCs studied in the works cited above (see also
Sec.~\ref{sec:mass_frac}).\\

\subsection{Accreted mass fraction vs total stellar mass}\label{sec:mass_frac}
\begin{figure*}
\centering
\hspace{-0.cm}
 \includegraphics[width=9.2cm, angle=-0]{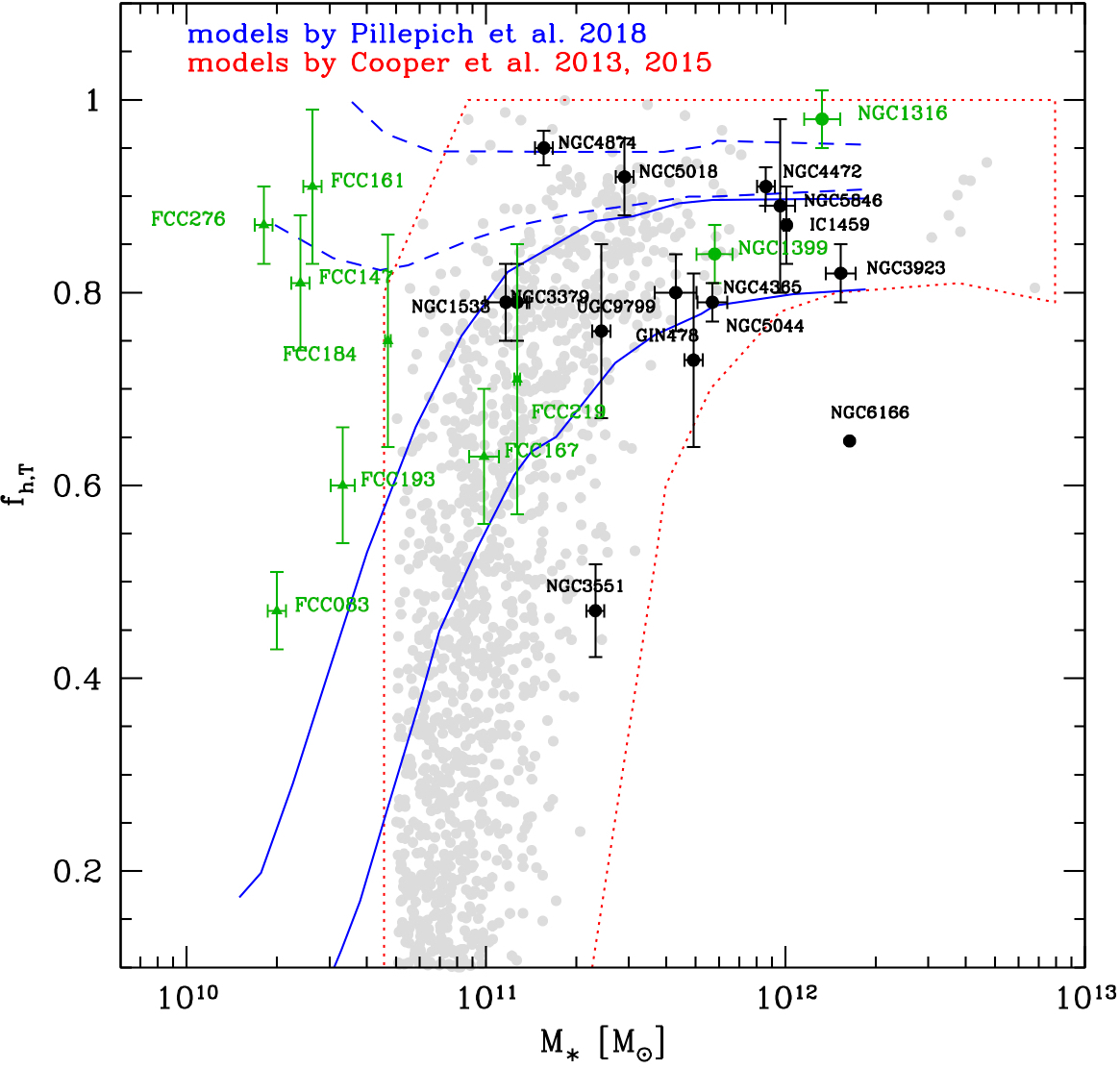}
 \includegraphics[width=9cm, angle=-0]{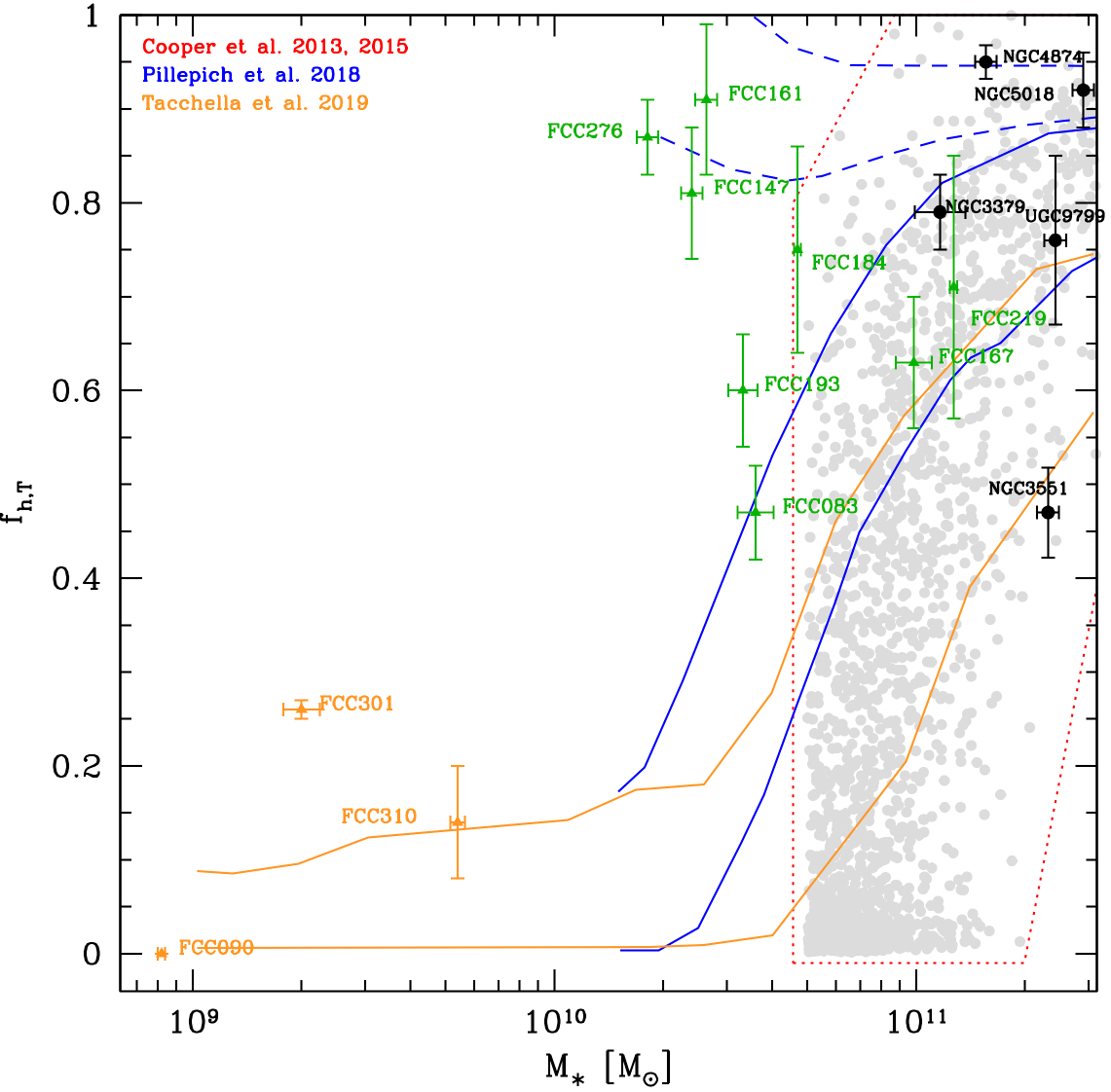}
\caption{{\it Left panel} - Accreted mass fraction vs. total stellar mass for ETGs. The measurements for FDS galaxies analysed in this work are given as green triangles. The green circles correspond to the FDS galaxies NGC1399 and NGC1316, published by \citet{Iodice2016, Iodice2017}. Black circles correspond to other BCGs from the literature
 \citep{Seigar2007, Bender2015, Spavone2017, Spavone2018, Iodice2019b, Cattapan2019}. The red region, enclosing grey dots,
   indicates the
 predictions of cosmological galaxy formation simulations by
 \citet{Cooper2013,Cooper2015}. Blue continuous and dashed regions
 indicate the accreted mass fraction measured within 30 kpc and
 outside 100 kpc, respectively, in Illustris simulations by
 \citealt{Pillepich2018} (see their Fig. 12). {\it Right panel} - The same as the right panel for less massive galaxies in the FDS sample (orange triangles). Orange regions indicate the accreted mass fraction for low mass galaxies in Illustris TNG simulations by \citet{Tacchella2019}}\label{halo}
\end{figure*}

In Fig. \ref{halo} (left panel) we compare the accreted mass ratios derived for the most massive FDS galaxies 
($1.8 \times\ 10^{10} M_{\odot} \leq\ M_{\star} \leq\ 12.7 \times\ 10^{10} M_{\odot}$), 
with the  theoretical predictions from semi-analytic particle-tagging
simulations by \citet{Cooper2013,Cooper2015}, and the Illustris TNG cosmological
hydrodynamical simulations by \citet{Pillepich2018}. 

Figure \ref{halo} also includes previous
estimates of the total accreted mass fraction for BCGs from the literature,
based on deep images and same fitting technique  
\citep{Seigar2007, Bender2015, Iodice2016, Iodice2017, Spavone2017, Spavone2018, Iodice2019b, Cattapan2019}. Since we are comparing the accreted mass fraction 
derived both for bright cluster members and for the brightest galaxies at the centre of less dense
environments as in group of galaxies, from this comparison there is a clear indication that the driving factor for the accreted mass is the total stellar mass of the galaxy, independently from the environment. 
This is in agreement with theoretical predictions, which are included in Figure~\ref{halo}. In the Fornax cluster, for massive galaxies ($\geq 10^{10} M_{\odot}$)
the stars accreted account for most of the total galaxy stellar mass.
A good agreement is also found in the lower-mass regime ($\leq 10^{10} M_{\odot}$), which remains 
still quite poorly explored on both theoretical and observational sides. In the right panel of 
Fig. \ref{halo}, we compare the estimate of the accreted mass fraction for the less massive galaxies in the Fornax cluster (FCC090, FCC301 and FCC310) with predictions from \citet{Tacchella2019} for Illustris TNG galaxies in the mass range $10^9-10^{11.5} M_{\odot}$. We find that in this range of stellar masses the accreted fraction is quite low ($f_{h,T} \leq 26\%$), as also predicted by simulations.\\

\subsection{Colour gradients in the stellar halos}
The three bins of stellar masses given in Sec.~\ref{color} and in 
Fig.~\ref{col_lowhigh} have been chosen in order to allow a direct comparison with the predictions based on Illustris TNG simulations by \cite{Tacchella2019}. They found that, at fixed $M_{\ast}$, galaxies with higher spheroid-to-total ratio (S/T) have an higher fraction of accreted mass, and that this morphological indicator is 
strongly correlated with galaxies stellar mass. They also found that the S/T ratio 
varies with the average ({\it g - r)} colour, with redder galaxies having higher S/T ratio, and that this is especially true for $M_{\ast} \simeq 10^{10.5} - 10^{11} 
M_{\odot}$. Compared to the Fornax cluster, this mass range includes all ETGs 
in the high-density region of the cluster, which are the redder objects of the sample and where the higher accreted mass fraction is found, consistently with the theoretical predictions cited above.
As expected, the most massive galaxies exhibit a colour gradient \citep{Labarbera2012}, from the central regions to the first transition radius ($R_{tr1}$), while beyond this radius, the average colour profile tends to flatten. For the less massive galaxies instead, this flattening is observed in the outskirts, beyond $\sim R_{tr2}$. A similar behaviour is also predicted for the metallicity in simulations by \citet{Cook2016}. This flattening is due to the ongoing accretion of stars coming from different satellite galaxies, which produces a mixing of different stellar populations.
The link between the transition radii and the variations in the colour trend, allows us to establish a correlation between the different components identified in the surface brightness profiles and their colours, i.e their stellar populations.

\subsection{Intracluster Light}
In this work we have provided the first accurate estimate of the ICL fraction inside the virial radius of the Fornax cluster, which amount to about 34\% (see Sec.~\ref{sec:results}).
By using deep imaging, \citet{Mihos2017} estimate an ICL fraction for the Virgo cluster ranging between 7\% and 15\%. For the Coma cluster, there are several studies
that quote the ICL, ranging from 25\% up to 50\% \citep{Melnick1977,Thuan1977,Bernstein1995,Adami2005,Teja2019}. 
Moreover, in a compilation of ICL estimates for nearby clusters, \citet{Ciardullo2004} found that the ICL fraction spans the range of 15\%-35\%.   
 \citet{Gonzalez2005} adopted a different approach in the study of 23 nearby galaxy clusters and groups, spanning a halo mass from $10^{13} - 10^{15}$~M$_\odot$. 
 They provided the fraction of ICL with respect to the luminosity of BCG+ICL and they found that ICL/BCG+ICL ratio is $\sim 50\%$ for clusters of galaxies with a
 halo mass comparable to that of the Fornax cluster, i.e. $\sim 10^{14}$~M$_\odot$. 
 For the Fornax cluster we derive ICL/BCG+ICL $\sim$ 65\%.

\begin{figure}
\hspace{-0.cm}
 \includegraphics[width=9.5cm,angle=-0]{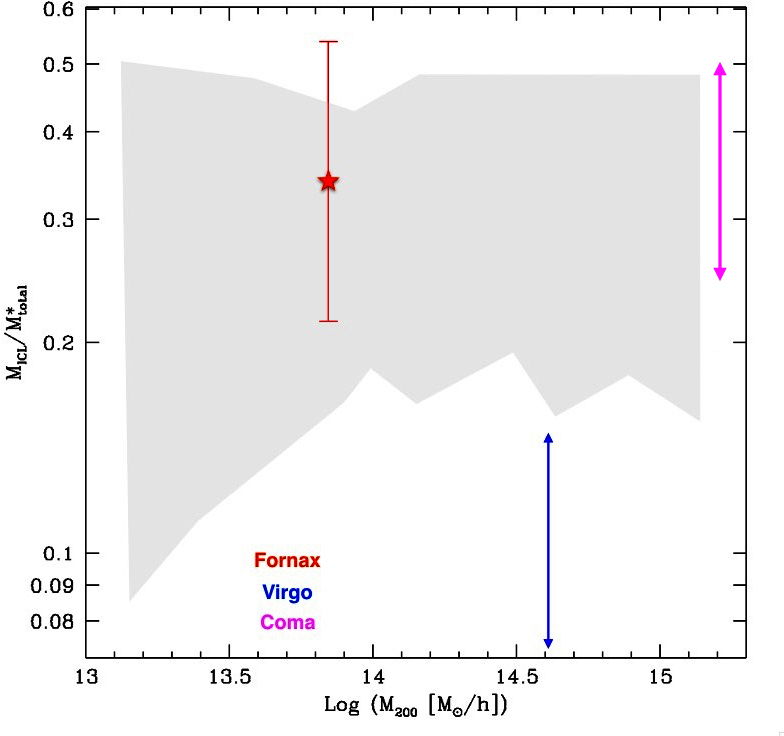}
\caption{ICL fraction as a function of halo mass for Fornax (red star). The ranges of ICL fractions for Virgo \citep{Mihos2017} and Coma  \citep{Melnick1977,Thuan1977,Bernstein1995,Adami2005,Teja2019} are given as blue and magenta arrows, respectively. The grey shaded region encloses all the median results from different models by \citealt{Contini2014} (see their Fig. 2). For the Fornax cluster, as well as for Virgo and Coma cluster, M$_{200}$ is the virial mass.}\label{ICL}
\end{figure}

Theoretical simulations tailored to studying the ICL component in clusters \citep{Rudick2011,Contini2014} found that this accounts for $\sim 10\%-40\%$ of the total light of the cluster, with no dependence from the halo mass. 
In Fig. \ref{ICL} we compare the ICL estimate derived for the Fornax cluster, as well as the 
available estimates for Coma and Virgo clusters, with theoretical predictions provided by 
\citet{Contini2014}. 
They measured the ICL fraction considering all galaxies inside the virial radius of the 
cluster and with stellar masses larger than $10^8$~M$_{\odot}$, so, consistent with the range 
of stellar masses for galaxies inside the virial radius of the Fornax cluster.
Despite the differing techniques used to estimate the total ICL fraction, 
they are reasonably in agreement with predicted values. 

As suggested by theoretical works cited above, the amount of ICL is an 
indicator of the evolutionary stage of a cluster, with more evolved ones having a higher ICL fraction. This would account for the different estimates derived for the Fornax cluster with respect to the Virgo and Coma clusters.
The higher ICL fraction found in Fornax is consistent with an evolved state, since most of the bright ($m_B<15$~mag) cluster members have transformed into ETGs (with an ETGs/LTGs ratio of 1.6 inside the virial radius) more than in the Virgo cluster, where the ETGs/LTGs $\sim 0.62$ \citep{Ferguson1989}.\\
The Coma cluster appears in a more evolved phase with respect to the Fornax cluster, 
since the majority of the 200 most luminous galaxies are ellipticals and S0s, with 
ETGs/LTGs ratio of 6 \citep{Colless2000}, and has a comparable ICL fraction to that in the Fornax cluster. On the other hand, the low ETGs/LTGs ratio of the Virgo cluster could 
consistently indicate an earlier evolutionary phase and therefore an expected lower fraction
of ICL, as shown in Fig.~\ref{ICL}. 


\section{Discussion: accreted mass fraction versus cluster-centric distance}\label{disc}

In this work we have studied the light and colour distribution of the brightest ETGs
inside the virial radius of the Fornax cluster, using the deep, multi-band 
images from the Fornax Deep Survey (FDS) with VST \citep{Iodice2019,Venhola2018}.
The large integration time and wide covered area of FDS data allow us to map the
light and colour distributions down to $\mu_g \geq 27$~mag/arcsec$^2$, therefore out 
to the regions of stellar halos and ICL regime. 
The main aim of the present work is to estimate the total accreted mass 
fraction in each galaxy of the sample and correlate it with the assembly history
of the Fornax cluster that emerged from previous studies. In addition, having a 
complete coverage of the Fornax cluster out to its virial radius down to such a 
low-surface brightness regime, we are also able to provide an accurate estimate of the ICL inside this region of the cluster.

Our main results can be summarised as follow:
\begin{itemize}
\item the highest accreted mass fraction (50\%-90\%) is found in the massive ($10^{10} \leq M \leq 10^{12}$~M$_\odot$) and reddest ETGs located in the high-density region of the cluster ($\leq 0.4 R_{vir}\sim 0.3$~Mpc). For all ETGs in the low density region, even for those with comparably high stellar masses as ETGs in the high density region, we found a lower accreted mass fraction (less than 50\%).
    \item the accreted mass fractions derived for the ETGs in Fornax are consistent with those from theoretical predictions, as well as with previous observational estimates;
    \item on average, the colour profiles of the ETGs with the largest accreted mass fraction tends to flatten in the galaxy's outskirts, i.e. beyond the transition radius from the central in-situ component to the ex-situ accreted one;
    \item the total luminosity of the ICL compared with the total luminosity of all cluster members (bright galaxies as well as dwarfs), inside the virial radius of the cluster, is about 34\%.
\end{itemize}

The new findings well fit the complex 
picture emerging for the assembly history of the Fornax cluster in the latest years.
By combining the structural properties of the galaxies from the FDS data and 
high-resolution MUSE data from F3D (i.e., morphology, colours, kinematics, and 
stellar population), we know that the cluster shows three well-defined groups of galaxies: the {\it core}, the {\it north-south clump} and 
the {\it infalling galaxies} (see Fig.~\ref{fig:Fornax_groups}). Galaxies in each group have different properties in their light and colour distributions, in their kinematics, 
and in their stellar populations \citep{Iodice2019a}. In particular, the clump 
galaxies are in the highest range of stellar masses, they are the reddest and more metal-rich galaxies of the sample.
Their outskirts have lower metallicity than the bright central regions, which is an indication that the mass assembly of metal-poor satellites proceeded in the outskirts.\\
The infalling galaxies  are  distributed nearly symmetrically around the core, 
in the low-density region of the cluster. The majority are LTGs with ongoing star formation and show signs of interaction with the environment and/or 
minor merging events in the form of tidal tails and disturbed molecular gas \citep{Zabel2019, Raj2019}. The ETGs belonging to this group are less massive and bluer than the ETGs in the clump. In this region, galaxies have on average lower metallicities with respect to galaxies in the clump.\\ 
The ETGs showing the largest accreted mass fraction and flat colour profiles in 
the outskirts are all the galaxies populating the clump.
The few ETGs with a small amount of accreted mass, in the low-density region of the cluster, belong to the infalling galaxies (see also Fig.~\ref{fig:Fornax_groups}). 

As suggested by \citet{Iodice2019a}, the clump may result from the early accretion of 
a group of galaxies during the gradual build-up of the cluster, 
which induced the observed asymmetry in the spatial distribution of the brightest 
galaxies. Due to the lower velocity dispersion in galaxy groups, 
the cluster members accreted as part of groups can undergo pre-processing and 
gravitational interactions that modified their structure and favoured the growth 
of the stellar halos. This would explain the high fraction of kinematically 
decoupled components in the galaxies belonging to the clump, the thick discs 
observed in the three edge-on S0 galaxies \citep{Pinna2019a,Pinna2019b}, and, as 
result of the present work, also the large fraction of the accreted mass in the 
galaxies in this group. Consistent with this picture, previous analysis of the 
FDS data showed that the region of the cluster where the clump is located is the 
only region inside the virial radius where intra-cluster baryons were found 
\citep{Iodice2016, Spiniello2018, Iodice2019}. Therefore, taking also 
into account that the ETGs in the clump have the largest accreted mass fraction, 
we can conclude that the major contribution to the total ICL inside the virial 
radius of the cluster comes from this region.

The observed properties of the infalling galaxies 
(late-type structures, bluer colours, 
active star formation, low metallicity and disturbed morphology) 
suggest that they can be considered as galaxies that are entering the cluster. 
The ETGs belonging to this group are locally isolated systems (FCC~083, FCC~090, 
FCC~301 and FCC~310 in Fig.~\ref{fig:Fornax_groups}), in the low-density region of the cluster. Compared to the galaxies in the clump, they do not undergo yet many galaxy-galaxy interactions or merging that might have triggered the mass assembly process and, as a consequence, they show the lowest fraction of accreted mass.
 
In conclusion, the main results of this work strongly suggest that, 
inside the Fornax cluster, there is a clear indication that the driving factor 
for the accretion process is the total stellar mass of the galaxy, in
agreement with the hierarchical accretion scenario, which also shapes the environment where the galaxies reside. In fact,  
the galaxies with the highest accreted mass are located in the high-density region of the cluster.
In this framework, the galaxies in the clump went through the mass accretion 
in small groups were they resided before merging into the cluster potential (see e.g. Raj et al. submitted, for the Fornax-A group). 
Therefore, at the present epoch of the Fornax assembly history, they 
are shaping the high-density region of the cluster, being the major 
contribution to the stellar density in the core of the cluster.

\begin{acknowledgements}
The authors are very grateful to the anonymous referee for his/her comments and suggestions which helped to improve and clarify the paper. This work is based on visitor mode observations taken at the ESO La
Silla Paranal Observatory within the VST Guaranteed Time Observations,
Programme IDs 094.B-0512(B), 094.B- 0496(A), 096.B-0501(B),
096.B-0582(A). MS and EI acknowledge financial support from the VST
project (P.I. P. Schipani), and wish to thank P. T. de Zeeuw, O. Gerhard, A. Pillepich, R. S. Remus and D. A. Forbes for useful discussions and suggestions. Authors wish to thank ESO for the
financial contribution given for the visitor mode runs at the ESO La Silla Paranal Observatory.
EI, MAR, RP, NRN and AV acknowledge financial support from the European Union
Horizon 2020 research and innovation programme under the Marie Skodowska-Curie grant agreement n. 721463 to the SUNDIAL ITN network. 
NRN acknowledges financial support from the ''One hundred top talent program of Sun Yat-sen University'' grant N. 71000-18841229.
GvdV acknowledges funding from the European Research Council (ERC) under the European Union's Horizon 2020 research and innovation programme under grant agreement No 724857 (Consolidator Grant ArcheoDyn). 
J.F-B acknowledges support through the RAVET project by the grant AYA2016-77237-C3-1-P from the Spanish Ministry of Science, Innovation and Universities (MCIU) and through the IAC project TRACES which is partially supported through the state budget and the regional budget of the Consejer\'\i a de Econom\'\i a, Industria, Comercio y Conocimiento of the Canary Islands Autonomous Community. AV thanks the Eemil Aaltonen foundation for the financial support. GvdV acknowledges funding from the European Research Council (ERC) under the European Union's Horizon 2020 research and innovation programme under grant agreement No 724857 (Consolidator Grant ArcheoDyn).
\end{acknowledgements}

\bibliographystyle{aa.bst}
  \bibliography{fornax}

\begin{thebibliography}{96}
\expandafter\ifx\csname natexlab\endcsname\relax\def\natexlab#1{#1}\fi

\bibitem[{{Adami} {et~al.}(2005){Adami}, {Slezak}, {Durret}, {Conselice},
  {Cuillandre}, {Gallagher}, {Mazure}, {Pell{\'o}}, {Picat}, \&
  {Ulmer}}]{Adami2005}
{Adami}, C., {Slezak}, E., {Durret}, F., {et~al.} 2005, \aap, 429, 39

\bibitem[{{Amorisco}(2017)}]{Amorisco2017}
{Amorisco}, N.~C. 2017, \mnras, 469, L48

\bibitem[{{Arnaboldi} {et~al.}(2012){Arnaboldi}, {Ventimiglia}, {Iodice},
  {Gerhard}, \& {Coccato}}]{Arnaboldi2012}
{Arnaboldi}, M., {Ventimiglia}, G., {Iodice}, E., {Gerhard}, O., \& {Coccato},
  L. 2012, \aap, 545, A37

\bibitem[{{Barbosa} {et~al.}(2018){Barbosa}, {Arnaboldi}, {Coccato}, {Gerhard},
  {Mendes de Oliveira}, {Hilker}, \& {Richtler}}]{Barbosa2018}
{Barbosa}, C.~E., {Arnaboldi}, M., {Coccato}, L., {et~al.} 2018, \aap, 609, A78

\bibitem[{{Bedregal} {et~al.}(2011){Bedregal}, {Cardiel},
  {Arag{\'o}n-Salamanca}, \& {Merrifield}}]{Bedregal2011}
{Bedregal}, A.~G., {Cardiel}, N., {Arag{\'o}n-Salamanca}, A., \& {Merrifield},
  M.~R. 2011, \mnras, 415, 2063

\bibitem[{{Bekki} {et~al.}(2003){Bekki}, {Forbes}, {Beasley}, \&
  {Couch}}]{Bekki2003}
{Bekki}, K., {Forbes}, D.~A., {Beasley}, M.~A., \& {Couch}, W.~J. 2003, \mnras,
  344, 1334

\bibitem[{{Bender} {et~al.}(2015){Bender}, {Kormendy}, {Cornell}, \&
  {Fisher}}]{Bender2015}
{Bender}, R., {Kormendy}, J., {Cornell}, M.~E., \& {Fisher}, D.~B. 2015, \apj,
  807, 56

\bibitem[{{Bernstein} {et~al.}(1995){Bernstein}, {Nichol}, {Tyson}, {Ulmer}, \&
  {Wittman}}]{Bernstein1995}
{Bernstein}, G.~M., {Nichol}, R.~C., {Tyson}, J.~A., {Ulmer}, M.~P., \&
  {Wittman}, D. 1995, \aj, 110, 1507

\bibitem[{{Borlaff} {et~al.}(2017){Borlaff}, {Eliche-Moral}, {Beckman},
  {Ciambur}, {P{\'e}rez-Gonz{\'a}lez}, {Barro}, {Cava}, \&
  {Cardiel}}]{Borlaff2017}
{Borlaff}, A., {Eliche-Moral}, M.~C., {Beckman}, J.~E., {et~al.} 2017, \aap,
  604, A119

\bibitem[{{Cantiello} {et~al.}(2018){Cantiello}, {D'Abrusco}, {Spavone},
  {Paolillo}, {Capaccioli}, {Limatola}, {Grado}, {Iodice}, {Raimondo},
  {Napolitano}, {Blakeslee}, {Brocato}, {Forbes}, {Hilker}, {Mieske},
  {Peletier}, {van de Ven}, \& {Schipani}}]{Cantiello2018}
{Cantiello}, M., {D'Abrusco}, R., {Spavone}, M., {et~al.} 2018, \aap, 611, A93

\bibitem[{{Caon} {et~al.}(1993){Caon}, {Capaccioli}, \& {D'Onofrio}}]{Caon1993}
{Caon}, N., {Capaccioli}, M., \& {D'Onofrio}, M. 1993, \mnras, 265, 1013

\bibitem[{{Capaccioli} \& {de Vaucouleurs}(1983)}]{Capaccioli1983}
{Capaccioli}, M. \& {de Vaucouleurs}, G. 1983, \apjs, 52, 465

\bibitem[{{Capaccioli} {et~al.}(2015){Capaccioli}, {Spavone}, {Grado},
  {Iodice}, {Limatola}, {Napolitano}, {Cantiello}, {Paolillo}, {Romanowsky},
  {Forbes}, {Puzia}, {Raimondo}, \& {Schipani}}]{Capaccioli2015}
{Capaccioli}, M., {Spavone}, M., {Grado}, A., {et~al.} 2015, \aap, 581, A10

\bibitem[{{Carollo} {et~al.}(1997){Carollo}, {Franx}, {Illingworth}, \&
  {Forbes}}]{Carollo1997}
{Carollo}, C.~M., {Franx}, M., {Illingworth}, G.~D., \& {Forbes}, D.~A. 1997,
  \apj, 481, 710

\bibitem[{{Cattapan} {et~al.}(2019){Cattapan}, {Spavone}, {Iodice}, {Rampazzo},
  {Ciroi}, {Ryan-Weber}, {Schipani}, {Capaccioli}, {Grado}, {Limatola},
  {Mazzei}, {Held}, \& {Marino}}]{Cattapan2019}
{Cattapan}, A., {Spavone}, M., {Iodice}, E., {et~al.} 2019, \apj, 874, 130

\bibitem[{{Ciardullo} {et~al.}(2004){Ciardullo}, {Mihos}, {Feldmeier},
  {Durrell}, \& {Sigurdsson}}]{Ciardullo2004}
{Ciardullo}, R., {Mihos}, J.~C., {Feldmeier}, J.~J., {Durrell}, P.~R., \&
  {Sigurdsson}, S. 2004, in IAU Symposium, Vol. 217, Recycling Intergalactic
  and Interstellar Matter, ed. P.-A. {Duc}, J.~{Braine}, \& E.~{Brinks}, 88

\bibitem[{{Coccato} {et~al.}(2013){Coccato}, {Arnaboldi}, \&
  {Gerhard}}]{Coccato2013}
{Coccato}, L., {Arnaboldi}, M., \& {Gerhard}, O. 2013, \mnras, 436, 1322

\bibitem[{{Coccato} {et~al.}(2010){Coccato}, {Gerhard}, {Arnaboldi}, {Das},
  {Douglas}, {Kuijken}, {Merrifield}, {Napolitano}, {Noordermeer},
  {Romanowsky}, {Capaccioli}, {Cortesi}, \& {de Lorenzi}}]{Coccato2010}
{Coccato}, L., {Gerhard}, O., {Arnaboldi}, M., {et~al.} 2010, Highlights of
  Astronomy, 15, 68

\bibitem[{{Coccato} {et~al.}(2011){Coccato}, {Gerhard}, {Arnaboldi}, \&
  {Ventimiglia}}]{Coccato2011}
{Coccato}, L., {Gerhard}, O., {Arnaboldi}, M., \& {Ventimiglia}, G. 2011, \aap,
  533, A138

\bibitem[{{Colless}(2000)}]{Colless2000}
{Colless}, M. 2000, {Coma Cluster}, ed. P.~{Murdin}, 2600

\bibitem[{{Contini} {et~al.}(2014){Contini}, {De Lucia}, {Villalobos}, \&
  {Borgani}}]{Contini2014}
{Contini}, E., {De Lucia}, G., {Villalobos}, {\'A}., \& {Borgani}, S. 2014,
  \mnras, 437, 3787

\bibitem[{{Contini} {et~al.}(2019){Contini}, {Yi}, \& {Kang}}]{Contini2019}
{Contini}, E., {Yi}, S.~K., \& {Kang}, X. 2019, \apj, 871, 24

\bibitem[{{Cook} {et~al.}(2016){Cook}, {Conroy}, {Pillepich},
  {Rodriguez-Gomez}, \& {Hernquist}}]{Cook2016}
{Cook}, B.~A., {Conroy}, C., {Pillepich}, A., {Rodriguez-Gomez}, V., \&
  {Hernquist}, L. 2016, \apj, 833, 158

\bibitem[{{Cooper} {et~al.}(2010){Cooper}, {Cole}, {Frenk}, {White}, {Helly},
  {Benson}, {De Lucia}, {Helmi}, {Jenkins}, {Navarro}, {Springel}, \&
  {Wang}}]{Cooper2010}
{Cooper}, A.~P., {Cole}, S., {Frenk}, C.~S., {et~al.} 2010, \mnras, 406, 744

\bibitem[{{Cooper} {et~al.}(2013){Cooper}, {D'Souza}, {Kauffmann}, {Wang},
  {Boylan-Kolchin}, {Guo}, {Frenk}, \& {White}}]{Cooper2013}
{Cooper}, A.~P., {D'Souza}, R., {Kauffmann}, G., {et~al.} 2013, \mnras, 434,
  3348

\bibitem[{{Cooper} {et~al.}(2015){Cooper}, {Parry}, {Lowing}, {Cole}, \&
  {Frenk}}]{Cooper2015}
{Cooper}, A.~P., {Parry}, O.~H., {Lowing}, B., {Cole}, S., \& {Frenk}, C. 2015,
  \mnras, 454, 3185

\bibitem[{{Cui} {et~al.}(2014){Cui}, {Murante}, {Monaco}, {Borgani}, {Granato},
  {Killedar}, {De Lucia}, {Presotto}, \& {Dolag}}]{Cui2014}
{Cui}, W., {Murante}, G., {Monaco}, P., {et~al.} 2014, \mnras, 437, 816

\bibitem[{{D'Abrusco} {et~al.}(2016){D'Abrusco}, {Cantiello}, {Paolillo},
  {Pota}, {Napolitano}, {Limatola}, {Spavone}, {Grado}, {Iodice}, {Capaccioli},
  {Peletier}, {Longo}, {Hilker}, {Mieske}, {Grebel}, {Lisker}, {Wittmann}, {van
  de Ven}, {Schipani}, \& {Fabbiano}}]{Dabrusco2016}
{D'Abrusco}, R., {Cantiello}, M., {Paolillo}, M., {et~al.} 2016, \apjl, 819,
  L31

\bibitem[{{Deason} {et~al.}(2013){Deason}, {Belokurov}, {Evans}, \&
  {Johnston}}]{Deason2013}
{Deason}, A.~J., {Belokurov}, V., {Evans}, N.~W., \& {Johnston}, K.~V. 2013,
  \apj, 763, 113

\bibitem[{{DeMaio} {et~al.}(2018){DeMaio}, {Gonzalez}, {Zabludoff}, {Zaritsky},
  {Connor}, {Donahue}, \& {Mulchaey}}]{DeMaio2018}
{DeMaio}, T., {Gonzalez}, A.~H., {Zabludoff}, A., {et~al.} 2018, \mnras, 474,
  3009

\bibitem[{{Donzelli} {et~al.}(2011){Donzelli}, {Muriel}, \&
  {Madrid}}]{Donzelli2011}
{Donzelli}, C.~J., {Muriel}, H., \& {Madrid}, J.~P. 2011, \apjs, 195, 15

\bibitem[{{Drinkwater} {et~al.}(2001){Drinkwater}, {Gregg}, \&
  {Colless}}]{Drinkwater2001}
{Drinkwater}, M.~J., {Gregg}, M.~D., \& {Colless}, M. 2001, \apjl, 548, L139

\bibitem[{{Duc}(2017)}]{Duc2017}
{Duc}, P.-A. 2017, in IAU Symposium, Vol. 321, Formation and Evolution of
  Galaxy Outskirts, ed. A.~{Gil de Paz}, J.~H. {Knapen}, \& J.~C. {Lee},
  180--182

\bibitem[{{Duc} {et~al.}(2015){Duc}, {Cuillandre}, {Karabal}, {Cappellari},
  {Alatalo}, {Blitz}, {Bournaud}, {Bureau}, {Crocker}, {Davies}, {Davis}, {de
  Zeeuw}, {Emsellem}, {Khochfar}, {Krajnovi{\'c}}, {Kuntschner}, {McDermid},
  {Michel-Dansac}, {Morganti}, {Naab}, {Oosterloo}, {Paudel}, {Sarzi}, {Scott},
  {Serra}, {Weijmans}, \& {Young}}]{Duc2015}
{Duc}, P.-A., {Cuillandre}, J.-C., {Karabal}, E., {et~al.} 2015, \mnras, 446,
  120

\bibitem[{{Ferguson}(1989)}]{Ferguson1989}
{Ferguson}, H.~C. 1989, \aj, 98, 367

\bibitem[{{Ferrarese} {et~al.}(2012){Ferrarese}, {C{\^o}t{\'e}}, {Cuillandre},
  {Gwyn}, {Peng}, {MacArthur}, {Duc}, {Boselli}, {Mei}, {Erben}, {McConnachie},
  {Durrell}, {Mihos}, {Jord{\'a}n}, {Lan{\c c}on}, {Puzia}, {Emsellem},
  {Balogh}, {Blakeslee}, {van Waerbeke}, {Gavazzi}, {Vollmer}, {Kavelaars},
  {Woods}, {Ball}, {Boissier}, {Courteau}, {Ferriere}, {Gavazzi},
  {Hildebrandt}, {Hudelot}, {Huertas-Company}, {Liu}, {McLaughlin}, {Mellier},
  {Milkeraitis}, {Schade}, {Balkowski}, {Bournaud}, {Carlberg}, {Chapman},
  {Hoekstra}, {Peng}, {Sawicki}, {Simard}, {Taylor}, {Tully}, {van Driel},
  {Wilson}, {Burdullis}, {Mahoney}, \& {Manset}}]{Ferrarese2012}
{Ferrarese}, L., {C{\^o}t{\'e}}, P., {Cuillandre}, J.-C., {et~al.} 2012, \apjs,
  200, 4

\bibitem[{{Forbes} {et~al.}(1998){Forbes}, {Grillmair}, {Williger}, {Elson}, \&
  {Brodie}}]{Forbes1998}
{Forbes}, D.~A., {Grillmair}, C.~J., {Williger}, G.~M., {Elson}, R.~A.~W., \&
  {Brodie}, J.~P. 1998, \mnras, 293, 325

\bibitem[{{Frank} {et~al.}(2013){Frank}, {Peterson}, {Andersson}, {Fabian}, \&
  {Sanders}}]{Frank2013}
{Frank}, K.~A., {Peterson}, J.~R., {Andersson}, K., {Fabian}, A.~C., \&
  {Sanders}, J.~S. 2013, \apj, 764, 46

\bibitem[{{Gonzalez} {et~al.}(2005){Gonzalez}, {Zabludoff}, \&
  {Zaritsky}}]{Gonzalez2005}
{Gonzalez}, A.~H., {Zabludoff}, A.~I., \& {Zaritsky}, D. 2005, \apj, 618, 195

\bibitem[{{Greene} {et~al.}(2019){Greene}, {Veale}, {Ma}, {Thomas},
  {Quenneville}, {Blakeslee}, {Walsh}, {Goulding}, \& {Ito}}]{Greene2019}
{Greene}, J.~E., {Veale}, M., {Ma}, C.-P., {et~al.} 2019, \apj, 874, 66

\bibitem[{{Hartke} {et~al.}(2018){Hartke}, {Arnaboldi}, {Gerhard}, {Agnello},
  {Longobardi}, {Coccato}, {Pulsoni}, {Freeman}, \& {Merrifield}}]{Hartke2018}
{Hartke}, J., {Arnaboldi}, M., {Gerhard}, O., {et~al.} 2018, \aap, 616, A123

\bibitem[{{Henden} {et~al.}(2019){Henden}, {Puchwein}, \&
  {Sijacki}}]{Henden2019}
{Henden}, N.~A., {Puchwein}, E., \& {Sijacki}, D. 2019, arXiv e-prints,
  arXiv:1911.12367

\bibitem[{{Hilker} {et~al.}(2018){Hilker}, {Richtler}, {Barbosa}, {Arnaboldi},
  {Coccato}, \& {Mendes de Oliveira}}]{Hilker2018}
{Hilker}, M., {Richtler}, T., {Barbosa}, C.~E., {et~al.} 2018, \aap, 619, A70

\bibitem[{{Iodice} {et~al.}(2016){Iodice}, {Capaccioli}, {Grado}, {Limatola},
  {Spavone}, {Napolitano}, {Paolillo}, {Peletier}, {Cantiello}, {Lisker},
  {Wittmann}, {Venhola}, {Hilker}, {D'Abrusco}, {Pota}, \&
  {Schipani}}]{Iodice2016}
{Iodice}, E., {Capaccioli}, M., {Grado}, A., {et~al.} 2016, \apj, 820, 42

\bibitem[{{Iodice} {et~al.}(2019{\natexlab{a}}){Iodice}, {Sarzi}, {Bittner},
  {Coccato}, {Costantin}, {Corsini}, {van de Ven}, {de Zeeuw},
  {Falc{\'o}n-Barroso}, {Gadotti}, {Lyubenova}, {Mart{\'\i}n-Navarro},
  {McDermid}, {Nedelchev}, {Pinna}, {Pizzella}, {Spavone}, \&
  {Viaene}}]{Iodice2019a}
{Iodice}, E., {Sarzi}, M., {Bittner}, A., {et~al.} 2019{\natexlab{a}}, \aap,
  627, A136

\bibitem[{{Iodice} {et~al.}(2017{\natexlab{a}}){Iodice}, {Spavone},
  {Cantiello}, {D'Abrusco}, {Capaccioli}, {Hilker}, {Mieske}, {Napolitano},
  {Peletier}, {Limatola}, {Grado}, {Venhola}, {Paolillo}, {Van de Ven}, \&
  {Schipani}}]{Iodice2017b}
{Iodice}, E., {Spavone}, M., {Cantiello}, M., {et~al.} 2017{\natexlab{a}},
  \apj, 851, 75

\bibitem[{{Iodice} {et~al.}(2017{\natexlab{b}}){Iodice}, {Spavone},
  {Capaccioli}, {Peletier}, {Richtler}, {Hilker}, {Mieske}, {Limatola},
  {Grado}, {Napolitano}, {Cantiello}, {D'Abrusco}, {Paolillo}, {Venhola},
  {Lisker}, {Van de Ven}, {Falcon-Barroso}, \& {Schipani}}]{Iodice2017}
{Iodice}, E., {Spavone}, M., {Capaccioli}, M., {et~al.} 2017{\natexlab{b}},
  \apj, 839, 21

\bibitem[{{Iodice} {et~al.}(2019{\natexlab{b}}){Iodice}, {Spavone},
  {Capaccioli}, {Peletier}, {van de Ven}, {Napolitano}, {Hilker}, {Mieske},
  {Smith}, {Pasquali}, {Limatola}, {Grado}, {Venhola}, {Cantiello}, {Paolillo},
  {Falcon-Barroso}, {D'Abrusco}, \& {Schipani}}]{Iodice2019}
{Iodice}, E., {Spavone}, M., {Capaccioli}, M., {et~al.} 2019{\natexlab{b}},
  \aap, 623, A1

\bibitem[{{Iodice} {et~al.}(2020){Iodice}, {Spavone}, {Cattapan}, {Bannikova},
  {Forbes}, {Rampazzo}, {Ciroi}, {Corsini}, {D'Ago}, {Oosterloo}, {Schipani},
  \& {Capaccioli}}]{Iodice2019b}
{Iodice}, E., {Spavone}, M., {Cattapan}, A., {et~al.} 2020, \aap, 635, A3

\bibitem[{{Jim{\'e}nez-Teja} {et~al.}(2019){Jim{\'e}nez-Teja}, {Dupke}, {Lopes
  de Oliveira}, {Xavier}, {Coelho}, {Chies-Santos}, {L{\'o}pez-Sanjuan},
  {Alvarez-Candal}, {Costa-Duarte}, {Telles}, {Hernandez-Jimenez},
  {Ben{\'\i}tez}, {Alcaniz}, {Cenarro}, {Crist{\'o}bal-Hornillos},
  {Ederoclite}, {Mar{\'\i}n-Franch}, {Mendes de Oliveira}, {Moles},
  {Sodr{\'e}}, {Varela}, \& {V{\'a}zquez Rami{\'o}}}]{Teja2019}
{Jim{\'e}nez-Teja}, Y., {Dupke}, R.~A., {Lopes de Oliveira}, R., {et~al.} 2019,
  \aap, 622, A183

\bibitem[{{Jord{\'a}n} {et~al.}(2007){Jord{\'a}n}, {Blakeslee}, {C{\^o}t{\'e}},
  {Ferrarese}, {Infante}, {Mei}, {Merritt}, {Peng}, {Tonry}, \&
  {West}}]{Jordan2007}
{Jord{\'a}n}, A., {Blakeslee}, J.~P., {C{\^o}t{\'e}}, P., {et~al.} 2007, \apjs,
  169, 213

\bibitem[{{Kuijken}(2011)}]{Kuijken2011}
{Kuijken}, K. 2011, The Messenger, 146, 8

\bibitem[{{La Barbera} {et~al.}(2012){La Barbera}, {Ferreras}, {de Carvalho},
  {Bruzual}, {Charlot}, {Pasquali}, \& {Merlin}}]{Labarbera2012}
{La Barbera}, F., {Ferreras}, I., {de Carvalho}, R.~R., {et~al.} 2012, \mnras,
  426, 2300

\bibitem[{{Longobardi} {et~al.}(2013){Longobardi}, {Arnaboldi}, {Gerhard},
  {Coccato}, {Okamura}, \& {Freeman}}]{Longobardi2013}
{Longobardi}, A., {Arnaboldi}, M., {Gerhard}, O., {et~al.} 2013, \aap, 558, A42

\bibitem[{{Lucy}(1974)}]{Lucy1974}
{Lucy}, L.~B. 1974, \aj, 79, 745

\bibitem[{{Ma} {et~al.}(2014){Ma}, {Greene}, {McConnell}, {Janish},
  {Blakeslee}, {Thomas}, \& {Murphy}}]{Ma2014}
{Ma}, C.-P., {Greene}, J.~E., {McConnell}, N., {et~al.} 2014, \apj, 795, 158

\bibitem[{{Mancillas} {et~al.}(2019){Mancillas}, {Duc}, {Combes}, {Bournaud},
  {Emsellem}, {Martig}, \& {Michel-Dansac}}]{Mancillas2019}
{Mancillas}, B., {Duc}, P.-A., {Combes}, F., {et~al.} 2019, \aap, 632, A122

\bibitem[{{Melnick} {et~al.}(1977){Melnick}, {White}, \&
  {Hoessel}}]{Melnick1977}
{Melnick}, J., {White}, S.~D.~M., \& {Hoessel}, J. 1977, \mnras, 180, 207

\bibitem[{{Merritt} {et~al.}(2016){Merritt}, {van Dokkum}, {Abraham}, \&
  {Zhang}}]{Merritt2016}
{Merritt}, A., {van Dokkum}, P., {Abraham}, R., \& {Zhang}, J. 2016, \apj, 830,
  62

\bibitem[{{Mihos}(2015)}]{Mihos2015}
{Mihos}, C. 2015, IAU General Assembly, 22, 2247903

\bibitem[{{Mihos} {et~al.}(2017){Mihos}, {Harding}, {Feldmeier}, {Rudick},
  {Janowiecki}, {Morrison}, {Slater}, \& {Watkins}}]{Mihos2017}
{Mihos}, J.~C., {Harding}, P., {Feldmeier}, J.~J., {et~al.} 2017, \apj, 834, 16

\bibitem[{{Monachesi} {et~al.}(2019){Monachesi}, {G{\'o}mez}, {Grand },
  {Simpson}, {Kauffmann}, {Bustamante}, {Marinacci}, {Pakmor}, {Springel},
  {Frenk}, {White}, \& {Tissera}}]{Monachesi2019}
{Monachesi}, A., {G{\'o}mez}, F.~A., {Grand }, R. J.~J., {et~al.} 2019, \mnras,
  485, 2589

\bibitem[{{Munoz} {et~al.}(2015){Munoz}, {Eigenthaler}, \& {Puzia}}]{Munoz2015}
{Munoz}, R.~P., {Eigenthaler}, P., \& {Puzia}, T.~H. e.~a. 2015, \apjl, 813,
  L15

\bibitem[{{Napolitano} {et~al.}(2002){Napolitano}, {Arnaboldi}, \&
  {Capaccioli}}]{Napolitano2002}
{Napolitano}, N.~R., {Arnaboldi}, M., \& {Capaccioli}, M. 2002, \aap, 383, 791

\bibitem[{{Napolitano} {et~al.}(2003){Napolitano}, {Pannella}, {Arnaboldi},
  {Gerhard}, {Aguerri}, {Freeman}, {Capaccioli}, {Ghigna}, {Governato},
  {Quinn}, \& {Stadel}}]{Napolitano2003}
{Napolitano}, N.~R., {Pannella}, M., {Arnaboldi}, M., {et~al.} 2003, \apj, 594,
  172

\bibitem[{{Oser} {et~al.}(2010){Oser}, {Ostriker}, {Naab}, {Johansson}, \&
  {Burkert}}]{Oser2010}
{Oser}, L., {Ostriker}, J.~P., {Naab}, T., {Johansson}, P.~H., \& {Burkert}, A.
  2010, \apj, 725, 2312

\bibitem[{{Paolillo} {et~al.}(2002){Paolillo}, {Fabbiano}, {Peres}, \&
  {Kim}}]{Paolillo2002}
{Paolillo}, M., {Fabbiano}, G., {Peres}, G., \& {Kim}, D.-W. 2002, \apj, 565,
  883

\bibitem[{{Peng} {et~al.}(2010){Peng}, {Lilly}, {Kova{\v c}}, {Bolzonella},
  {Pozzetti}, {Renzini}, {Zamorani}, {Ilbert}, {Knobel}, {Iovino}, {Maier},
  {Cucciati}, {Tasca}, {Carollo}, {Silverman}, {Kampczyk}, {de Ravel},
  {Sanders}, {Scoville}, {Contini}, {Mainieri}, {Scodeggio}, {Kneib}, {Le
  F{\`e}vre}, {Bardelli}, {Bongiorno}, {Caputi}, {Coppa}, {de la Torre},
  {Franzetti}, {Garilli}, {Lamareille}, {Le Borgne}, {Le Brun}, {Mignoli},
  {Perez Montero}, {Pello}, {Ricciardelli}, {Tanaka}, {Tresse}, {Vergani},
  {Welikala}, {Zucca}, {Oesch}, {Abbas}, {Barnes}, {Bordoloi}, {Bottini},
  {Cappi}, {Cassata}, {Cimatti}, {Fumana}, {Hasinger}, {Koekemoer},
  {Leauthaud}, {Maccagni}, {Marinoni}, {McCracken}, {Memeo}, {Meneux}, {Nair},
  {Porciani}, {Presotto}, \& {Scaramella}}]{Peng2010}
{Peng}, Y.-j., {Lilly}, S.~J., {Kova{\v c}}, K., {et~al.} 2010, \apj, 721, 193

\bibitem[{{Pillepich} {et~al.}(2018){Pillepich}, {Nelson}, {Hernquist},
  {Springel}, {Pakmor}, {Torrey}, {Weinberger}, {Genel}, {Naiman}, {Marinacci},
  \& {Vogelsberger}}]{Pillepich2018}
{Pillepich}, A., {Nelson}, D., {Hernquist}, L., {et~al.} 2018, \mnras, 475, 648

\bibitem[{{Pinna} {et~al.}(2019{\natexlab{a}}){Pinna}, {Falc{\'o}n-Barroso},
  {Martig}, {Coccato}, {Corsini}, {de Zeeuw}, {Gadotti}, {Iodice}, {Leaman},
  {Lyubenova}, {Mart{\'\i}n-Navarro}, {Morelli}, {Sarzi}, {van de Ven},
  {Viaene}, \& {McDermid}}]{Pinna2019b}
{Pinna}, F., {Falc{\'o}n-Barroso}, J., {Martig}, M., {et~al.}
  2019{\natexlab{a}}, \aap, 625, A95

\bibitem[{{Pinna} {et~al.}(2019{\natexlab{b}}){Pinna}, {Falc{\'o}n-Barroso},
  {Martig}, {Sarzi}, {Coccato}, {Iodice}, {Corsini}, {de Zeeuw}, {Gadotti},
  {Leaman}, {Lyubenova}, {McDermid}, {Minchev}, {Morelli}, {van de Ven}, \&
  {Viaene}}]{Pinna2019a}
{Pinna}, F., {Falc{\'o}n-Barroso}, J., {Martig}, M., {et~al.}
  2019{\natexlab{b}}, \aap, 623, A19

\bibitem[{{Pop} {et~al.}(2018){Pop}, {Pillepich}, {Amorisco}, \&
  {Hernquist}}]{Pop2018}
{Pop}, A.-R., {Pillepich}, A., {Amorisco}, N.~C., \& {Hernquist}, L. 2018,
  \mnras, 480, 1715

\bibitem[{{Pota} {et~al.}(2018){Pota}, {Napolitano}, {Hilker}, {Spavone},
  {Schulz}, {Cantiello}, {Tortora}, {Iodice}, {Paolillo}, {D'Abrusco},
  {Capaccioli}, {Puzia}, {Peletier}, {Romanowsky}, {van de Ven}, {Spiniello},
  {Norris}, {Lisker}, {Munoz}, {Schipani}, {Eigenthaler}, {Taylor},
  {S{\'a}nchez-Janssen}, \& {Ordenes-Brice{\~n}o}}]{Pota2018}
{Pota}, V., {Napolitano}, N.~R., {Hilker}, M., {et~al.} 2018, \mnras, 481, 1744

\bibitem[{{Prole} {et~al.}(2019){Prole}, {Hilker}, {van der Burg}, {Cantiello},
  {Venhola}, {Iodice}, {van de Ven}, {Wittmann}, {Peletier}, {Mieske},
  {Capaccioli}, {Napolitano}, {Paolillo}, {Spavone}, \&
  {Valentijn}}]{Prole2019}
{Prole}, D.~J., {Hilker}, M., {van der Burg}, R.~F.~J., {et~al.} 2019, \mnras,
  484, 4865

\bibitem[{{Raj} {et~al.}(2019){Raj}, {Iodice}, {Napolitano}, {Spavone}, {Su},
  {Peletier}, {Davis}, {Zabel}, {Hilker}, {Mieske}, {Falcon Barroso},
  {Cantiello}, {van de Ven}, {Watkins}, {Salo}, {Schipani}, {Capaccioli}, \&
  {Venhola}}]{Raj2019}
{Raj}, M.~A., {Iodice}, E., {Napolitano}, N.~R., {et~al.} 2019, \aap, 628, A4

\bibitem[{{Richardson}(1972)}]{Richardson1972}
{Richardson}, W.~H. 1972, Journal of the Optical Society of America
  (1917-1983), 62, 55

\bibitem[{{Rudick} {et~al.}(2011){Rudick}, {Mihos}, \& {McBride}}]{Rudick2011}
{Rudick}, C.~S., {Mihos}, J.~C., \& {McBride}, C.~K. 2011, \apj, 732, 48

\bibitem[{{Sarzi} {et~al.}(2018){Sarzi}, {Iodice}, {Coccato}, {Corsini}, {de
  Zeeuw}, {Falc{\'o}n-Barroso}, {Gadotti}, {Lyubenova}, {McDermid}, {van de
  Ven}, {Fahrion}, {Pizzella}, \& {Zhu}}]{Sarzi2018}
{Sarzi}, M., {Iodice}, E., {Coccato}, L., {et~al.} 2018, \aap, 616, A121

\bibitem[{{Scharf} {et~al.}(2005){Scharf}, {Zurek}, \& {Bureau}}]{Scharf2005}
{Scharf}, C.~A., {Zurek}, D.~R., \& {Bureau}, M. 2005, \apj, 633, 154

\bibitem[{{Schipani} {et~al.}(2012){Schipani}, {Noethe}, {Arcidiacono},
  {Argomedo}, {Dall'Ora}, {D'Orsi}, {Farinato}, {Magrin}, {Marty}, {Ragazzoni},
  \& {Umbriaco}}]{Schipani2012}
{Schipani}, P., {Noethe}, L., {Arcidiacono}, C., {et~al.} 2012, Journal of the
  Optical Society of America A, 29, 1359

\bibitem[{{Seigar} {et~al.}(2007){Seigar}, {Graham}, \& {Jerjen}}]{Seigar2007}
{Seigar}, M.~S., {Graham}, A.~W., \& {Jerjen}, H. 2007, \mnras, 378, 1575

\bibitem[{{Serra} {et~al.}(2016){Serra}, {de Blok}, {Bryan}, {Colafrancesco},
  {Dettmar}, {Frank}, {Govoni}, {Jozsa}, {Kraan-Korteweg}, {Maccagni},
  {Loubser}, {Murgia}, {Oosterloo}, {Peletier}, {Pizzo}, {Richter},
  {Ramatsoku}, {Smith}, {Trager}, {van Gorkom}, \& {Verheijen}}]{Serra2016}
{Serra}, P., {de Blok}, W.~J.~G., {Bryan}, G.~L., {et~al.} 2016, in Proceedings
  of MeerKAT Science: On the Pathway to the SKA. 25-27 May, 2016 Stellenbosch,
  South Africa (MeerKAT2016). Online at <A
  href=``href=''>href=``https://pos.sissa.it/cgi-bin/reader/conf.cgi?confid=277</A>,
  id.8, 8

\bibitem[{{S{\'e}rsic}(1963)}]{Sersic}
{S{\'e}rsic}, J.~L. 1963, Boletin de la Asociacion Argentina de Astronomia La
  Plata Argentina, 6, 41

\bibitem[{{Sheardown} {et~al.}(2018){Sheardown}, {Roediger}, {Su}, {Kraft},
  {Fish}, {ZuHone}, {Forman}, {Jones}, {Churazov}, \& {Nulsen}}]{Sheardown2018}
{Sheardown}, A., {Roediger}, E., {Su}, Y., {et~al.} 2018, \apj, 865, 118

\bibitem[{{Spavone} {et~al.}(2017){Spavone}, {Capaccioli}, {Napolitano},
  {Iodice}, {Grado}, {Limatola}, {Cooper}, {Cantiello}, {Forbes}, {Paolillo},
  \& {Schipani}}]{Spavone2017}
{Spavone}, M., {Capaccioli}, M., {Napolitano}, N.~R., {et~al.} 2017, \aap, 603,
  A38

\bibitem[{{Spavone} {et~al.}(2018){Spavone}, {Iodice}, {Capaccioli}, {Bettoni},
  {Rampazzo}, {Brosch}, {Cantiello}, {Napolitano}, {Limatola}, {Grado}, \&
  {Schipani}}]{Spavone2018}
{Spavone}, M., {Iodice}, E., {Capaccioli}, M., {et~al.} 2018, \apj, 864, 149

\bibitem[{{Spiniello} {et~al.}(2018){Spiniello}, {Napolitano}, {Arnaboldi},
  {Tortora}, {Coccato}, {Capaccioli}, {Gerhard}, {Iodice}, {Spavone},
  {Cantiello}, {Peletier}, {Paolillo}, \& {Schipani}}]{Spiniello2018}
{Spiniello}, C., {Napolitano}, N.~R., {Arnaboldi}, M., {et~al.} 2018, \mnras,
  477, 1880

\bibitem[{{Tacchella} {et~al.}(2019){Tacchella}, {Diemer}, {Hernquist},
  {Genel}, {Marinacci}, {Nelson}, {Pillepich}, {Rodriguez-Gomez}, {Sales},
  {Springel}, \& {Vogelsberger}}]{Tacchella2019}
{Tacchella}, S., {Diemer}, B., {Hernquist}, L., {et~al.} 2019, \mnras, 487,
  5416

\bibitem[{{Thuan} \& {Kormendy}(1977)}]{Thuan1977}
{Thuan}, T.~X. \& {Kormendy}, J. 1977, \pasp, 89, 466

\bibitem[{{Trujillo} \& {Fliri}(2016)}]{Trujillo2016}
{Trujillo}, I. \& {Fliri}, J. 2016, \apj, 823, 123

\bibitem[{{van der Kruit} \& {Freeman}(2011)}]{vdKruit2011}
{van der Kruit}, P.~C. \& {Freeman}, K.~C. 2011, \araa, 49, 301

\bibitem[{{van Dokkum} {et~al.}(2014){van Dokkum}, {Abraham}, \&
  {Merritt}}]{vanDokkum2014}
{van Dokkum}, P.~G., {Abraham}, R., \& {Merritt}, A. 2014, \apjl, 782, L24

\bibitem[{{Veale} {et~al.}(2018){Veale}, {Ma}, {Greene}, {Thomas}, {Blakeslee},
  {Walsh}, \& {Ito}}]{Veale2018}
{Veale}, M., {Ma}, C.-P., {Greene}, J.~E., {et~al.} 2018, \mnras, 473, 5446

\bibitem[{{Venhola} {et~al.}(2018){Venhola}, {Peletier}, {Laurikainen}, {Salo},
  {Iodice}, {Mieske}, {Hilker}, {Wittmann}, {Lisker}, {Paolillo}, {Cantiello},
  {Janz}, {Spavone}, {D'Abrusco}, {Ven}, {Napolitano}, {Kleijn}, {Maddox},
  {Capaccioli}, {Grado}, {Valentijn}, {Falc{\'o}n-Barroso}, \&
  {Limatola}}]{Venhola2018}
{Venhola}, A., {Peletier}, R., {Laurikainen}, E., {et~al.} 2018, \aap, 620,
  A165

\bibitem[{{Venhola} {et~al.}(2017){Venhola}, {Peletier}, {Laurikainen}, {Salo},
  {Lisker}, {Iodice}, {Capaccioli}, {Kleijn}, {Valentijn}, {Mieske}, {Hilker},
  {Wittmann}, {van de Ven}, {Grado}, {Spavone}, {Cantiello}, {Napolitano},
  {Paolillo}, \& {Falc{\'o}n-Barroso}}]{Venhola2017}
{Venhola}, A., {Peletier}, R., {Laurikainen}, E., {et~al.} 2017, \aap, 608,
  A142

\bibitem[{{Zabel} {et~al.}(2019){Zabel}, {Davis}, {Smith}, {Maddox}, {Bendo},
  {Peletier}, {Iodice}, {Venhola}, {Baes}, {Davies}, {de Looze}, {Gomez},
  {Grossi}, {Kenney}, {Serra}, {van de Voort}, {Vlahakis}, \&
  {Young}}]{Zabel2019}
{Zabel}, N., {Davis}, T.~A., {Smith}, M. W.~L., {et~al.} 2019, \mnras, 483,
  2251

\end{thebibliography}

%
%
\begin{appendix}
\section{Results of multi-component fits}\label{res_fit}

\begin{figure*}[htb]
    \begin{minipage}[t]{.45\textwidth}
        \centering
        \includegraphics[width=\textwidth]{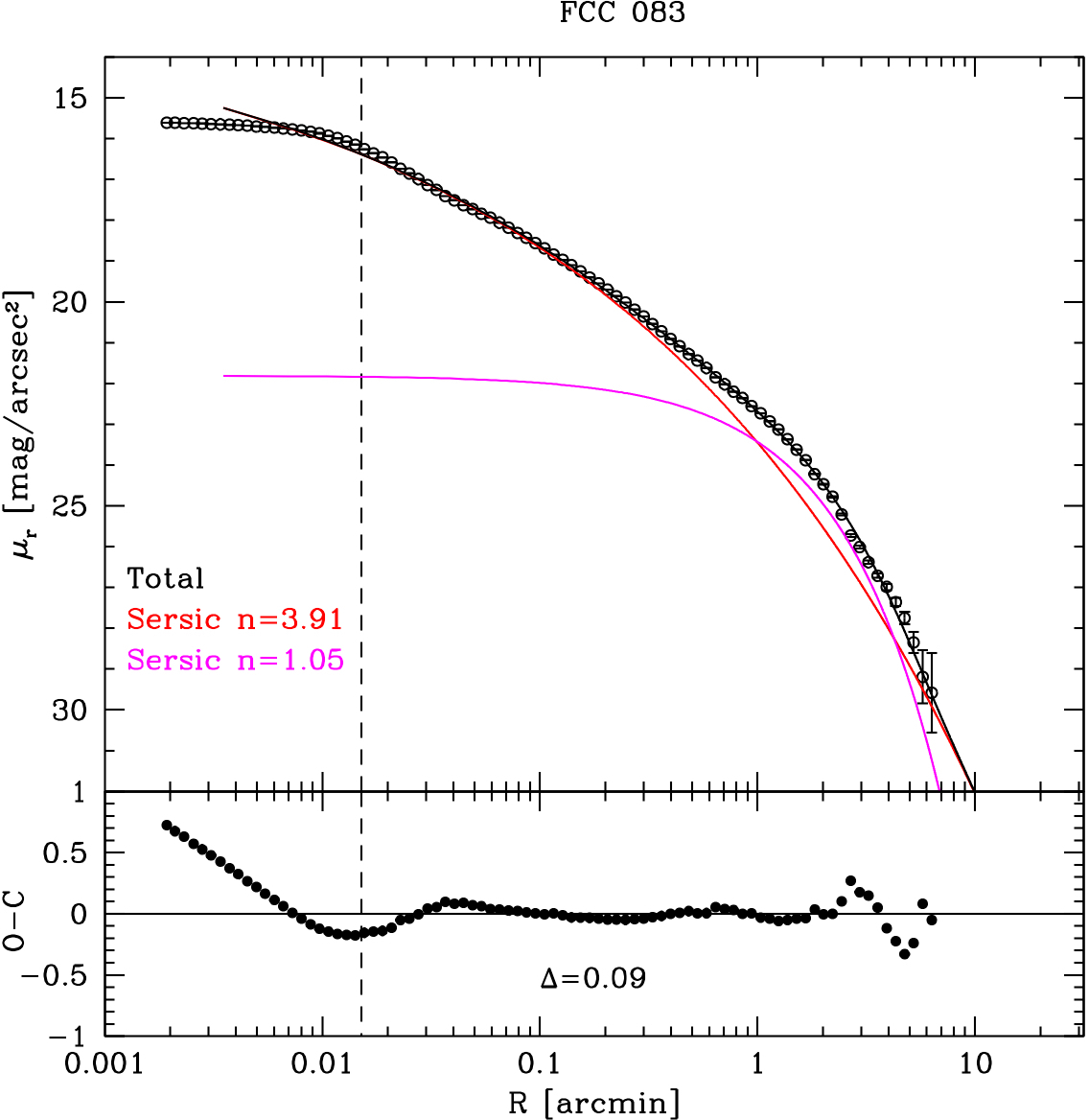}
    \end{minipage}
    \hfill
    \begin{minipage}[t]{.45\textwidth}
        \centering
        \includegraphics[width=\textwidth]{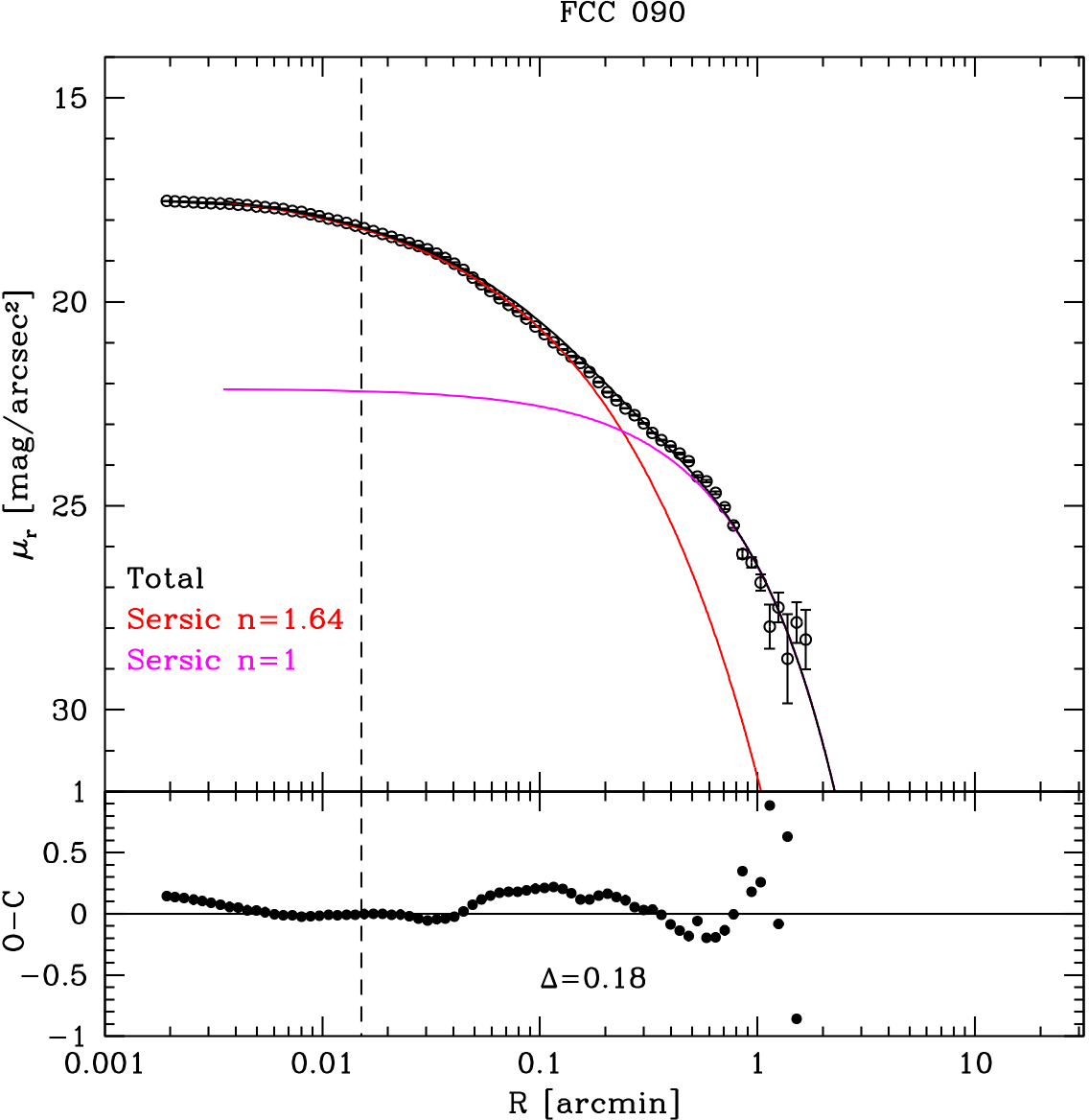}
    \end{minipage}  

    \begin{minipage}[t]{.45\textwidth}
        \centering
        \includegraphics[width=\textwidth]{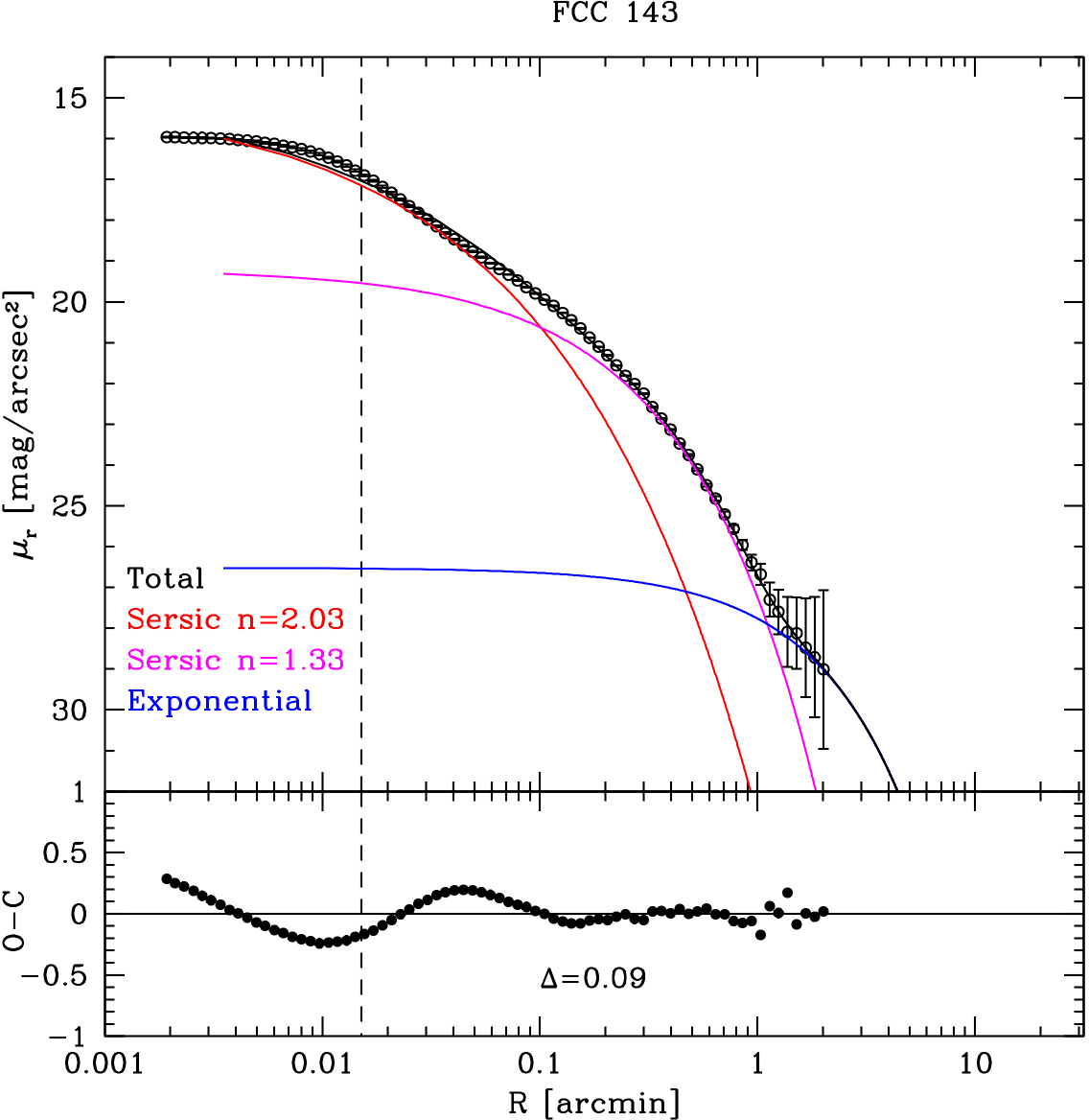}
    \end{minipage}
    \hfill
    \begin{minipage}[t]{.45\textwidth}
        \centering
        \includegraphics[width=\textwidth]{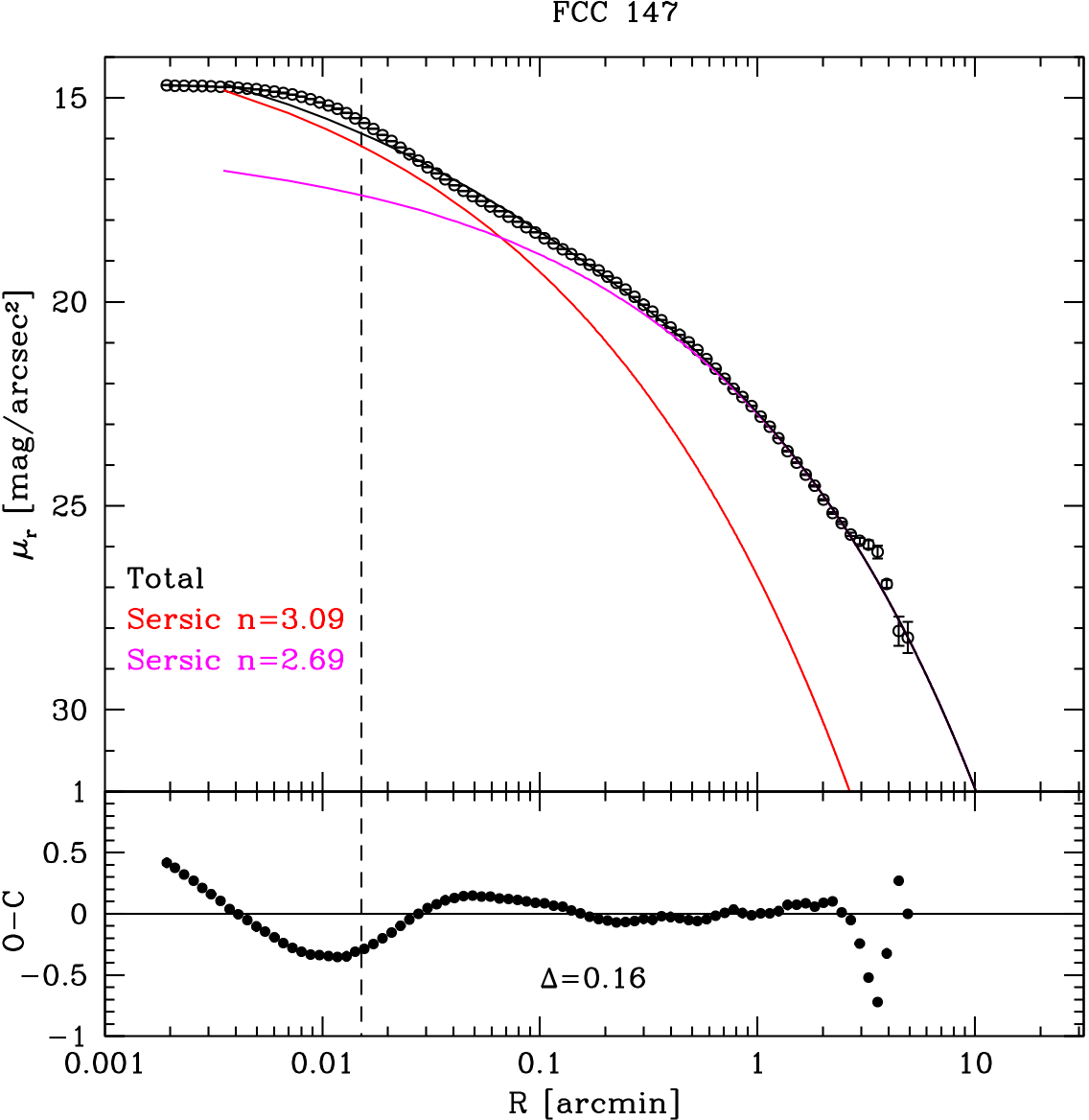}
    \end{minipage}  
    
    \begin{minipage}[t]{.45\textwidth}
        \centering
        \includegraphics[width=\textwidth]{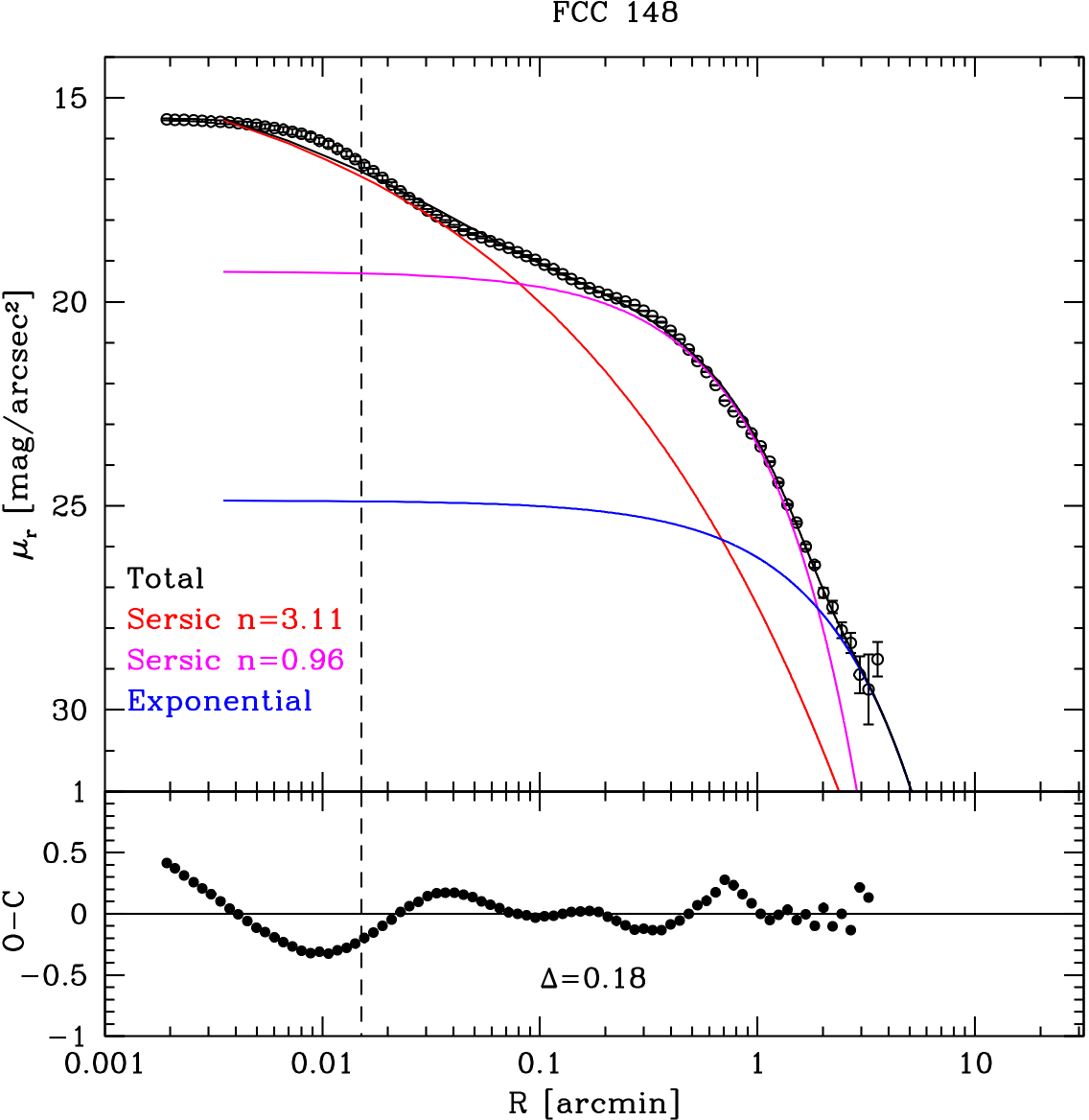}
    \end{minipage}
    \hfill
    \begin{minipage}[t]{.45\textwidth}
        \centering
        \includegraphics[width=\textwidth]{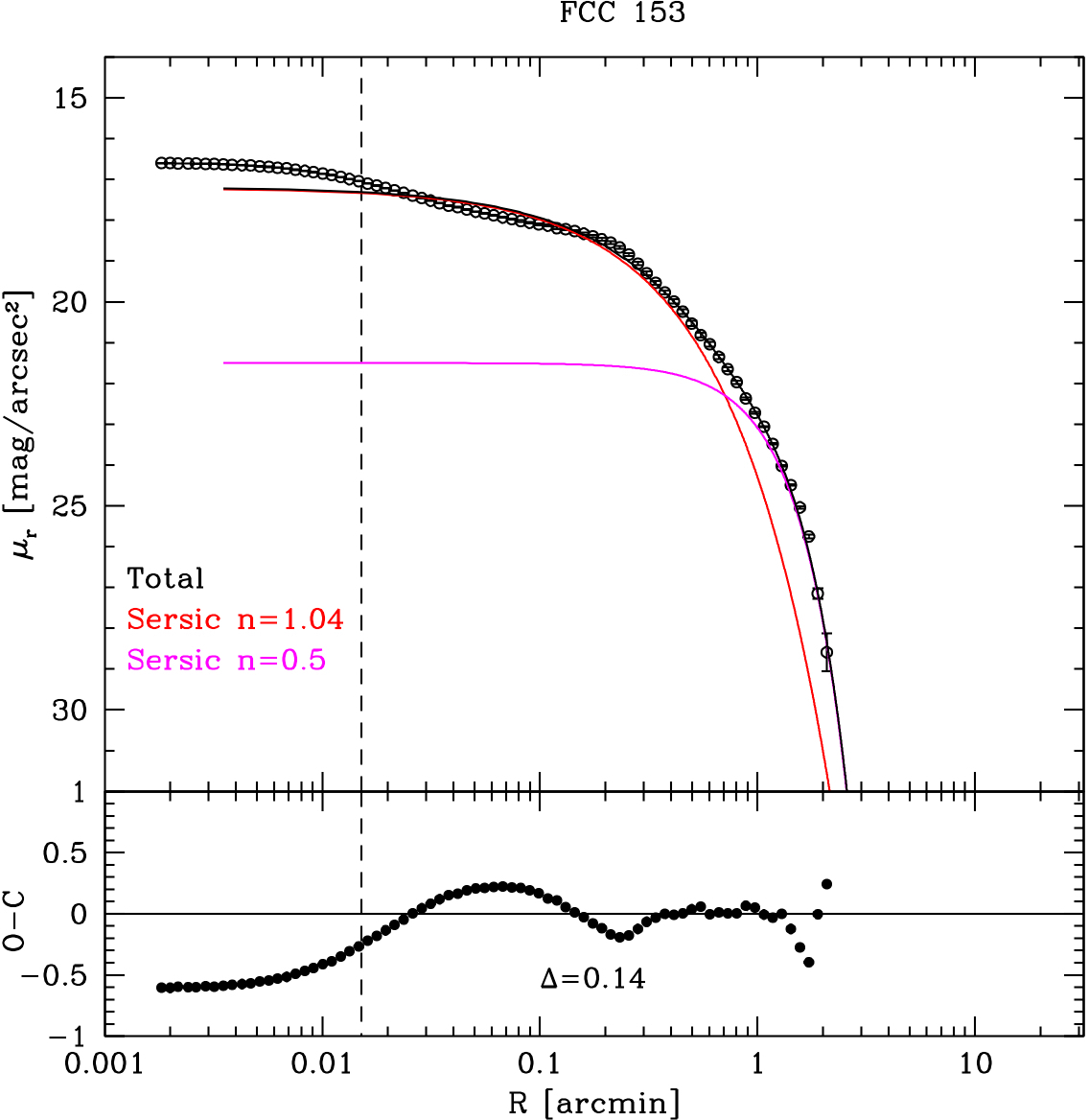}
    \end{minipage}
    \end{figure*}
    
    \begin{figure*}[htb]
        \begin{minipage}[t]{.45\textwidth}
        \centering
        \includegraphics[width=\textwidth]{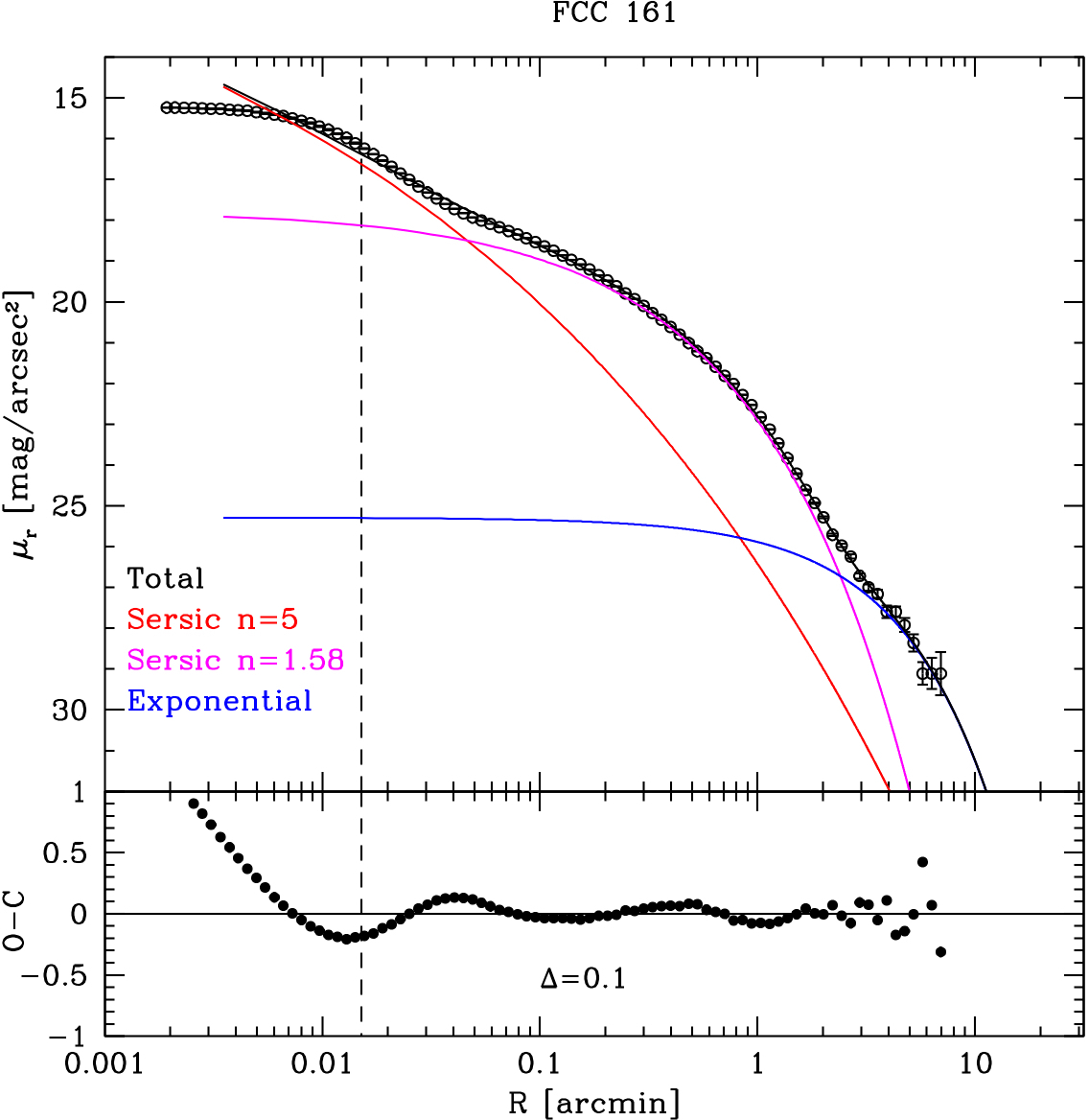}
    \end{minipage}
    \hfill
    \begin{minipage}[t]{.45\textwidth}
        \centering
        \includegraphics[width=\textwidth]{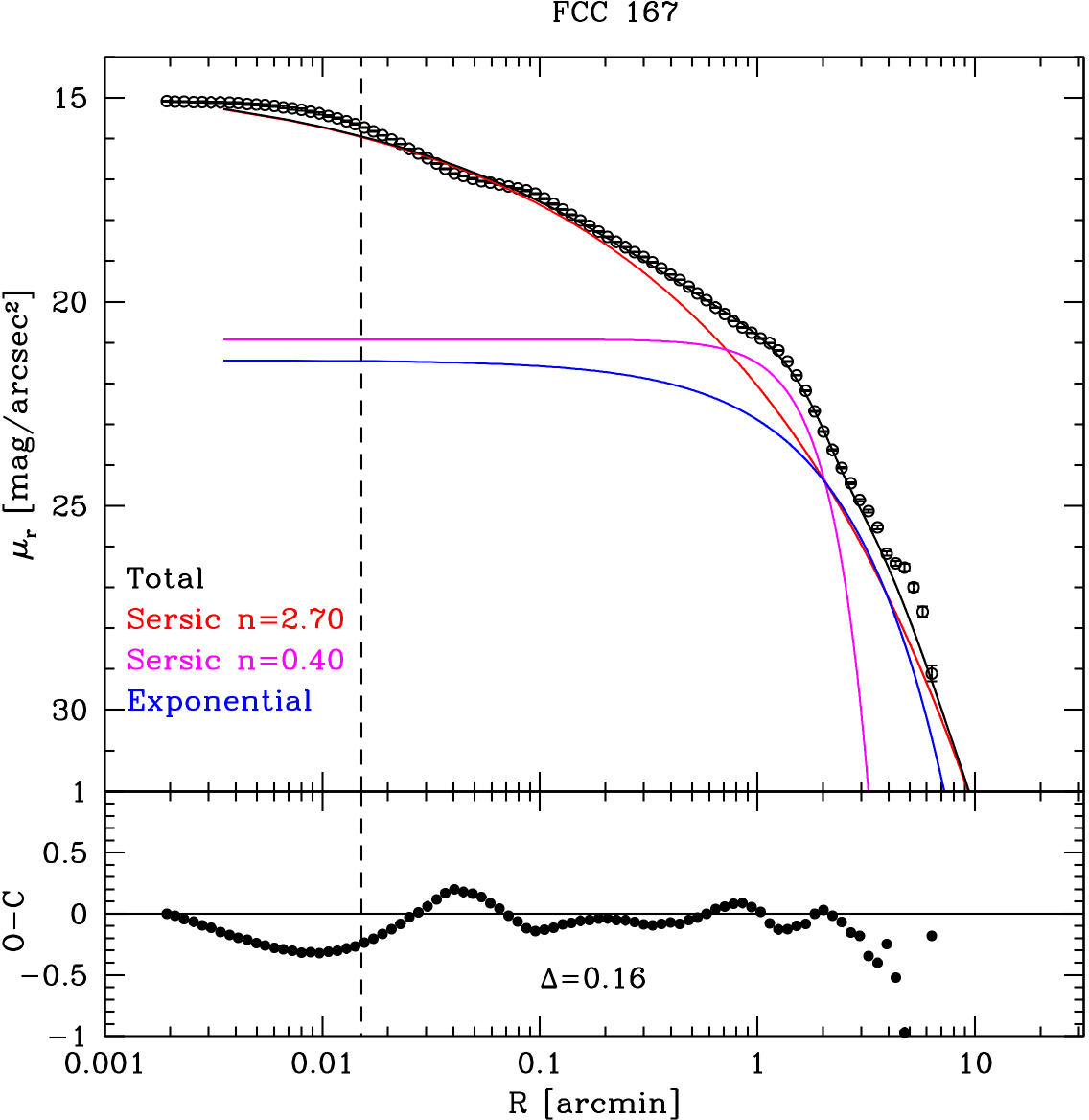}
    \end{minipage}

    \begin{minipage}[t]{.45\textwidth}
        \centering
        \includegraphics[width=\textwidth]{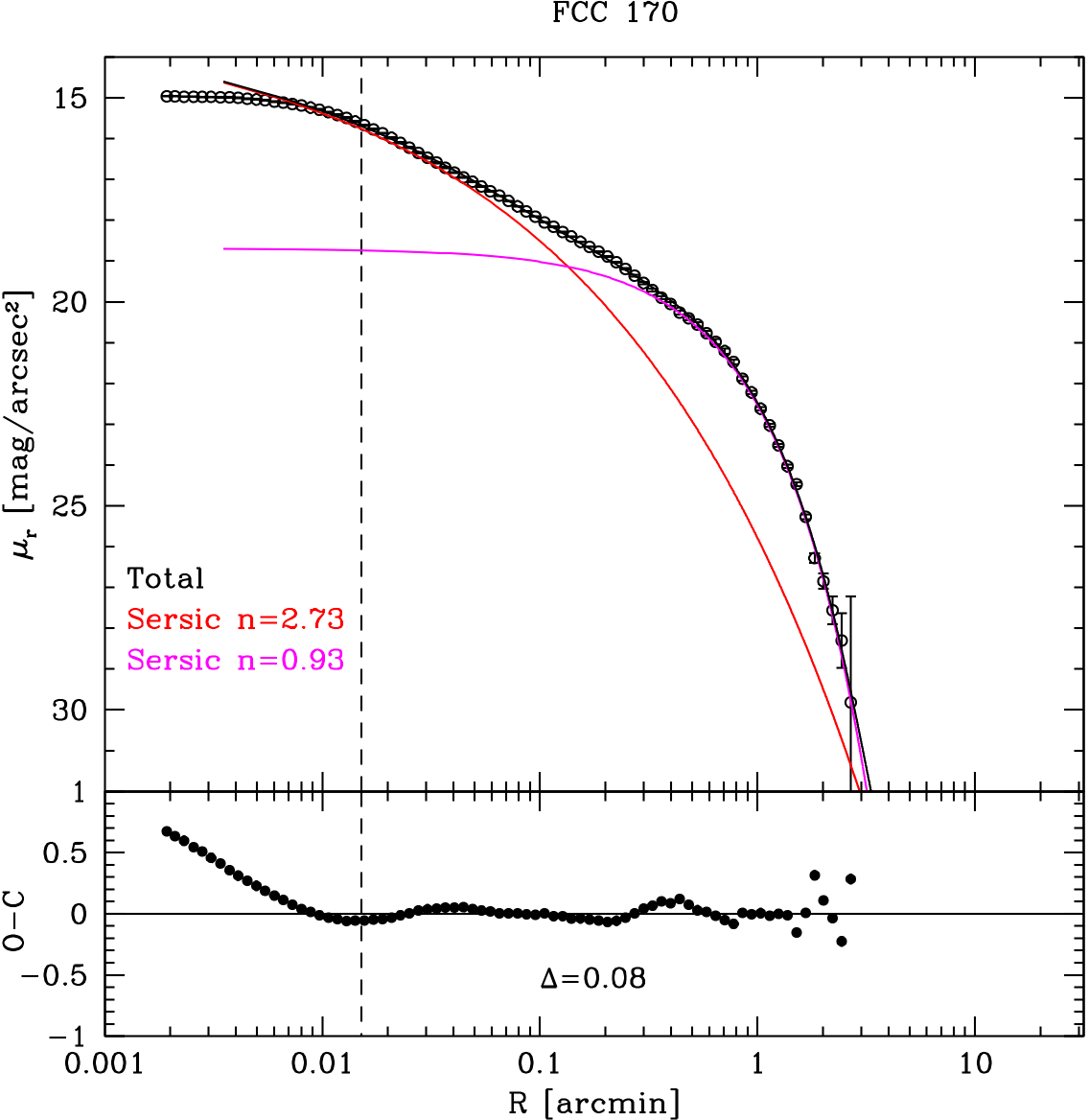}
    \end{minipage}
    \hfill
    \begin{minipage}[t]{.45\textwidth}
        \centering
        \includegraphics[width=\textwidth]{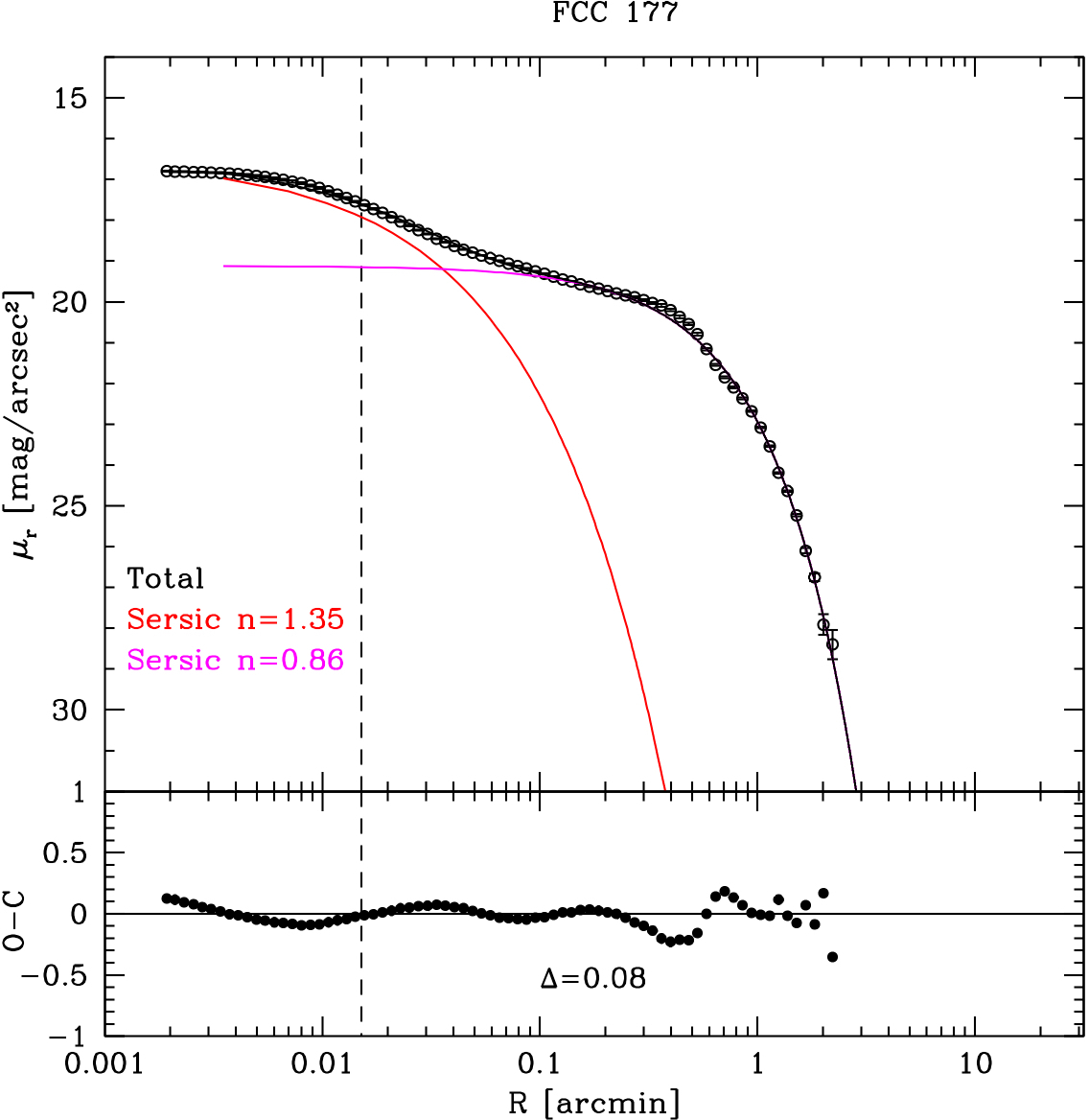}
    \end{minipage}
    
    \begin{minipage}[t]{.45\textwidth}
        \centering
        \includegraphics[width=\textwidth]{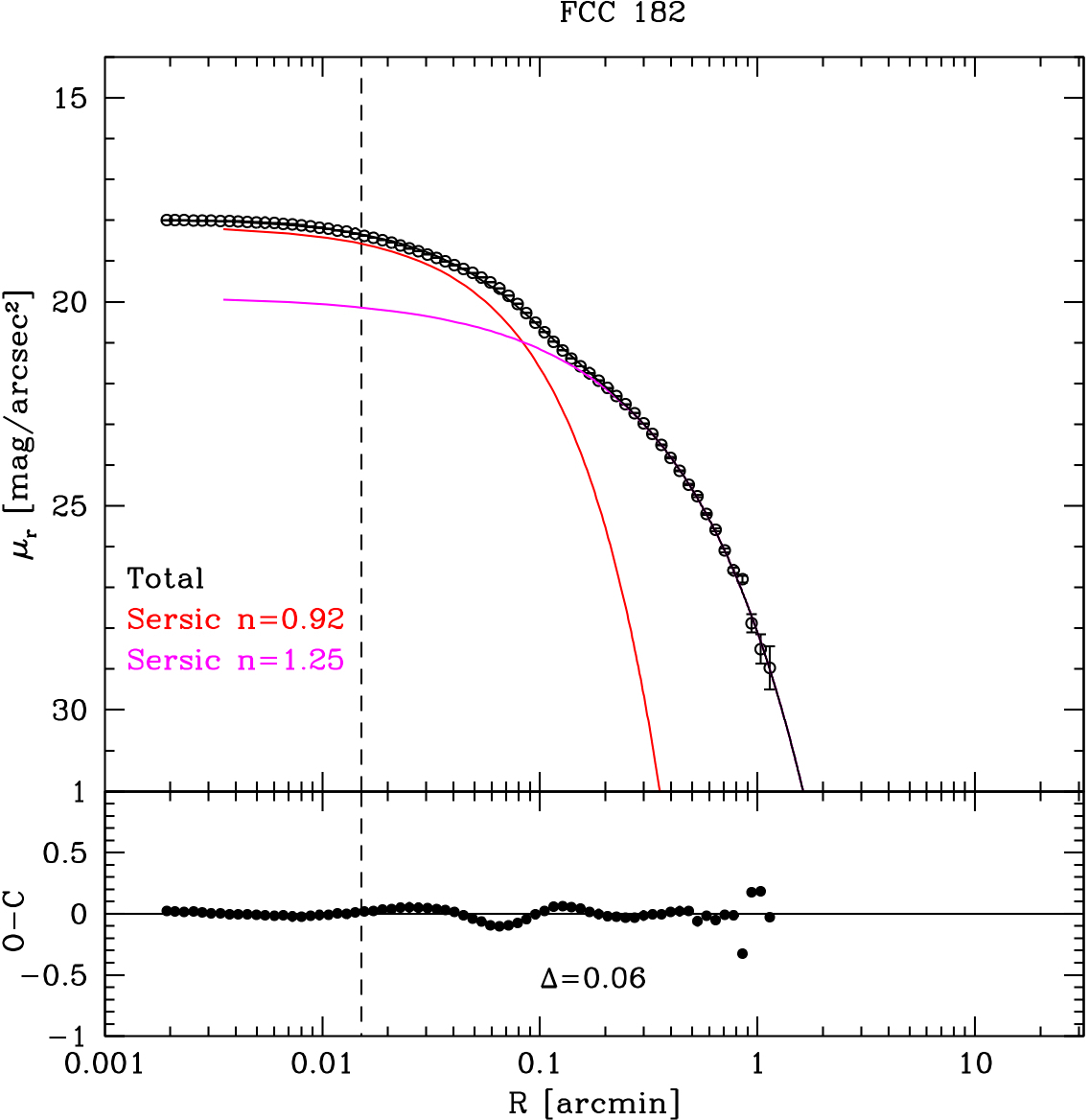}
    \end{minipage}
    \hfill
    \begin{minipage}[t]{.45\textwidth}
        \centering
        \includegraphics[width=\textwidth]{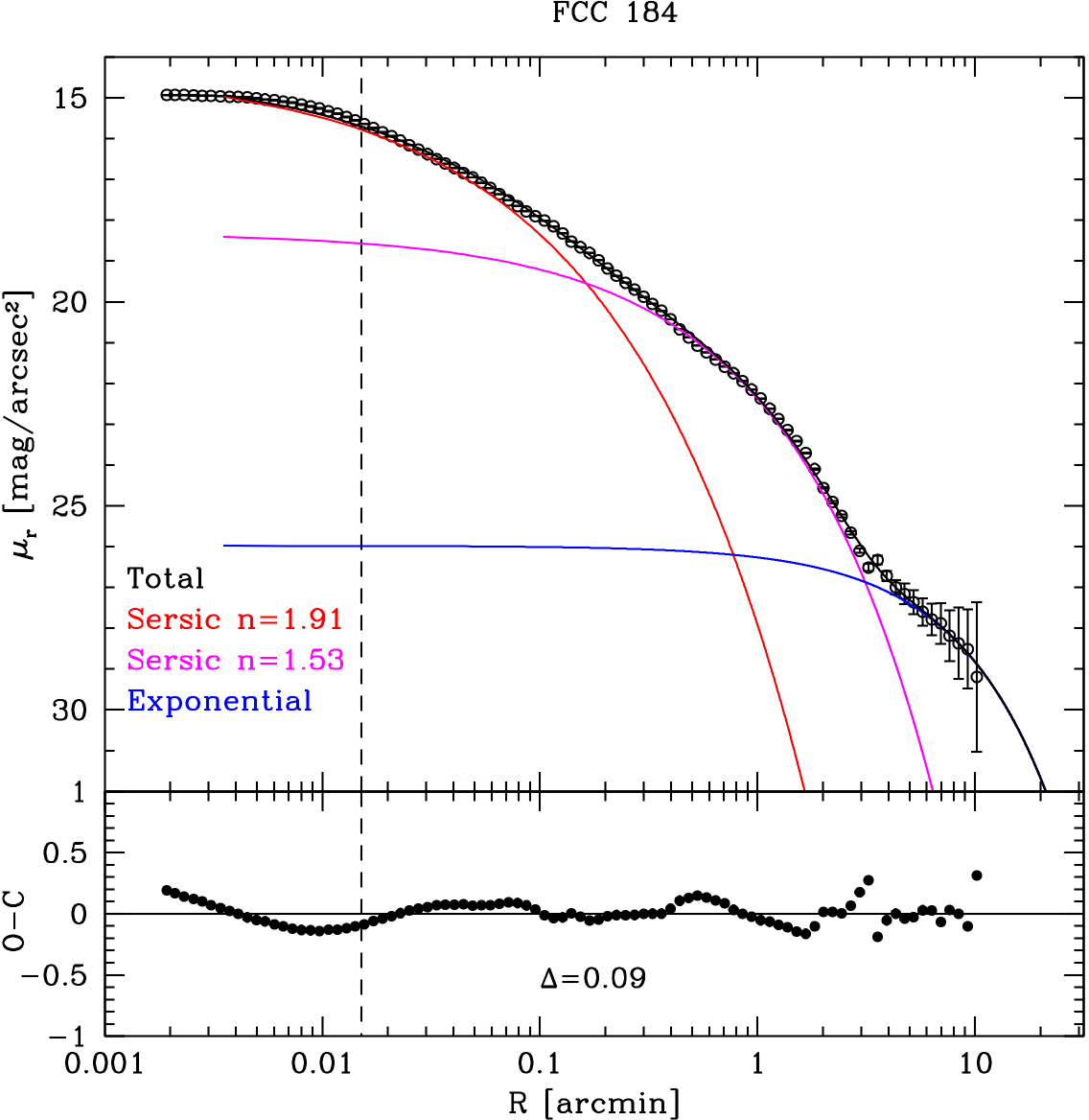}
    \end{minipage}
\end{figure*}
    
\begin{figure*}[htb]   
    \begin{minipage}[t]{.45\textwidth}
        \centering
        \includegraphics[width=\textwidth]{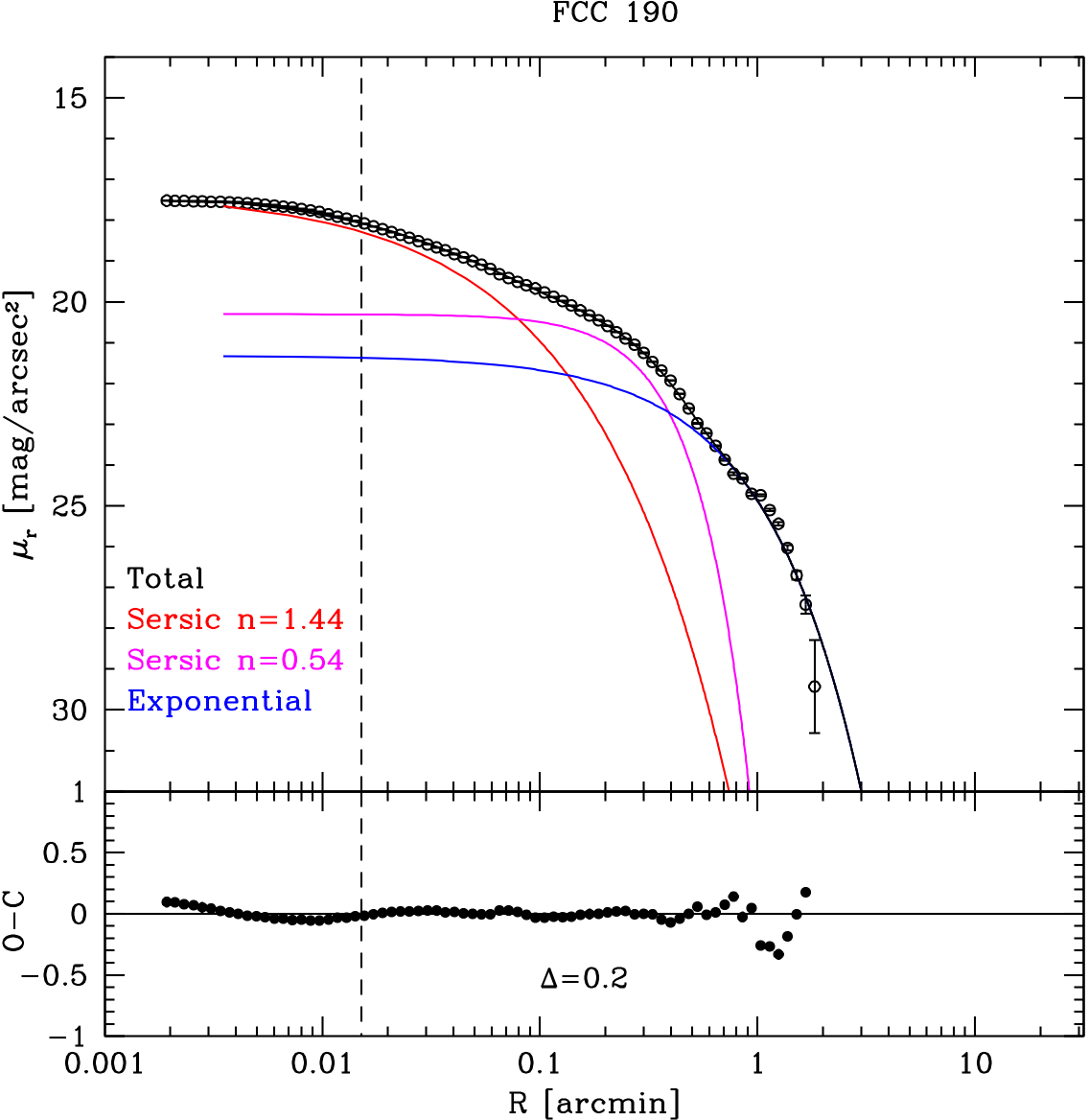}
    \end{minipage}
    \hfill
    \begin{minipage}[t]{.45\textwidth}
        \centering
        \includegraphics[width=\textwidth]{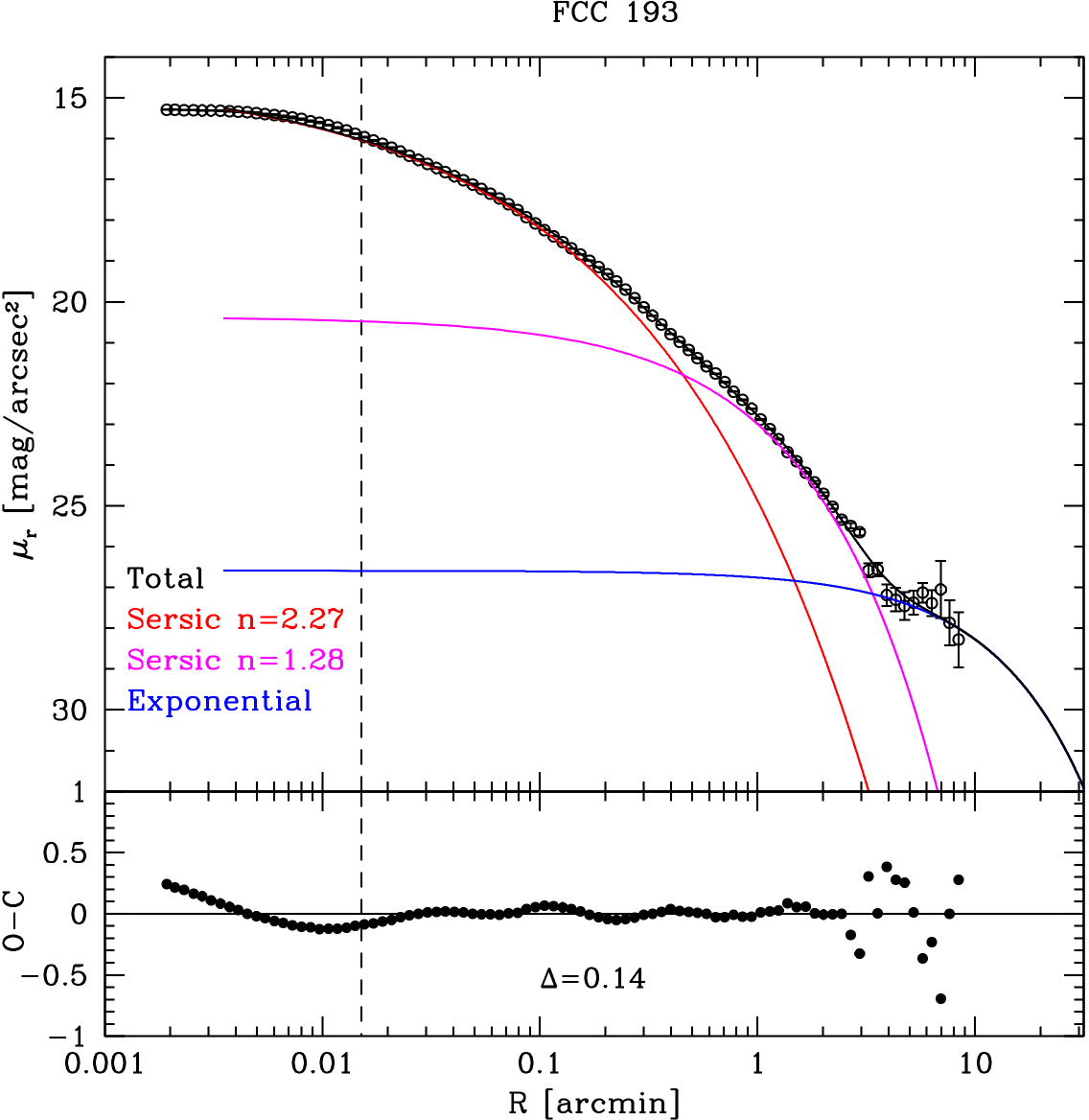}
    \end{minipage}
    
    \begin{minipage}[t]{.45\textwidth}
        \centering
        \includegraphics[width=\textwidth]{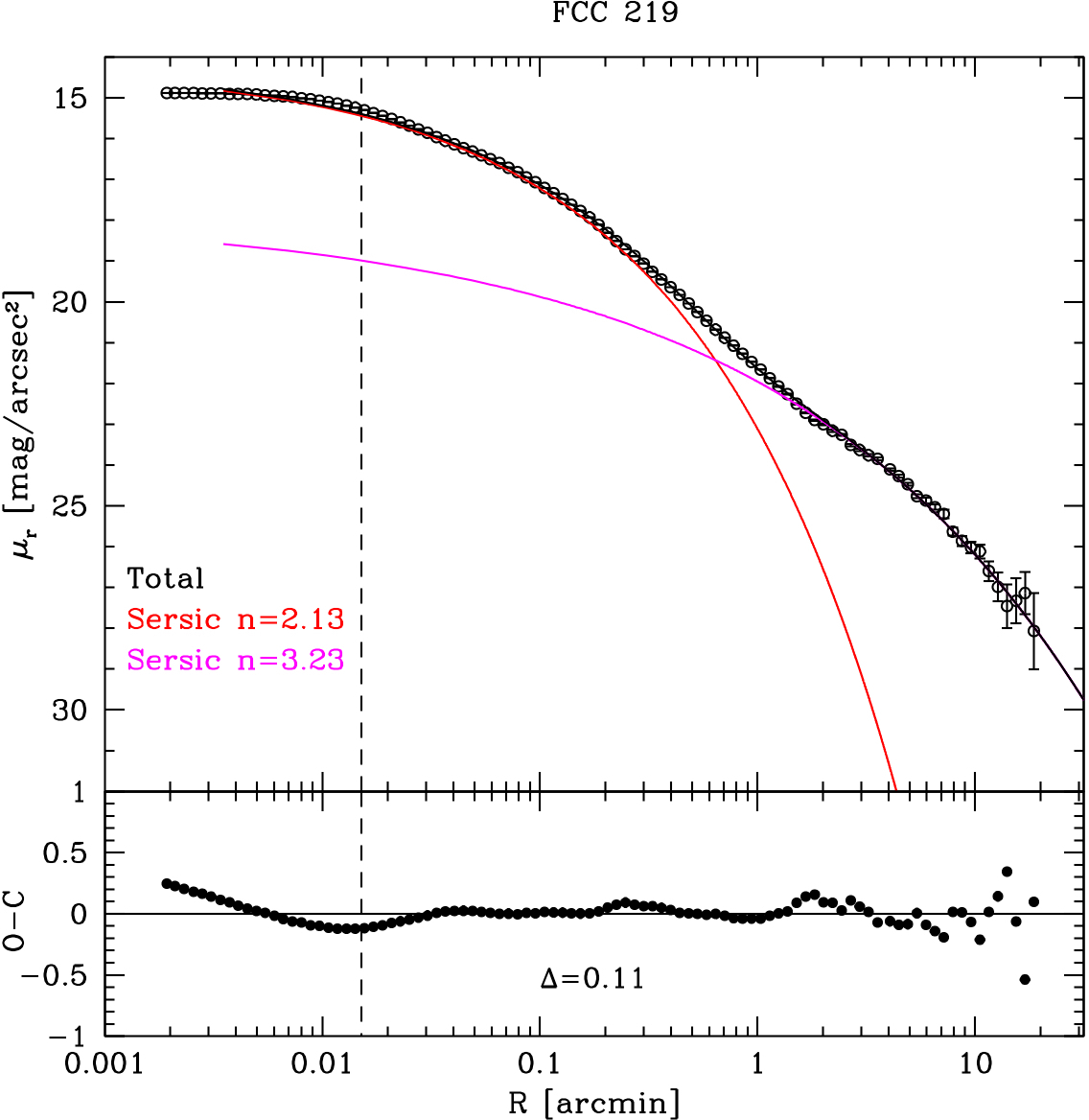}
    \end{minipage}
    \hfill
    \begin{minipage}[t]{.45\textwidth}
        \centering
        \includegraphics[width=\textwidth]{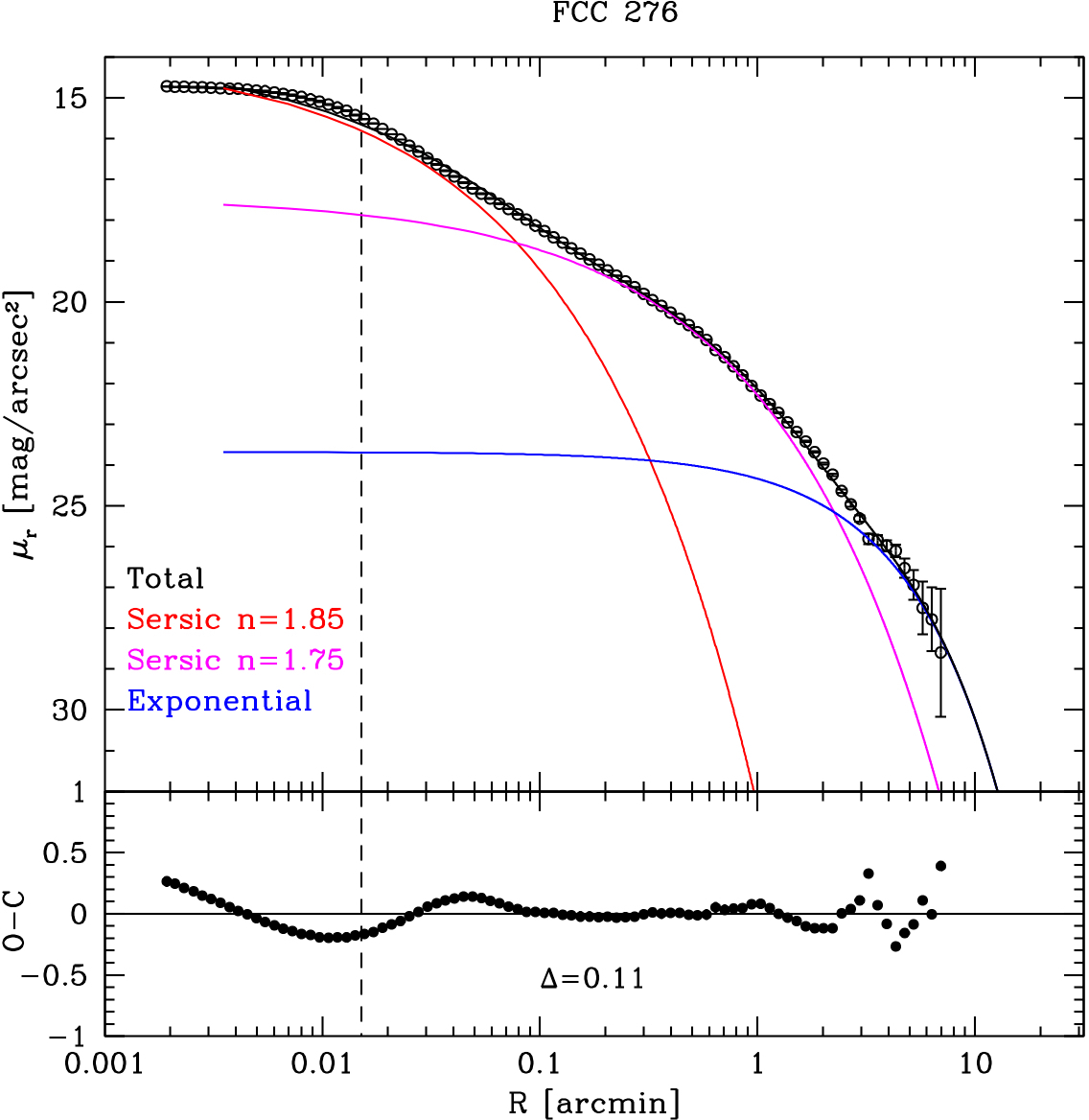}
    \end{minipage}
    
        \begin{minipage}[t]{.45\textwidth}
        \centering
        \includegraphics[width=\textwidth]{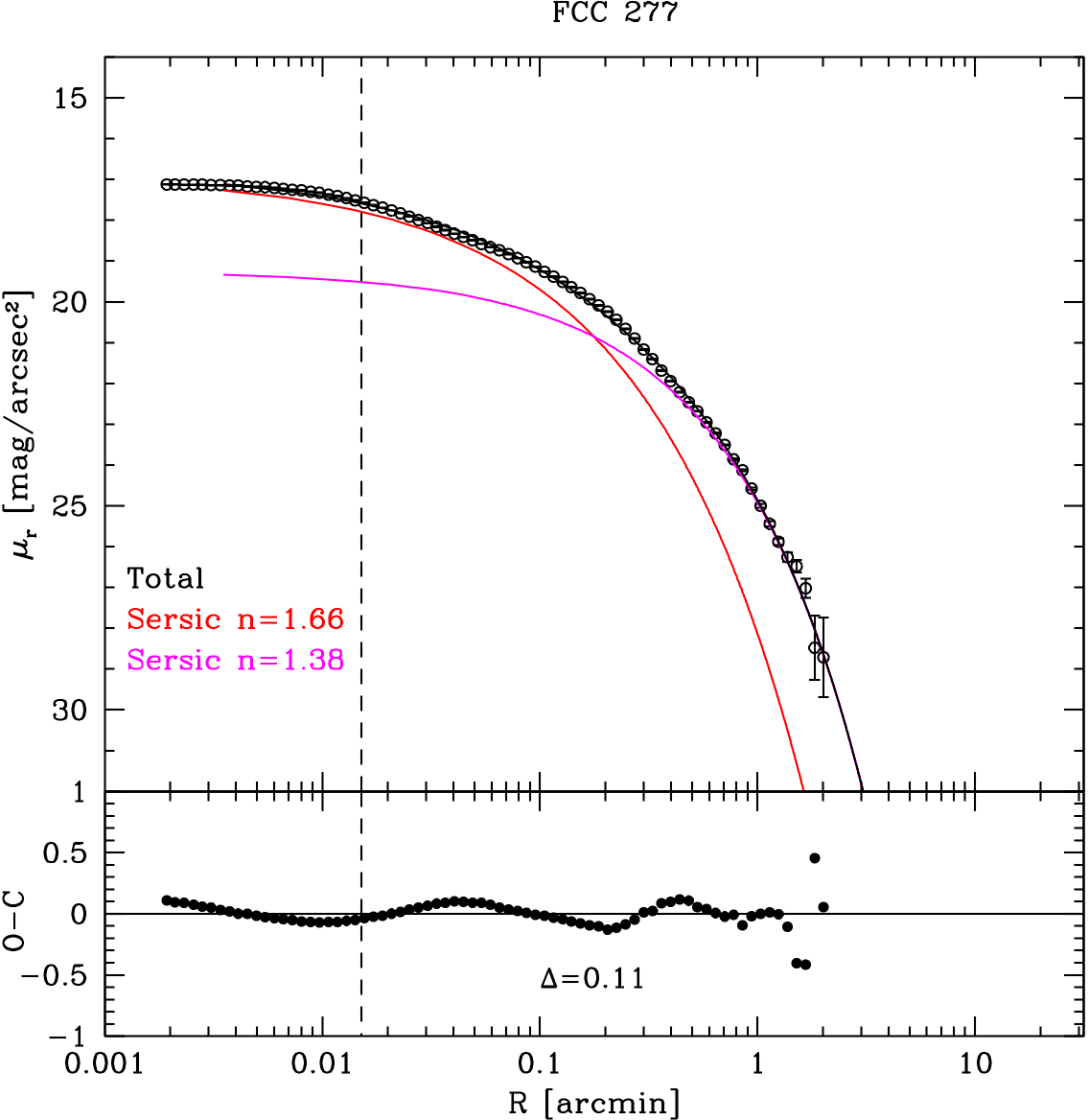}
    \end{minipage}
    \hfill
    \begin{minipage}[t]{.45\textwidth}
        \centering
        \includegraphics[width=\textwidth]{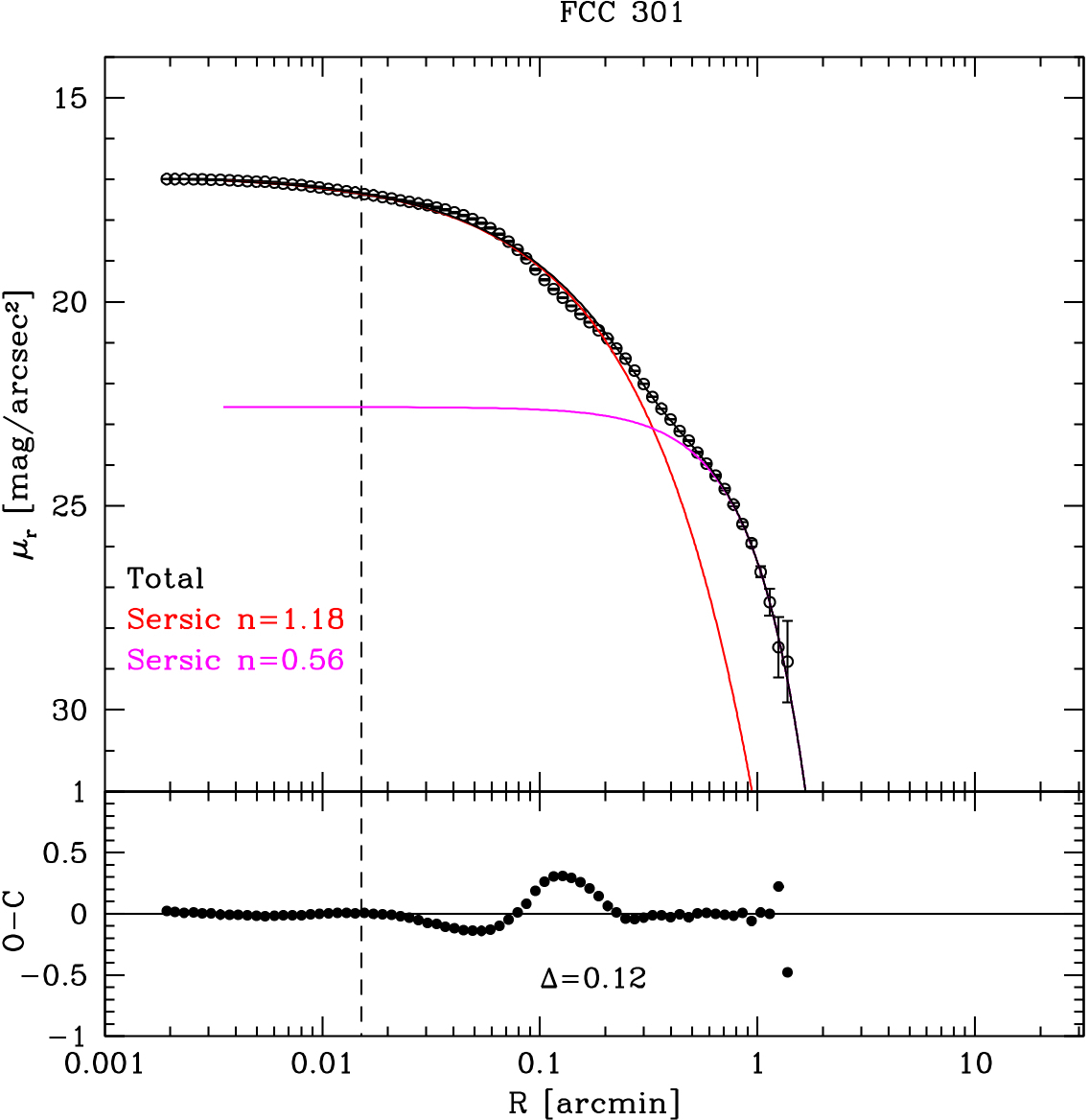}
    \end{minipage}
 \end{figure*}
 \begin{figure*}[htb]   
         \centering

    \begin{minipage}[t]{.45\textwidth}
        \includegraphics[width=\textwidth]{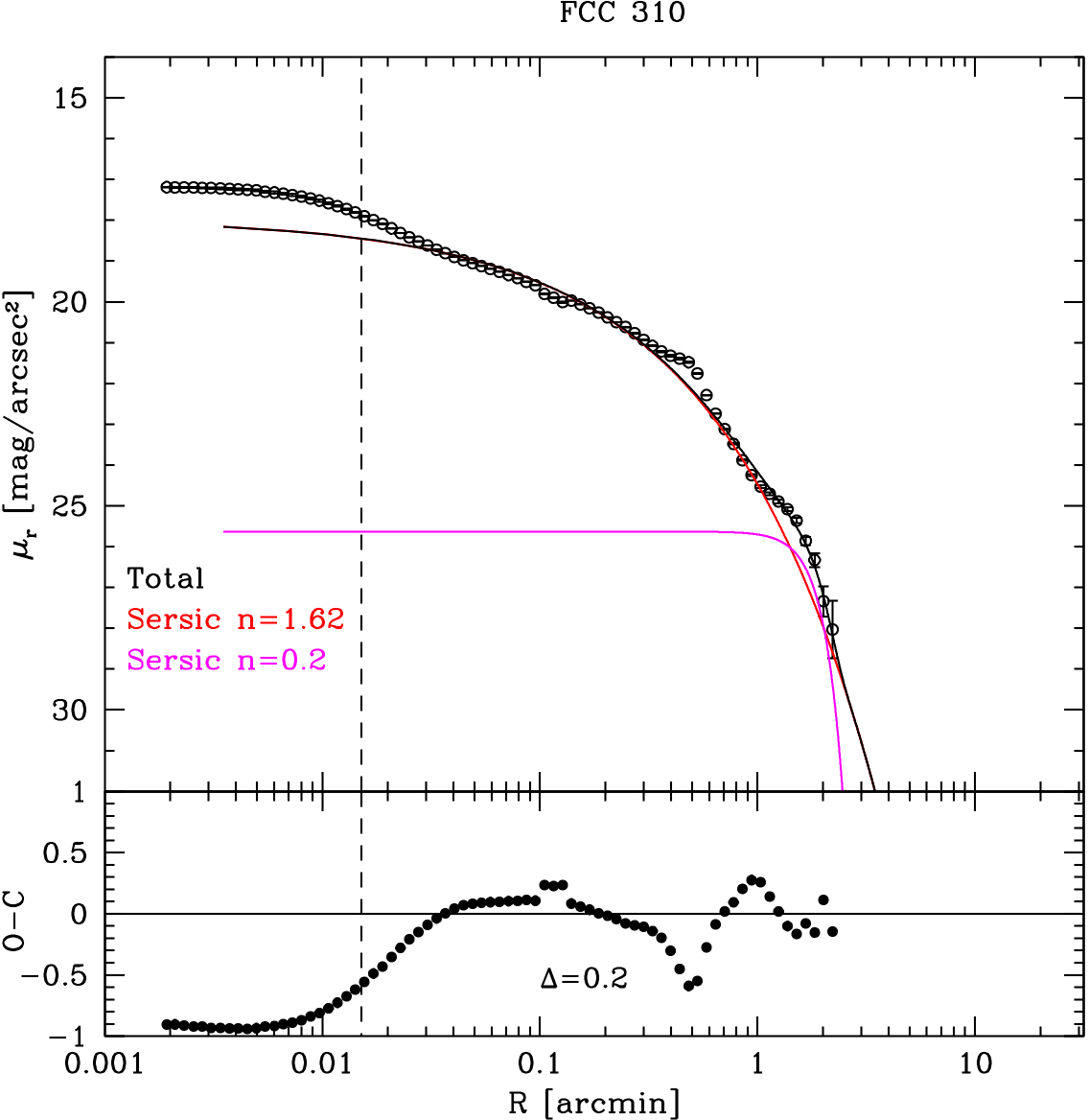}
    \end{minipage}
    
    \caption{{\it Top Panels}: Deconvolved VST r-band profiles of ETGs in FDS plotted on a logarithmic scale. The
blue line indicates a fit to the outer regions. The red and magenta
lines indicate a fit to the inner and middle regions with a S{\'e}rsic
profile, and the black line indicates the sum of the components from
each fit. The dotted line indicates the core of the galaxy, which was excluded in the fit. {\it Bottom Panels}: $\Delta$ rms scatter (see text for details).}\label{fig:fit_all}
    \end{figure*} 

\end{appendix}

\end{document}